\documentclass[aps,prd,superscriptaddress,groupedaddress,nofootinbib,nobibnotes]{revtex4}

\usepackage{graphicx}
\usepackage{dcolumn}
\usepackage{bm}
\usepackage{amssymb}
\usepackage{epstopdf}
\usepackage{amsmath}
\usepackage{amsfonts}
\usepackage{color}
\usepackage{mathrsfs}
\usepackage{wick}
\usepackage{feynmp}
\usepackage{simplewick}

\newcommand{\pd}{\partial}

\newcommand{\be}{\begin{equation}}
\newcommand{\ee}{\end{equation}}
\newcommand{\ba}{\begin{eqnarray}}
\newcommand{\ea}{\end{eqnarray}}
\newcommand{\nn}{\nonumber}
\newcommand{\barr}{\begin{array}}
\newcommand{\earr}{\end{array}}

\newcommand{\bigoh}{\mathcal{O}}

\newcommand\lsim{\mathrel{\rlap{\lower4pt\hbox{\hskip1pt$\sim$}}
        \raise1pt\hbox{$<$}}}
\newcommand\gsim{\mathrel{\rlap{\lower4pt\hbox{\hskip1pt$\sim$}}
        \raise1pt\hbox{$>$}}}

\def\threej#1#2#3#4#5#6{\left( \begin{array}{ccc} #1 & #2 & #3 \\ #4 & #5 & #6 \end{array} \right) }

\def\hkappa{{\widehat\kappa}}
\def\Var{\mbox{Var}}
\def\k{{\bf k}}
\def\q{{\bf q}}
\def\hk{\widehat{\bf k}}

\def\r{{\bf r}}
\def\hr{\widehat{\bf r}}
\def\n{\widehat{\bf n}}
\def\hC{\widehat C}
\def\tT{\widetilde T}
\def\Cov{\mbox{Cov}}
\def\E{\widehat {\mathcal E}}

\def\ellmax{\ell_{\rm max}}

\def\x{{\bf x}}
\def\fnlloc{f_{NL}^{\rm loc}}
\def\fnleq{f_{NL}^{\rm eq}}
\def\fnlorth{f_{NL}^{\rm orth}}
\def\gnlloc{g_{NL}^{\rm loc}}
\def\gnldotpi4{g_{NL}^{\dot\sigma^4}}
\def\gnldpi4{g_{NL}^{(\partial\sigma)^4}}
\def\gnlB{g_{NL}^{\dot\sigma^2 (\partial\sigma)^2}}
\def\taunl{\tau_{NL}}
\def\Nfact{N_{\rm fact}}
\def\fsky{f_{\rm sky}}

\def\ta{\tilde a}
\def\tb{\tilde b}
\def\ts{\tilde s}
\def\tn{\tilde n}
\def\Nmc{N_{\rm mc}}
\def\Ntr{N_{\rm tr}}
\def\Nin{N_{\rm in}}
\def\Nout{N_{\rm out}}

\def\sraise{\;\raise1.0pt\hbox{$'$}\hskip-6pt\partial}
\def\slower{\;\overline{\raise1.0pt\hbox{$'$}\hskip-6pt\partial}}
\def\e{{\hat e}}
\def\hF{{\hat F}}
\def\hSigma{{\hat\Sigma}}
\def\fc{{\rm f.c.}}

\begin{document}

% feynmp
\unitlength = 1mm

% for simplewick appendix
\def\pT{\phantom{T}}

\title{Optimal analysis of the CMB trispectrum}

\author{Kendrick M. Smith}
\affiliation{Perimeter Institute for Theoretical Physics, Waterloo, ON N2L 2Y5, Canada}

\author{Leonardo Senatore}
\affiliation{Stanford Institute for Theoretical Physics, Stanford University, Stanford, CA 94305, USA}
\affiliation{Kavli Institute for Particle Astrophysics and Cosmology, SLAC and Stanford University, Menlo Park, CA 94025, USA}

\author{Matias Zaldarriaga}
\affiliation{Institute for Advanced Study, Einstein Drive, Princeton, NJ 08540, USA}

\date{\today}

\begin{abstract}
We develop a general framework for data analysis and phenomenology of the CMB four-point
function or trispectrum.  To lowest order in the derivative expansion, the inflationary action admits
three quartic operators consistent with symmetry: $\dot\sigma^4$, $\dot\sigma^2 (\partial\sigma^2)$,
and $(\partial\sigma)^4$.
In single field inflation, only the first of these operators can be the leading non-Gaussian signal.
A Fisher matrix analysis shows that there is one near-degeneracy
among the three CMB trispectra, so we parameterize the trispectrum with two
coefficients $\gnldotpi4$ and $\gnldpi4$, in addition to the coefficient $\gnlloc$ of
$\zeta^3$-type local non-Gaussianity.  This three-parameter space is analogous to the
parameter space $(\fnlloc, \fnleq, \fnlorth)$ commonly used to parameterize the CMB
three-point function. We next turn to data analysis and show how to represent these
trispectra in a factorizable form which leads to computationally fast
operations such as evaluating a CMB estimator or simulating a non-Gaussian CMB.
We discuss practical issues in CMB analysis pipelines, and perform an optimal
analysis of WMAP data.
Our minimum-variance estimates are $\gnlloc = (-3.80 \pm 2.19) \times 10^5$,
$\gnldotpi4 = (-3.20 \pm 3.09) \times 10^6$,
and $\gnldpi4 = (-10.8 \pm 6.33) \times 10^5$
after correcting for the effects of CMB lensing.
No evidence of a nonzero inflationary four-point function is seen.
\end{abstract}
\pacs{}

\maketitle

\section{Introduction and Main Results}

Uncovering the nature of inflation is one of the most important open questions in our current cosmological model.
Non-Gaussianity of the primordial density perturbations probes the interaction structure of the inflationary Lagrangian.
Since interactions contain most of the information on the dynamics of the fields,
the search for primordial non-Gaussianity has played a central role in constraining the physics of inflation.

So far, the search for non-Gaussianity has been focused mainly on the bispectrum, or 3-point function.
Limits on the inflationary 3-point function have been obtained following two different approaches. 
The first is based on providing templates for 3-point functions that are matched against the data,
while the second approach attempts to reconstruct a generic 3-point function from the data. 
The first method has the advantage that it can be restricted to theoretically motivated models 
over which one can perform an optimal analysis, but the disadvantage of potentially missing a signal
in the data simply because it was not looked for.
It has been used to search for the (now) well-known
local~\cite{Lyth:2002my,Zaldarriaga:2003my,Komatsu:2003iq,Senatore:2010wk}, 
equilateral~\cite{Creminelli:2005hu}, 
and orthogonal~\cite{Senatore:2009gt} template bispectra,
plus some very recently identified higher-derivative bispectra~\cite{Behbahani:2014upa}.
The second approach (e.g.~\cite{Fergusson:2009nv}) has the advantage of being sensitive to any potential signal,
but the disadvantage that significance of a signal can be diluted away as many independent shapes are matched to the data.

At present, the most constraining search for non-Gaussianity is provided by Planck~\cite{Ade:2013ydc}, 
which finds no evidence of non-Gaussianity. 
While this analysis is a huge observational achievement, 
it should be stressed that from a particle physics point of view the limit is still rather weak.
The skewness of the distribution of the primordial fluctuations 
is constrained to be smaller than about $10^{-3}$. 
This constrains inflation to be more or less as interacting 
as the electron in quantum electrodynamics, or as the pion at energies of order of its mass. It would be clearly very interesting 
to further constrain the level of non-Gaussianity by one or two orders of magnitude, a sensitivity that the recently 
developed effective field theory of large scale structure~\cite{Carrasco:2012cv} has shown the potential to achieve with surveys 
in the next decade.

The observational interest in non-Gaussianity is not just due to the fact that it is related to the dynamics of the theory. 
It additionally represents a very non-trivial signal. Because of translation and rotation invariance, the two point function 
of the primordial density perturbation is described by a scalar function of the modulus of the wavenumber $k$. 
Once we impose approximate scale invariance, this function can be described by a number, the amplitude, and another number, 
the slight deviation from scale invariance, the tilt. Instead, after assuming the same symmetries 
({\it i.e}. translation, rotation and scale invariance), the bispectrum is described by a scalar function of  
two scalar variables~\cite{Babich:2004gb}. We pass from one single number to a full function of two variables. 
Clearly, a detection of such a signal would be an extremely non-trivial signature in the sky. It is the same information that 
describes 2-to-2 scattering in a collider. When we pass to the trispectrum, the same symmetries make the trispectrum a scalar 
function of five variables. This is a fantastically non-trivial function that if we were so lucky to be able to see it in the sky, 
it would offer tremendous constraining power on the physics of inflation.

The analysis of the trispectrum, or four-point function, of the primordial density perturbation is less developed than the bispectrum.
We briefly summarize existing results in the literature.
The so-called $\gnlloc$ trispectrum is generated if
the primordial curvature perturbation $\zeta(\vec x)$ can be expressed in the form:
\be
\zeta(\vec x)=\zeta_g(\vec x) + \frac{9}{25} \gnlloc \zeta_g(\vec x)^3  \label{eq:gnlloc_def}
\ee
where $\zeta_g$ is a Gaussian field.
This leads to the following $\zeta$-trispectrum:
\be
\langle \zeta_{\k_1} \zeta_{\k_2} \zeta_{\k_3} \zeta_{\k_4} \rangle
  = \frac{54}{25} \gnlloc \Big( P_\zeta(k_1) P_\zeta(k_2) P_\zeta(k_3) + \mbox{3 perm.} \Big) 
     (2\pi)^3 \delta^3\Big( \sum \k_i \Big)  \label{eq:gnlloc_zeta_trispectrum}
\ee
While it is impossible to obtain such a signal in single field inflation, 
as Maldacena's consistency condition~\cite{Maldacena:2002vr} generalized to the four-point function
(e.g.~\cite{Senatore:2012wy}) shows~\footnote{The fact that Maldacena's consistency condition gives a non-zero bispectrum or trispectrum 
in the squeezed limit should not be regarded as predicting a non-vanishing physical signal in that limit. It is indeed the way to write in comoving 
coordinates the fact that there is no physical correlation among modes of different wavelengths: a local observer can obtain the same result for a 
local experiment by starting with vanishing super-Hubble correlations. See discussion about this in~\cite{Baldauf:2011bh}.}, there are technically natural multifield inflationary models that generate this signal without generating an observationally larger bispectrum~\cite{Senatore:2010wk}. If we call the additional light field $\sigma$'s, a measurable $\gnlloc$ can be enforced by imposing, just as an example, an approximate $Z_2$ symmetry of the $\sigma$'s or protecting them with an approximate supersymmetry~\cite{Senatore:2010wk}.
Several groups have constrained $\gnlloc$ from WMAP 
data~\cite{Vielva:2009jz,Smidt:2010ra,Fergusson:2010gn,Hikage:2012bs},
and most recently~\cite{Sekiguchi:2013hza} who use the optimal estimator.

Another ``local'' four-point function is the $\taunl$-trispectrum, defined by:
\be
\langle \zeta_{\k_1} \zeta_{\k_2} \zeta_{\k_3} \zeta_{\k_4} \rangle
  = \taunl P_\zeta(k_2) P_\zeta(k_4) P_\zeta(|\k_1+\k_2|) (2\pi)^3 \delta^3\Big(\sum\k_i\Big) + \mbox{(11 perm.)}  \label{eq:taunl_def}
\ee
The $\taunl$-trispectrum can arise if $\zeta$ is a local quadratic combination of multiple uncorrelated fields.
For example, suppose
\be
\zeta(\x) = \zeta_G(\x) + A \zeta_G(\x) \sigma(\x)
\ee
where $A$ is a free
parameter and $\zeta_G, \sigma$ are uncorrelated Gaussian fields with equal power spectra.
In this model, the three-point function is zero and the four-point function takes the form~(\ref{eq:taunl_def})
with $\taunl = A^2$.
The parameter $\taunl$ has been constrained from WMAP~\cite{Smidt:2010ra,Hikage:2012bs} and Planck~\cite{Ade:2013ydc} data.

Going beyond the local-type signals $g_{NL}^{\rm loc}$ and $\tau_{NL}$, the
only primordial trispectrum which has been constrained is an ``equilateral''
trispectrum, which we will denote $\gnldotpi4$ and define by:
\ba
\langle \zeta_{\k_1} \zeta_{\k_2} \zeta_{\k_3} \zeta_{\k_4} \rangle
   &=& \frac{9216}{25} \gnldotpi4 A_\zeta^3 \int_{-\infty}^0 d\tau_E \, \tau_E^4 
              \left( \prod_{i=1}^4 \frac{e^{k_i\tau_E}}{k_i} \right)  
        (2\pi)^3 \delta^3\Big(\sum\k_i\Big) \nn \\
   &=& \frac{221184}{25} \gnldotpi4\, A_\zeta^3
          \frac{1}{k_1k_2k_3k_4 (k_1+k_2+k_3+k_4)^5}
        (2\pi)^3 \delta^3\Big(\sum\k_i\Big)  \label{eq:gnldotpi4}
\ea
where $A_\zeta$ is the amplitude of the $\zeta$ power spectrum,
defined by $P_\zeta(k) = A_\zeta/k^3$.
In the effective field theory description of inflation, this trispectrum
arises from a quartic operator of the schematic form $\dot\sigma^4$ (we
will be more precise in the next section).
WMAP5 constraints on this trispectrum were reported in~\cite{Fergusson:2010gn}.

Now is a good time to explain our normalization convention
in Eq.~(\ref{eq:gnldotpi4}).
Recall that in the bispectrum case, 
the bispectrum parameters $(\fnleq, \fnlorth)$ are normalized by fixing
the bispectrum amplitude on equilateral triangles to have the same value as the
local bispectrum with $\fnlloc=1$.
Analogously, we normalize trispectra so that 
$\langle \zeta_{\k_1} \zeta_{\k_2} \zeta_{\k_3} \zeta_{\k_4} \rangle = (216/25) g_{NL} A_\zeta^3 / k^9$
for tetrahedral 4-point configurations with $|\k_i|=k$ and $\k_i\cdot\k_j=-k^2/3$ for $i\ne j$.
This convention fixes all trispectra to have the same value on tetrahedrons as the local
trispectrum with $\gnlloc=1$.
Another detail: in Eq.~(\ref{eq:gnldotpi4}),
and in Eqs.~(\ref{eq:gnlB}),~(\ref{eq:gnldpi4}) below,
we write the trispectrum in two forms, either with a time integral which is
unevaluated, or after evaluation of the integral.
We do this because the first form will be directly useful when obtaining
factorizable representations for the trispectra, as we will explain later.

In this paper, we will introduce two new trispectrum shapes which
correspond to quartic operators of the form $\dot\sigma^2 (\partial_i \sigma)^2$
and $(\partial_i \sigma)^2 (\partial_j \sigma)^2$ in the effective field
theory of inflation.
Following our normalization convention above, we define parameters $\gnlB$ and $\gnldpi4$ by:
\ba
\langle \zeta_{\k_1} \zeta_{\k_2} \zeta_{\k_3} \zeta_{\k_4} \rangle
  &=& -\frac{13824}{325} \gnlB A_\zeta^3 \int_{-\infty}^0 d\tau_E \, \tau_E^2
        \left( \frac{(1-k_3\tau_E)(1-k_4\tau_E)}{k_1k_2k_3^3k_4^3} (\k_3\cdot\k_4) e^{\sum k_i\tau_E} + \mbox{(5 perm.)} \right) 
      \nn \\
       &&  \hspace{2cm} \times
         (2\pi)^3 \delta^3\Big(\sum\k_i\Big)
      \nn \\
  &=& 
    -\frac{27648}{325} 
     \gnlB A_\zeta^3
        \left( \frac{K^2 + 3(k_3+k_4)K + 12k_3k_4}{k_1 k_2 k_3^3 k_4^3 K^5} (\k_3\cdot\k_4) + \mbox{5 perm.} \right) \,
         (2\pi)^3 \delta^3\Big(\sum\k_i\Big)  \label{eq:gnlB}  \\
\langle \zeta_{\k_1} \zeta_{\k_2} \zeta_{\k_3} \zeta_{\k_4} \rangle
   &=& \frac{82944}{2575} \gnldpi4 A_\zeta^3 \int_{-\infty}^0 d\tau_E\,
              \left( \prod_{i=1}^4 \frac{(1-k_i\tau_E)e^{k_i\tau_E}}{k_i^3} \right) 
             \Big( (\k_1\cdot\k_2)(\k_3\cdot\k_4) + \mbox{(2 perm.)} \Big) \,
         (2\pi)^3 \delta^3\Big(\sum\k_i\Big)
  \nn \\
  &=& 
    \frac{165888}{2575}  \gnldpi4 A_\zeta^3
        \left( \frac{2K^4 - 2K^2\sum k_i^2 + K \sum k_i^3 + 12 k_1 k_2 k_3 k_4}{k_1^3 k_2^3 k_3^3 k_4^3 K^5} \right) \nn \\
   && \hspace{1cm} \times
             \Big( (\k_1\cdot\k_2)(\k_3\cdot\k_4) + \mbox{(2 perm.)} \Big)  \,
         (2\pi)^3 \delta^3\Big(\sum\k_i\Big)
\label{eq:gnldpi4}
\ea
where $K=k_1+k_2+k_3+k_4$.
In this paper, we will implement the optimal CMB estimator for four trispectra:
$\gnlloc$, $\gnldotpi4$, $\gnlB$, and $\gnldpi4$.
(The $\taunl$ trispectrum requires slightly different techniques for reasons
that will be apparent later, so we have omitted it in this paper.)
Searching for these four trispectra is analogous to searching for the standard bispectra
$\fnlloc$, $\fnleq$, and $\fnlorth$.\footnote{Recall that the space of bispectra spanned by 
$\fnleq$, $\fnlorth$ is equal, by a linear transformation, to the space generated by the
cubic operators $\dot\sigma^3$ and $\dot\sigma (\partial\sigma)^2$~\cite{Senatore:2009gt}.}

There is a basic computational problem which arises for computational
operations with trispectra, for example applying an estimator to CMB maps, or computing a Fisher
matrix.
Naively, these operations have computational cost $\bigoh(\ellmax^7)$, which is prohibitive
for a large experiment like WMAP or Planck with $\ellmax \sim 10^3$.
The same computational problem arises for the bispectrum, where it has been solved using the
idea of {\em factorizability}~\cite{Wang:1999vf,Komatsu:2003iq,Creminelli:2005hu,Smith:2006ud,Fergusson:2009nv,Bucher:2009nm}.
If a bispectrum can be represented as a sum of terms which satisfy a suitable factorizability
condition (the precise condition is given in Eq.~(\ref{eq:bispectrum_fact}) below), then 
computational cost is dramatically reduced.
A variety of general strategies have been proposed for making bispectrum data analysis
computationally feasible (e.g.~\cite{Smith:2006ud,Fergusson:2009nv,Bucher:2009nm,Donzelli:2012ts,Byun:2013jba}); 
while the details of these proposals are very different, they can all be viewed as different strategies 
for representing a bispectrum as a sum of factorizable terms.
Analogously for the trispectrum, we will formulate a suitable definition of factorizability, 
show that it leads to dramatically reduced computational cost, and give a physically motivated,
Feynman diagram based prescription for representing inflationary trispectra in factorizable form.
This will allow us to analyze the local, $\dot\sigma^4$, $\dot\sigma^2 (\partial\sigma)^2$, and 
$(\partial\sigma)^4$ trispectra.

Among other things, factorizability means that we can do a Fisher
matrix analysis of correlations between trispectra.
We will show that there is one near-degeneracy among the four trispectra.
To quantify this, the $\dot\sigma^2 (\partial\sigma)^2$ trispectrum is 99.2\% correlated with
a suitably chosen linear combination of the $\dot\sigma^4$ and $(\partial\sigma)^4$ trispectra.
Therefore, we will eliminate the parameter $\gnlB$, and reduce our set of trispectra
to three: $\gnlloc$, $\gnldotpi4$, and $\gnldpi4$.

We will construct trispectrum estimators and present details of analysis pipelines
which are suitable for realistic experiments such as WMAP or Planck.
We would like to emphasize three technical issues from the outset.

First, the trispectrum estimator is potentially very sensitive to modeling errors in the
two-point function due to slightly incorrect cosmological parameters, detector noise
properties, or beams.
Suppose the trispectrum is estimated assuming covariance matrix $C_0$, but the true
covariance is $C_{\rm true} = C_0 + \Delta C$.
The trispectrum estimators we use will have the property that the resulting bias
is parametrically $\bigoh((\Delta C)^2)$ rather than $\bigoh(\Delta C)$.
This property turns out to be critical in practice.

The second issue is that many technical tricks are necessary to reduce the number of Monte Carlo
simulations in the trispectrum estimation pipeline to a reasonable level.
We will find several situations where an ``obvious'' Monte Carlo procedure is slow,
but there is an alternate Monte Carlo procedure which is faster (examples include
Eqs.~(\ref{eq:fisher_mc}),~(\ref{eq:optimal_pipeline_F_final}), and~(\ref{eq:fast_FV})).
% \kms{I considered elaborating more here with a toy example of the fast MC phenomenon,
% but when I tried writing it out, I decided it was too long of a digression to be
% worth it pedagogically}

Third, gravitational lensing and other secondary effects (such as contamination
by residual infrared sources) generate a nonzero trispectrum which must be subtracted.
The lensing trispectrum has been measured in ACT~\cite{Das:2011ak,Das:2013zf},
SPT~\cite{vanEngelen:2012va,Story:2014dwa}, 
and Planck~\cite{Ade:2013tyw}, 
with recent measurements approaching 40$\sigma$!
Although lensing is an interesting source of cosmological information,
in this paper our focus will be on the primordial trispectrum, so we will
treat lensing as a large contaminant whose bias must be subtracted when
estimating other trispectrum shapes.

We will conclude by performing an optimal analysis of WMAP data.
We find the following constraints (all 95\% CL):
\ba
(-8.18 \times 10^5) < \gnlloc < (0.58 \times 10^5) && \hspace{1cm} \nn \\
(-9.38 \times 10^6) < \gnldotpi4 < (2.98 \times 10^6) && \hspace{1cm}  \\
(-2.34 \times 10^6) < \gnldpi4 <(0.19 \times 10^6) && \hspace{1cm} \nn
\ea
We find no evidence of primordial trispectra, and the error bars agree
with Fisher matrix forecasts. 

\section{Mini-Review of Effective Field Theories of Single and Multifield Inflation}

In this section we briefly review the particle physics motivation for studying the trispectra that we analyze. 
We do this by using the effective field theory of inflation~\cite{Cheung:2007st} and of multifield inflation~\cite{Senatore:2010wk}.

We start from single field inflation. By assuming that inflation is an early phase of the universe characterized by a 
spontaneous breaking of time diffeomorphisms, it is possible to construct a model independent Lagrangian for the fluctuations. 
Furthermore, in inflation we are interested in computing correlation functions at an energy scale around the Hubble scale 
during the early quasi de Sitter phase. Often, this energy scale is high enough to write the action in the so-called decoupling limit, 
where the Lagrangian takes a very simple form: 
\ba
&&S_{\rm \pi} =\int d^4 x   \sqrt{- g} \bigg[ -M^2_{\rm Pl}\dot{H} \left(\partial_\mu \pi\right)^2
+2 M^4_2
\left(\dot\pi^2+\dot{\pi}^3-\dot\pi\frac{(\partial_i\pi)^2}{a^2}+(\pd_\mu\pi)^2(\pd_\nu\pi)^2
\right) \nn \\
&& \hspace{3cm}
- \frac{M_3^4}{3!} \left( 8\,\dot{\pi}^3+12 \dot\pi^2(\pd_\mu\pi)^2+\cdots \right)
+ \frac{M_4^4}{4!} \left( 16\,\dot\pi^4+32\dot\pi^3(\pd_\mu\pi)^2 + \cdots \right)+\cdots \bigg] \ .  \label{eq:Spi}
\ea
where `$\cdots$' represents higher order terms, higher derivative terms, and slow-roll suppressed terms.
Here spatial indexes $i$ are contracted with the $\delta^{ij}$-tensor, while space-time indexes $\mu$ are contracted with the FRW metric $g^{\mu\nu}$.
The field $\pi$ represents the Goldstone boson of time translations. It is related to the curvature perturbation $\zeta$ as
\be\label{eq:zeta-pi}
\zeta=-H \pi + \mbox{(higher-order terms)}.
\ee
The non-linear realization of time-diffeomorphisms 
forces the appearance of~$\pi$ into non-linear blocks. Simple inspection of the action shows that 
it is impossible to have a four-point function induced by operators of the 
form $\dot\pi^2(\pd_\mu\pi)^2$ and $(\pd_\mu\pi)^2(\pd_\nu\pi)^2$ that is the leading 
non-Gaussian signal: when these operators are turned on, there is always a cubic operator 
that induces a bispectrum with much higher signal-to-noise ratio~\cite{Senatore:2010jy}. 
At the level of the leading derivative operators, the only term that has a chance of 
producing a trispectrum without a bispectrum with a large signal-to-noise ratio is the 
operator $\dot\pi^4$. 
One should be careful about radiative corrections though.
The non-linear realization of time-diffeomorphisms forces the presence of a quintic 
operator $\dot\pi^3(\pd_\mu\pi)^2$ together with $\dot\pi^4$. One is naturally lead 
to wonder if this operator will induce, under radiative corrections, a cubic operator 
that dominates the signal. It turns out that the relative coefficients of $\dot\pi^4$ 
and $\dot\pi^3(\pd_\mu\pi)^2$ are fixed by time-diffeomorphism invariance in such a 
way that, when the signal-to-noise in $\dot\pi^4$ is large ($g_{NL}^{\dot\pi^4}\gg 10^5$), 
the radiative corrections induced by the quintic operator generate at most a cubic 
operator $\dot\pi^3$ and $\dot\pi(\pd_\mu\pi)^2$ with an $f_{NL}\sim 1$, and therefore 
subleading~\cite{Senatore:2010jy}.
This can be interpreted as an approximate $Z_2$ symmetry of the inflaton.~\footnote{The fact that the coefficient of $\dot\pi^4$ is 
unrelated to the coefficients of the cubic terms had been already noticed 
in \cite{Chen:2009bc,Arroja:2009pd} for a subclass of the models we consider with the 
effective field theory consisting of scalar field Lagrangians of the 
form $P((\pd\phi)^2,\phi)$. However, without the identification of a mechanism 
protecting the generation of cubic terms, it is hard to imagine why one should 
concentrate on the particular Lagrangian allowing for a large quartic operator 
and small cubic ones.} We therefore conclude that it is possible to have a trispectrum 
induced by $\dot\pi^4$ as the leading non-Gaussian signal. A constraint 
on this trispectrum can be directly mapped, in the context of single field inflation, 
into a constraint of the coefficient $M_4^4$ of (\ref{eq:Spi}), as already done by the 
WMAP and the Planck experiments for the coefficients $M_2^4$ and $M_3^4$ from analysis 
of $f_{NL}^{\rm equil}$ and $f_{NL}^{\rm orthog}$~\cite{Senatore:2009gt,Bennett:2012zja,Ade:2013ydc}.

It is also possible to have higher derivative interactions leading to large non-Gaussianities directly 
in the form of a trispectrum from interactions with more than four overall derivatives, very schematically 
of the form $(\partial^2\pi)^4$~\cite{Senatore:2010jy,Bartolo:2010di,Behbahani:2014upa,Creminelli:2010qf,Bartolo:2013eka,Arroja:2013dya}. 
As for the case of the three-point function, where the same phenomenon appears, the signal can be made detectable 
only by lowering enough the unitarity bound of the theory, and it is furthermore  possible 
to have strong degeneracies with shapes with fewer derivatives.
This makes the prospects of detection somewhat more unlikely. We leave the study of these shapes to future work.

\begin{figure}
\begin{center}
\includegraphics[width=9cm]{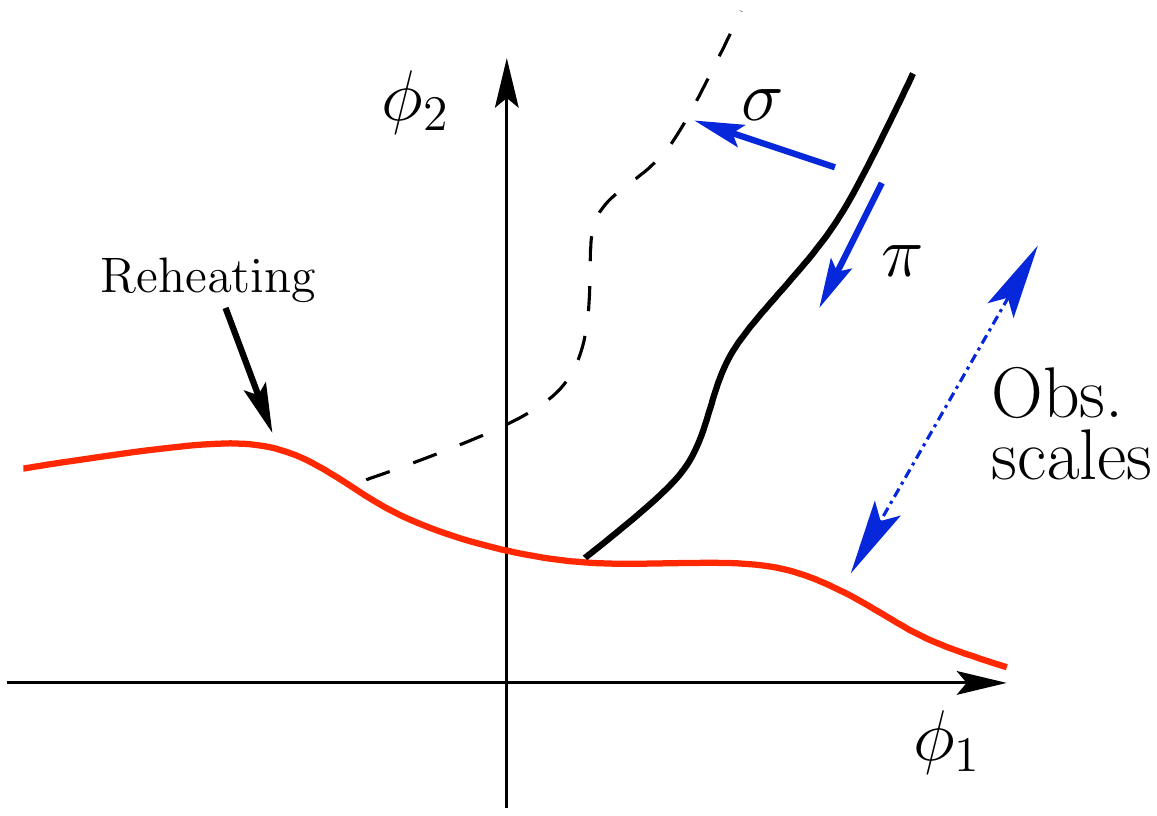}
\caption{\label{fig:multifield_potential} \small Representation of a typical multifield potential. Modes of interest for observation cross the horizon about sixty $e$-foldings before the end of inflation. Therefore, effects coming from the evolution of the fields after horizon crossing can be treated locally in real space. The effective theory is more general than this example, as it does not assume that the inflaton is a scalar field. This example is however interesting in helping in visualizing the different scales in the problem.}
\end{center}
\end{figure}

We now pass to multifield inflation. An effective field theory description of multifield inflation can be constructed after realizing that the predictions of multifield inflation largely do not depend on the background solution, with scalar fields developing possibly complicated trajectories in field space, but simply on the Lagrangian of the fluctuations. This Lagrangian can be simply constructed by coupling additional light degrees of freedom to the Goldstone boson of time-translations $\pi$. The main difference between single field and multifield inflation is that while in single field inflation the relationship between the Goldstone boson $\pi$ and the curvature perturbation~$\zeta$ is fixed by the background cosmology as in Eq.~(\ref{eq:zeta-pi}), the same is not true for the effect of the additional inflationary fields on $\zeta$. How much a given fluctuation of the additional fields contributes to the curvature perturbations depends on the whole trajectory of the fields from the time a mode crosses the horizon to reheating, and also on the details of the reheating epoch. However, the fact that these effects happen when all the modes of interest are outside of the horizon and gradients are therefore negligible (see Fig.~\ref{fig:multifield_potential}) permits a crucial simplification~\cite{Senatore:2010wk}: the relationship between the fluctuations of additional scalar fields, $\sigma_I$, and $\zeta$, must be local in space, and since fluctuations are quasi-Gaussian, the relationship can be Taylor expanded. We are therefore led to
\ba
\zeta(x) &=&
  - H\, \pi(x)
  + \left( \frac{\pd\zeta}{\pd\sigma_I} \right)_0 \sigma_I(x)
  + \left(\frac{\pd^2\zeta}{\pd\pi\pd\sigma_I} \right)_0 \pi(x) \sigma_I(x)
  + \frac{1}{2!} \left( \frac{\pd^2\zeta}{\pd\sigma_I\pd\sigma_J} \right)_0 \sigma_I(x) \sigma_J(x)
\nn \\ && \hspace{1cm}
  + \frac{1}{3!} \left( \frac{\pd^3\zeta}{\pd\sigma_I\pd\sigma_J\pd\sigma_K} \right)_0 
          \sigma_I(x) \sigma_J(x) \sigma_K(x) +...\ ,  \label{eq:zeta_relation}
\ea
where $(\pd^n\zeta/\sigma_{I_1}\ldots\pd\sigma_{I_n})_0$
are numbers representing the Taylor expansion of the generic relation, local in real space, 
between $\zeta$ and $\sigma_I$, $\zeta(\vec x)=f(\sigma_I(\vec x))$, around the point $\sigma_I=0$. 
This relationship, developed in~\cite{Senatore:2010wk}, generalizes in a non-trivial way the 
so-called $\delta N$ formalism of~\cite{Sasaki:1995aw,Starobinsky:1986fxa,Sasaki:1998ug,Lee:2005bb}.

At this point the problem of writing the effective field theory of multifield inflation is reduced to 
writing a Lagrangian for the additional light fields present during inflation, possibly coupled to the 
Goldstone boson $\pi$. Clearly, there is some freedom in what kind of fields we decide to include. 
In this paper we will content ourselves with the fields studied in~\cite{Senatore:2010wk},
although it would be interesting to study additional possibilities.
There, the additional fields that were included were scalar fields $\sigma_I$
which generate curvature perturbations after horizon crossing, and not just through their effect on $\pi$.~\footnote{This 
means that our discussion does not include models of the class of the so-called quasi-single field inflation~\cite{Chen:2009we,Chen:2009zp}}
To ensure that quantum corrections are small and do not make the mass of these additional fields large, 
it was postulated that these fields were either the Goldstone bosons of some global symmetry, 
abelian or non-abelian, or they were protected by an approximate supersymmetry.~\footnote{Supersymmetry is 
broken during inflation minimally only by the Hubble scale $H$, which means that radiative corrections to 
the superpotential vanish above the scale. For weakly coupled theories, where loops are suppressed by a 
weak coupling parameter, this makes radiative corrections perturbatively small~\cite{Senatore:2010wk}.}

These different mechanisms that protect the lightness of the additional scalar fields from quantum corrections 
can lead to distinguishable signals that, if detected, might allow us to infer the mechanism protecting the
lightness of these fields, as described in detail in~\cite{Senatore:2010wk}, to which we refer for details. 
Unfortunately, these mechanism-specific signals appear only as either subleading signals that have lower 
signal-to-noise ratio than other ones that should be detected first, or as signals appearing in correlation 
functions involving isocurvature fluctuations. Unfortunately, the leading signal is not able to distinguish 
among the various mechanisms protecting the lightness of the additional fields. Since in this paper we will 
restrict to adiabatic fluctuations, and since we are just trying to detect the leading signal, we can neglect 
all these distinctions, and we can focus on the following Lagrangian, which is common to all three mechanisms 
above (Abelian Goldstone bosons, non-Abelian Goldstone bosons, and supersymmetry):
\be\label{eq:multi}
S_\sigma=\int d^4x\, \sqrt{-g}\; \left[ \frac{1}{2}(\pd_\mu\sigma)^2+\frac{1}{\Lambda^4_1}\dot\sigma^4+\frac{1}{\Lambda^4_2}\dot\sigma^2 (\pd_i\sigma)^2+\frac{ 1}{\Lambda^4_3}(\pd_i\sigma)^2 (\pd_j\sigma)^2+\frac{\mu^4}{\Lambda^4}\sigma^4+\dots \right]\ .
\ee
This Lagrangian reproduces the relevant features that are contained in the models in~\cite{Senatore:2010wk}. 
First, notice that we did not write any cubic terms that would give rise to a bispectrum signature. 
These can be suppressed with some symmetry, such as for example a $Z_2$ symmetry $\sigma\to-\sigma$ 
or by imposing a Lorentz invariance in the theory, as described in~\cite{Senatore:2010wk}. 
The quartic couplings are suppressed by scales $\Lambda_{1,2,3}$, the smallest of which 
represents the unitarity bound of the theory. 
% $\Lambda$ can be taken as the smallest of the $\Lambda_{1,2,3}$ and $\mu^4\ll\Lambda^4$.
The first three interactions are compatible with a shift symmetry of the $\sigma$ field, and their coefficients are all independent. 
This means that they can generate observable templates associated to the operators $\dot\sigma^4$,
$\dot\sigma^2 (\pd_i\sigma)^2$ and $(\pd_i\sigma)^2 (\pd_j\sigma)^2$. 
Note that the operator $\dot\sigma^4$ generates the same trispectrum as the operator $\dot\pi^4$ considered previously in the single field case.
The operator $\sigma^4$ is 
present in the case the $\sigma$ fields are supersymmetric, or when the the symmetry that the Goldstone bosons $\sigma$'s 
non-linearly realize is softly broken. This operator gives rise to a bispectrum of the local form which, again, cannot be 
generated in the single field case. 
The signal produced by the $\sigma^4$ term in the Lagrangian can give rise to a much larger signal than
the one associated to an $f_{NL}^{\rm loc}$ of order unity. 
On top of these contributions, there are the ones associated to the non-linear 
relation between $\zeta$ and $\sigma$'s in Eq.~(\ref{eq:zeta_relation}). They give rise to bispectra and trispectra of local type.

Finally, we notice that one can enforce a particular symmetry 
in the case of multifield inflation, where non-Gaussianities are generated in a theory where Lorentz invariance in the 
multifield sector is left unbroken~\cite{Senatore:2010wk}. In this case, only two operators survive, 
$(\pd_\mu\sigma)^2(\pd_\nu\sigma)^2$ and $\sigma^4$. This symmetry can be explicitly checked to be mapped 
into a conformal symmetry of the three-dimensional templates~\cite{Creminelli:2011mw,Maldacena:2011nz}. 
There are finally additional trispectra, as for example $\sigma^2(\partial\sigma)^2$, associated to soft breaking 
of the some internal symmetries or to supersymmetry~\cite{Senatore:2010wk} whose analysis we defer to a subsequent work. 
Additional interesting studies, including some very early ones, for the inflationary trispectrum, both in single field 
and multifield inflation, can be found in~\cite{Bernardeau:2002jy,Bernardeau:2002jf,Seery:2006vu,Huang:2006eha,Seery:2006js,Byrnes:2006vq,Bernardeau:2007xi,Arroja:2008ga,Seery:2008ax,Engel:2008fu,Huang:2009xa,Gao:2009gd,Kawakami:2009iu,Mizuno:2009mv,Bartolo:2009kg,ValenzuelaToledo:2009nq,Huang:2010ab,Izumi:2010wm,Gao:2010xk,Leblond:2010yq,Langlois:2010fe,Meyers:2011mm,Agullo:2011aa,Elliston:2012wm,Anderson:2012em,Renaux-Petel:2013wya,Abolhasani:2013zya,Renaux-Petel:2013ppa,Leung:2013rza,Fasiello:2013dla,Byrnes:2013qjy}.

We conclude this section by relating parameters in the above Lagrangians
to the $g_{NL}$ coefficients defined in the introduction.
For the case of the single field Lagrangian (Eq.~(\ref{eq:Spi})), a short calculation 
using the in-in formalism~\cite{Maldacena:2002vr} shows:
\be
\gnldotpi4 = \frac{25}{288} \frac{M_4^4}{H^4} A_\zeta c_s^3 \ .  \label{eq:gnldotpi4_sf}
\ee
For the multifield Lagrangian (Eq.~(\ref{eq:multi})), we find:
\be
\gnldotpi4 A_\zeta
  = \frac{25}{768}\frac{H^4}{ \Lambda_1^4}\ , \qquad 
\gnlB A_\zeta
  = -\frac{325}{6912}\frac{H^4}{ \Lambda_2^4}\ , \qquad 
\gnldpi4 A_\zeta 
  = \frac{2575}{20736} \frac{H^4}{\Lambda_3^4}\ .   \label{eq:gnl_multifield}
\ee
Notice that, as expected, $(\gnldotpi4 A_\zeta)$, $(\gnlB A_\zeta)$, and $(\gnldpi4 A_\zeta)$ 
scale as $(H/\Lambda)^4$, being generated by dimension eight operators. 
The Lorentz invariant trispectrum generated by the 
operator $(\partial_\mu\sigma)^2(\partial_\nu\sigma)^2$ is obtained 
by setting $\Lambda_1^4=\Lambda_3^4=-2\Lambda_2^4$. 

Finally, for the local trispectrum, we get either:
\be
\gnlloc A_\zeta = -\frac{50}{27} \frac{\mu^4}{\Lambda^4} N_e \left(1 + \bigoh\left(\frac{1}{N_e}\right) \right)
\ee
in the case where the trispectrum is generated by a $\sigma^4$ interaction in the
multifield action (Eq.~(\ref{eq:multi})), or
\be
\gnlloc =  \frac{25}{54} \frac{H^6}{A_\zeta^3} \left(  \frac{\partial^3\zeta}{\partial\sigma_I \partial\sigma_J \partial\sigma_K} \right)_0
      \left( \frac{\partial\zeta}{\partial\sigma_I} \right)_0
      \left( \frac{\partial\zeta}{\partial\sigma_J} \right)_0
      \left( \frac{\partial\zeta}{\partial\sigma_K} \right)_0
\ee
in the case where the local trispectrum is generated by the conversion mechanism in Eq.~(\ref{eq:zeta_relation}).

\section{The CMB trispectrum and its optimal estimator}
\label{sec:estimator}

\subsection{Toy model}
\label{ssec:toy_model}

Before diving into the complications of the CMB,
it may be illuminating to construct the optimal trispectrum
estimator for the following toy model.
Let $x_1, \cdots, x_N$ be independent identically distributed random variables
whose distribution is nearly Gaussian, with mean zero, known variance $\sigma^2$,
and small kurtosis $\kappa \ll \sigma^4$ which we would like to estimate.
Thus the two-point and four-point functions are:
\be
\langle x_i x_j \rangle = \sigma^2 \delta_{ij} 
  \hspace{1cm} 
\langle x_i x_j x_k x_l \rangle = \sigma^4 (\delta_{ij}\delta_{kl} + \delta_{ik}\delta_{jl} + \delta_{il}\delta_{jk}) + \kappa \delta_{ij} \delta_{jk} \delta_{kl}
\ee
It may seem natural to estimate $\kappa$ using the simple estimator:
\be
\hkappa_{\rm naive} = \frac{1}{N} \left( \sum_{i=1}^N x_i^4 \right) - 3\sigma^4
\ee
However, this estimator is suboptimal!
The optimal (minimum variance) estimator turns out to be
\be
\hkappa_{\rm opt} = \frac{1}{N} \left( \sum_{i=1}^N x_i^4 \right) - 6 \sigma^2 \frac{1}{N} \left( \sum_{i=1}^N x_i^2 \right) + 3 \sigma^4  \label{eq:opt_toy}
\ee
A short calculation using Wick's theorem shows that
\be
\Var(\hkappa_{\rm naive}) = \frac{96 \sigma^8}{N} 
  \hspace{1cm}
\Var(\hkappa_{\rm opt}) = \frac{24 \sigma^8}{N}
\ee
so the naive estimator $\hkappa_{\rm naive}$ is significantly suboptimal.

In addition to having lower variance, the optimal estimator has another property which
is crucial in practice.  Suppose that the variance $\sigma^2$ is not precisely known in
advance, but has been estimated with some error 
$\Delta\sigma^2 = \sigma^2_{\rm est} - \sigma^2_{\rm true}$.
Let us compute the ``two-point bias'' in our estimate of $\kappa$ due to the
incorrectly estimated variance.  A short calculation gives:
\be
\langle \hkappa_{\rm naive} \rangle = \kappa -6 \sigma^2 (\Delta\sigma^2) + \bigoh(\Delta\sigma^2)^2
  \hspace{1.5cm}
\langle \hkappa_{\rm opt} \rangle = \kappa + 3 (\Delta\sigma^2)^2
\ee
In other words, the optimal estimator is {\em parametrically more robust} (by one power
of $\Delta\sigma^2$) to errors in our estimates of the two-point function.
This extra robustness will be important when we generalize to the CMB, where beams, noise, and residual foregrounds
all contribute to the two-point function and are notoriously difficult to estimate
precisely.

\subsection{CMB estimator}

Let us first establish some notation.
We denote the angular four-point function or trispectrum of the CMB by:
\ba
T^{\ell_1\ell_2\ell_3\ell_4}_{m_1m_2m_3m_4} 
   &=& \langle a_{\ell_1m_1} a_{\ell_2m_2} a_{\ell_3m_3} a_{\ell_4m_4} \rangle_c \nn \\
   &=& \langle a_{\ell_1m_1} a_{\ell_2m_2} a_{\ell_3m_3} a_{\ell_4m_4} \rangle
         - \langle a_{\ell_1m_1} a_{\ell_2m_2} \rangle \langle a_{\ell_3m_3} a_{\ell_4m_4} \rangle \nn \\
    && \hspace{1cm}
         - \langle a_{\ell_1m_1} a_{\ell_3m_3} \rangle \langle a_{\ell_2m_2} a_{\ell_4m_4} \rangle
         - \langle a_{\ell_1m_1} a_{\ell_4m_4} \rangle \langle a_{\ell_2m_2} a_{\ell_3m_3} \rangle  \nn \\
  &=& \langle a_{\ell_1m_1} a_{\ell_2m_2} a_{\ell_3m_3} a_{\ell_4m_4} \rangle 
        - \bigg[ (-1)^{m_1+m_3} C_{\ell_1} C_{\ell_3} \delta_{\ell_1\ell_2} \delta_{\ell_3\ell_4} \delta_{m_1,-m_3} \delta_{m_2,-m_4} + \mbox{(2 perm.)} \bigg]
\ea
The trispectrum $T^{\ell_1\ell_2\ell_3\ell_4}_{m_1m_2m_3m_4}$ is invariant under the $4!$ permutations
of its indices $(\ell_i,m_i)$, and satisfies the reality condition
\be
T^{\ell_1\ell_2\ell_3\ell_4*}_{m_1m_2m_3m_4} = (-1)^{m_1+m_2+m_3+m_4} T^{\ell_1\ell_2\ell_3\ell_4}_{(-m_1)(-m_2)(-m_3)(-m_4)}  \label{eq:T_reality}
\ee
We will only consider CMB trispectra which are rotationally invariant, i.e.~$T^{\ell_1\ell_2\ell_3\ell_4}_{m_1m_2m_3m_4}$
is unchanged if a common rotation is applied to all four pairs of indices $(\ell_i,m_i)$.
This means that the trispectrum has fewer degrees of freedom than the index notation $T^{\ell_1\ell_2\ell_3\ell_4}_{m_1m_2m_3m_4}$
would suggest.  (It is possible to devise alternate notation which makes this more explicit~\cite{Hu:2001fa},
but we will not do so in this paper.)

Now consider a CMB experiment in which the instrumental response is linear and
the noise is Gaussian.
The observed CMB $a_{\ell m}$ is a sum of signal and noise contributions:
we have $a_{\ell m} = s_{\ell m} + n_{\ell m}$, where
$s_{\ell m}$ is the true CMB and $n_{\ell m}$ is Gaussian noise.
Let $C_{\ell_1m_1,\ell_2m_2} = \langle a_{\ell_1m_1} a_{\ell_2m_2} \rangle$
be the total (signal + noise) covariance of the observed CMB.
Note that although the signal contribution to $C$ will be diagonal in $(\ell,m)$,
the noise contribution will generally be nondiagonal.

Armed with the above notation, the optimal trispectrum estimator can be written in the
following general form~\cite{Regan:2010cn}:
\ba
\E &=& \frac{1}{24F} \sum_{\ell_im_i} T^{\ell_1\ell_2\ell_3\ell_4*}_{m_1m_2m_3m_4} 
   \Bigg[ (C^{-1}a)_{\ell_1m_1} (C^{-1}a)_{\ell_2m_2} (C^{-1}a)_{\ell_3m_3} (C^{-1}a)_{\ell_4m_4}  \nn \\
   && \hspace{4cm} - 6\, C^{-1}_{\ell_1m_1,\ell_2m_2} (C^{-1}a)_{\ell_3m_3} (C^{-1}a)_{\ell_4m_4} \nn \\
   && \hspace{4cm} + 3\, C^{-1}_{\ell_1m_1,\ell_2m_2} C^{-1}_{\ell_3m_3,\ell_4m_4} \Bigg]  \label{eq:optimal_estimator}
\ea
where the normalizing constant $F$ is given by
\be
F = \frac{1}{4!} \sum_{\ell_im_i\ell'_im'_i} T^{\ell_1\ell_2\ell_3\ell_4*}_{m_1m_2m_3m_4}
  C^{-1}_{\ell_1m_1,\ell'_1m'_1}
  C^{-1}_{\ell_2m_2,\ell'_2m'_2}
  C^{-1}_{\ell_3m_3,\ell'_3m'_3}
  C^{-1}_{\ell_4m_4,\ell'_4m'_4}
T^{\ell'_1\ell'_2\ell'_3\ell'_4}_{m'_1m'_2m'_3m'_4}   \label{eq:F_harmonic}
\ee
Let us now interpret the terms in the optimal estimator above.
The first (quartic) term on the RHS of Eq.~(\ref{eq:optimal_estimator})
is a sum over 4-tuples $(\ell_i,m_i)$ in which
each 4-tuple is weighted by the template signal
$T^{\ell_1\ell_2\ell_3\ell_4}_{m_1m_2m_3m_4}$,
and inversely weighted by the total covariance $C$.
This type of weighting appears in a variety of optimal CMB estimators,
for example optimal estimators for the power spectrum or bispectrum.
The second (quadratic) and third (constant) terms in Eq.~(\ref{eq:optimal_estimator}) parallel the terms found
previously for the toy model in Eq.~(\ref{eq:opt_toy}).
We note that a similar structure occurs in the optimal estimator for the
three-point function, where there is a one-point term in addition to the leading
three-point term~\cite{Creminelli:2005hu}.

As in the toy model, the additional terms in Eq.~(\ref{eq:optimal_estimator})
reduce the variance,
and also make the estimator more robust to errors in the two-point function.
To make this last point more precise,
if the total covariance $C$ is estimated incorrectly with nonzero error $\Delta C$,
then it is easy to show that
the bias in the estimator is parametrically $\bigoh((\Delta C)^2)$ rather than
$\bigoh(\Delta C)$.
This property is critical in practice.
If we used an estimator whose bias is parametrically $\bigoh(\Delta C)$, we would
need to model beams, noise bias, residual foregrounds, etc. with fractional accuracy 
$1/\ellmax \approx 0.1\%$.
This level of accuracy is extremely difficult to achieve for an experiment as
complex as Planck.
On the other hand, with an estimator whose bias is parametrically $\bigoh(\Delta C^2)$,
the required fractional accuracy is $\approx 1/\ellmax^{1/2}$ or a few percent, which
is easily achieved in practice.

A short calculation shows that the variance of the optimal estimator is
\be
\Var(\E) = \frac{1}{F}
\ee
i.e.~$F$ determines both the normalization of the estimator and its variance.

\subsection{The $Q$-symbol}

We now define notation which will be used ubiquitously throughout the paper.
Given a CMB trispectrum $T^{\ell_1\ell_2\ell_3\ell_4}_{m_1m_2m_3m_4}$ and CMB realization $a_{\ell m}$,
we define the ``$Q$-symbol'' $Q_T[a]$ by:
\be
Q_T[a] = \frac{1}{4!} \sum_{\ell_im_i} T^{\ell_1\ell_2\ell_3\ell_4*}_{m_1m_2m_3m_4} a_{\ell_1m_1} a_{\ell_2m_2} a_{\ell_3m_3} a_{\ell_4m_4}   \label{eq:Q_def}
\ee
The reality condition~(\ref{eq:T_reality}) for the trispectrum, 
together with the reality condition $a_{\ell m}^* = (-1)^m a_{\ell(-m)}$, 
implies that $Q_T[a]$ is real.
Note that a similar notation $T[a]$ was defined for the bispectrum in~\cite{Smith:2006ud}.

This notation is useful since most of the machinery in this paper can be written
purely in terms of the $Q$-symbol.
Therefore, our machinery applies to a trispectrum if a fast algorithm exists
for evaluating its $Q$-symbol.
For example, the optimal estimator from the previous section can be written as the
following Monte Carlo average:
\be
\E[a] = \frac{1}{F} \left( Q[C^{-1}a, C^{-1}a, C^{-1}a, C^{-1}a] 
           - 6 \Big\langle Q[C^{-1}a, C^{-1}a, C^{-1}b, C^{-1}b] \Big\rangle_b
           + \Big\langle Q[C^{-1}b, C^{-1}b, C^{-1}b, C^{-1}b] \Big\rangle_b \right)  \label{eq:optimal_estimator_mc}
\ee
where $\langle \cdot \rangle_b$ denotes an average over Gaussian random realizations $b$
with covariance matrix $C$.
In~\S\ref{sec:fisher} we will develop fast algorithms for computing Fisher matrices,
and in~\S\ref{sec:pipelines} we will present detailed data analysis pipelines,
under the assumption that $Q_T[a]$ is computable.
We will also present an algorithm for simulating a non-Gaussian map with specified
trispectrum, although we defer this to Appendix~\ref{app:ngsim} since it is somewhat
peripheral to our goal of analyzing WMAP data.

We define the gradient $\partial_{\ell m} Q_T[a]$ by:
\be
\partial_{\ell m} Q_T[a] = \frac{\partial Q_T[a]}{\partial a_{\ell m}^*} = \frac{1}{3!} \sum_{\ell_im_i} T^{\ell\ell_1\ell_2\ell_3}_{mm_1m_2m_3} a_{\ell_1m_1}^* a_{\ell_2m_2}^* a_{\ell_3m_3}^*
\ee
The object $\partial_{\ell m} Q_T[a]$ is a harmonic-space map, as the index notation suggests.
It transforms covariantly under rotations, in the sense that $\partial Q[R \cdot a] = R \cdot \partial Q[a]$,
where $(R\cdot a)$ denotes the action of a rotation $R \in SO(3)$ on a harmonic-space map $a_{\ell m}$.

We will sometimes omit the subscript $T$, and simply write $Q[a]$ or $\partial_{\ell m} Q[a]$, if the trispectrum is understood.
It will also be convenient to define the following generalizations of $Q$ and $\partial Q$, which are functions of four CMB realizations
$(a,b,c,d)$ and three realizations $(a,b,c)$ respectively:
\ba
Q_T[a,b,c,d] &=& \frac{1}{4!} \sum_{\ell_im_i} T^{\ell_1\ell_2\ell_3\ell_4*}_{m_1m_2m_3m_4} a_{\ell_1m_1} b_{\ell_2m_2} c_{\ell_3m_3} d_{\ell_4m_4} \nn \\
\partial_{\ell m} Q_T[a,b,c] &=& \frac{1}{3!} \sum_{\ell_im_i} T^{\ell\ell_1\ell_2\ell_3}_{mm_1m_2m_3} a_{\ell_1m_1}^* b_{\ell_2m_2}^* c_{\ell_3m_3}^*
\ea

\subsection{An alternate approach?}
\label{ssec:alternate_approach}

Let us temporarily return to the toy model from~\ref{ssec:toy_model}.
We construct an interesting near-optimal trispectrum estimator as follows.
Suppose we use the naive estimator $\hkappa_{\rm naive}$, but estimate the variance $\sigma^2$ internally from data,
rather than assuming a priori knowledge of $\sigma^2$.  In other words, consider the pure four-point estimator:
\be
\hkappa_{\rm alt} = \frac{1}{N} \left( \sum_i x_i^4 \right) - \frac{3}{N(N-1)} \left( \sum_{i \ne j} x_i^2 x_j^2 \right)
\ee
It is not hard to show that the variance is
\be
\Var(\hkappa_{\rm alt}) = \frac{24 \sigma^8}{N} \left( 1 + \frac{3}{N-1} \right)
\ee
Comparing with the result for the optimal estimator (Eq.~(\ref{eq:opt_toy})) we see that
$\hkappa_{\rm alt}$ is near-optimal, in the sense that its variance agrees with the
optimal estimator to leading order in $1/N$.
This estimator also has the property that its two-point bias (due to incorrectly
estimated $\sigma^2$) is zero!
The estimator $\hkappa_{\rm alt}$ is only sensitive to the four-point signal $\kappa$, 
with no dependence on the variance $\sigma^2$.
We note that this statement does assume that the covariance matrix of the $x_i$
is proportional to the identity matrix, and there is no estimator which has zero
bias for an arbitrary covariance matrix $C_{ij}$.
Nevertheless it is interesting that a zero-bias estimator exists for a restricted
form of covariance matrix, and natural to ask whether this generalizes to the CMB context.

Ideally we would like to construct a CMB trispectrum estimator which is unbiased if either
(1) the isotropic signal power spectrum $C_\ell$, or (2) the noise covariance
is estimated incorrectly.
We speculate that it is possible to give a general construction of such an estimator.
Noise bias can be eliminated by dividing the data into subsets with uncorrelated noise,
making maps $(a_1)_{\ell m}$, $(a_2)_{\ell m}$, $\cdots$ from one subset at a time, and allowing
only ``cross'' terms $Q[a_i,a_j,a_k,a_l]$ with $(i,j,k,l)$ distinct.
Signal bias can be eliminated by estimating $C_\ell$ directly from the data and subtracting
a term which is quadratic in the {\em estimated} $C_\ell$'s, by analogy with the toy
model case above.
Such an estimator would be very useful e.g.~for the gravitational lensing four-point function,
where a variety of noise bias cancelling schemes have been 
proposed~\cite{Sherwin:2010ge,Das:2011ak,Plaszczynski:2012ej,Namikawa:2012pe}.
However, we defer this topic for future work.

\section{3D $\rightarrow$ 2D projection}

We will often be interested in ``primordial'' trispectra, that is, CMB trispectra which arise
by linearly evolving a physically motivated four-point function in the 3D adiabatic initial curvature $\zeta$.
The $\zeta$-trispectrum is defined by:
\ba
\langle \zeta_{\k_1} \zeta_{\k_2} \zeta_{\k_3} \zeta_{\k_4} \rangle_c 
  &=& \langle \zeta_{\k_1} \zeta_{\k_2} \zeta_{\k_3} \zeta_{\k_4} \rangle
      - \langle \zeta_{\k_1} \zeta_{\k_2} \rangle \langle \zeta_{\k_3} \zeta_{\k_4} \rangle
      - \langle \zeta_{\k_1} \zeta_{\k_3} \rangle \langle \zeta_{\k_2} \zeta_{\k_4} \rangle
      - \langle \zeta_{\k_1} \zeta_{\k_4} \rangle \langle \zeta_{\k_2} \zeta_{\k_3} \rangle \nn \\
  &=& \langle \zeta_{\k_1} \zeta_{\k_2} \zeta_{\k_3} \zeta_{\k_4} \rangle 
      - \bigg[ P(k_1) P(k_3) (2\pi)^6 \delta^3(\k_1+\k_2) \delta^3(\k_3+\k_4) + \mbox{(2 perm.)} \bigg]
\ea
We will also use the ``primed'' notation $\langle \zeta_{\k_1} \zeta_{\k_2} \zeta_{\k_3} \zeta_{\k_4} \rangle'$
to denote the $\zeta$-trispectrum without its momentum-conserving delta function, i.e.
\be
\langle \zeta_{\k_1} \zeta_{\k_2} \zeta_{\k_3} \zeta_{\k_4} \rangle_c 
  = \langle \zeta_{\k_1} \zeta_{\k_2} \zeta_{\k_3} \zeta_{\k_4} \rangle' \,
     (2\pi)^3 \delta^3\left( \sum \k_i \right)
\ee
We can project a $\zeta$-trispectrum to an angular CMB trispectrum as follows.
Recall that the CMB multipoles $a_{\ell m}$ are related to the initial curvature $\zeta$ by:
\be
a_{\ell m} = 4\pi i^\ell \int \frac{d^3\k}{(2\pi)^3} \Delta_\ell(k) \zeta(\k) Y_{\ell m}^*(\hk)
\ee
where the transfer function $\Delta_\ell(k)$ defined by this equation can be computed numerically
using CAMB~\cite{Lewis:1999bs}.
The following general trispectrum projection formula follows immediately:
\be
T^{\ell_1\ell_2\ell_3\ell_4}_{m_1m_2m_3m_4}
  = \int \frac{d^3\k_1\,d^3\k_2\,d^3\k_3\,d^3\k_4}{(2\pi)^{12}} \, 
      \langle \zeta_{\k_1} \zeta_{\k_2} \zeta_{\k_3} \zeta_{\k_4} \rangle_c
      \prod_{i=1}^4 \left( 4\pi i^{\ell_i} \Delta_{\ell_i}(k_i) Y_{\ell_im_i}^*(\hk_i) \right)  \label{eq:T_projection}
\ee
Plugging this into the definition~(\ref{eq:Q_def}) of the $Q$-symbol,
we get the following expression for $Q_T[a]$:
\be
Q_T[a] = \frac{1}{4!} \int \frac{d^3\k_1\,d^3\k_2\,d^3\k_3\,d^3\k_4}{(2\pi)^{12}} \, 
      \langle \zeta_{\k_1} \zeta_{\k_2} \zeta_{\k_3} \zeta_{\k_4} \rangle_c^* \,
 \prod_{i=1}^4 \Bigg( \sum_{\ell_im_i} 4\pi (-i)^{\ell_i} \Delta_{\ell_i}(k_i) a_{\ell_i m_i} Y_{\ell_i m_i}(\hk_i) \Bigg)  \label{eq:Q_projection}
\ee
Since the functional form of $Q_T[a]$ uniquely determines $T^{\ell_1\ell_2\ell_3\ell_4}_{m_1m_2m_3m_4}$,
this expression for $Q_T$ is equivalent to the projection formula~(\ref{eq:T_projection}) for $T$.
In fact, throughout the paper we will often find it more convenient to specify a trispectrum $T$
by giving a formula for $Q_T[a]$ than by an explicit expression for $T^{\ell_1\ell_2\ell_3\ell_4}_{m_1m_2m_3m_4}$.

\section{Factorizability}

Evaluating $Q_T[a]$ directly from its harmonic-space definition~(\ref{eq:Q_def})
is computationally intractable, since the number of terms in the sum is $\bigoh(\ellmax^7)$,
where $\ellmax = \bigoh(10^3)$ for WMAP or Planck.
An analogous computational problem arises in analysis of the CMB bispectrum,
where it has been solved using the idea of
finding a {\em factorizable representation} of the bispectrum.
We start by briefly reviewing factorizability for the bispectrum,
in notation which will set the stage for the trispectrum discussion to follow.

\subsection{Review of factorizability for the bispectrum}

\par\noindent
A CMB three-point function is said to be {\em factorizable} if it is a sum of terms of the form:
\be
\langle a_{\ell_1m_1} a_{\ell_2m_2} a_{\ell_3m_3} \rangle 
  = \sum_{I=1}^{\Nfact} A^I_{\ell_1} B^I_{\ell_2} C^I_{\ell_3} \, {\mathcal G}^{\ell_1\ell_2\ell_3}_{m_1m_2m_3} 
      + \mbox{(5 perm.)} \label{eq:bispectrum_fact}
\ee
where $A^I_{\ell}, B^I_{\ell}, C^I_{\ell}$ are $\Nfact$-by-$\ellmax$ real-valued matrices, and
${\mathcal G}^{\ell_1\ell_2\ell_3}_{m_1m_2m_3}$ is the Gaunt symbol, defined by:
\ba
{\mathcal G}^{\ell_1\ell_2\ell_3}_{m_1m_2m_3} &=& \int d^2\n\,\, Y_{\ell_1m_1}(\n) \, Y_{\ell_2m_2}(\n) \, Y_{\ell_3m_3}(\n)  \nn \\
  &=& \sqrt{\frac{(2\ell_1+1)(2\ell_2+1)(2\ell_3+1)}{4\pi}} \threej{\ell_1}{\ell_2}{\ell_3}{0}{0}{0}
\ea
The significance of the factorizability condition is that it makes the bispectrum estimator computationally feasible.
First recall~\cite{Smith:2006ud} that the bispectrum estimator can be written in terms of the $T$-symbol, defined by:
\be
T[a] = \frac{1}{6} \sum_{\ell_im_i} \langle a_{\ell_1m_1} a_{\ell_2m_2} a_{\ell_3m_3} \rangle^* \, a_{\ell_1m_1} a_{\ell_2m_2} a_{\ell_3m_3}
\ee
This harmonic-space sum is computationally infeasible, but
if the bispectrum satisfies the factorizability condition~(\ref{eq:bispectrum_fact})
then $T[a]$ can be rewritten:
\be
T[a] = \sum_{I=1}^{\Nfact} \int d^2\n\,
  \left( \sum_{\ell_1m_1} A^I_{\ell_1} a_{\ell_1m_1} Y_{\ell_1m_1}(\n) \right)
  \left( \sum_{\ell_2m_2} B^I_{\ell_2} a_{\ell_2m_2} Y_{\ell_2m_2}(\n) \right)
  \left( \sum_{\ell_3m_3} C^I_{\ell_3} a_{\ell_3m_3} Y_{\ell_3m_3}(\n) \right)
\ee
which is straightforward to evaluate efficiently using fast spherical transforms.
For this reason, finding a factorizable representation for a given bispectrum is
the key to making data analysis practically feasible.

Many CMB bispectra of interest arise from 3D$\rightarrow$2D projection of a
$\zeta$-bispectrum $\langle \zeta_{\k_1} \zeta_{\k_2} \zeta_{\k_3} \rangle$.
There is also a useful notion of factorizability for a $\zeta$-bispectrum as follows.
A $\zeta$-bispectrum $\langle \zeta_{\k_1} \zeta_{\k_2} \zeta_{\k_3} \rangle$
is said to be factorizable if:
\be
\langle \zeta_{\k_1} \zeta_{\k_2} \zeta_{\k_3} \rangle' = \frac{1}{6} \sum_{I=1}^{\Nfact} \alpha_I(k_1) \beta_I(k_2) \gamma_I(k_3) 
   + \mbox{(5 perm.)} \label{eq:bispectrum_fact3d}
\ee
where $\alpha_I(k), \beta_I(k), \gamma(k)$ are arbitrary functions.
A general 3D$\rightarrow$2D projection formula for bispectra, very similar to the one
given for trispectra in the last section, shows that the corresponding CMB bispectrum is:
\ba
\langle a_{\ell_1m_1} a_{\ell_2m_2} a_{\ell_3m_3} \rangle
 &=& \frac{1}{6} \sum_{I=1}^{\Nfact}
     \int r^2 dr\, 
        \left( \int \frac{2k_1^2dk_1}{\pi} j_{\ell_1}(k_1r) \Delta_{\ell_1}(k_1) \alpha_I(k_1) \right) \nn \\
&& \hspace{1.5cm} \times
        \left( \int \frac{2k_2^2dk_2}{\pi} j_{\ell_2}(k_2r) \Delta_{\ell_2}(k_2) \beta_I(k_2) \right) \nn \\
&& \hspace{1.5cm} \times
        \left( \int \frac{2k_3^2dk_3}{\pi} j_{\ell_3}(k_3r) \Delta_{\ell_3}(k_3) \gamma_I(k_3) \right)
   {\mathcal G}^{\ell_1\ell_2\ell_3}_{m_1m_2m_3} + \mbox{(5 perm.)}
\ea
This equation shows that
if we approximate the $r$ integral by a finite sum, we obtain a CMB bispectrum which is factorizable
in the sense defined by Eq.~(\ref{eq:bispectrum_fact}).
Thus a $\zeta$-bispectrum which is factorizable gives rise to a CMB bispectrum 
which is also factorizable, although the number of terms will increase by a large factor,
since many points will be needed to approximate the $r$-integral.

A variety of general schemes have been proposed in the literature for making bispectrum data analysis
computationally feasible (e.g.~\cite{Smith:2006ud,Fergusson:2009nv,Bucher:2009nm,Donzelli:2012ts,Byun:2013jba}).
These schemes can all be viewed as different proposals for representing a bispectrum as
a sum of factorizable terms.
Some methods operate directly on the CMB bispectrum, for example
the binned estimator in~\cite{Bucher:2009nm}
uses bandpowers in $\ell$ to define basis functions $A^I_\ell, B^I_\ell, C^I_\ell$.
Other methods operate on the $\zeta$-bispectrum before 3D$\rightarrow$2D
projection, for example by expanding the $\zeta$-bispectrum in
a set of orthogonal basis functions~\cite{Fergusson:2009nv,Byun:2013jba}.
Finally, in some cases it is possible to find an approximate factorizable representation as
a pure ansatz.  
The canonical example is the equilateral bispectrum~\cite{Creminelli:2005hu}, where the factorizable template:
\be
\langle \zeta_{\k_1} \zeta_{\k_2} \zeta_{\k_3} \rangle' = \frac{(k_1+k_2-k_3)(k_2+k_3-k_1)(k_3+k_1-k_2)}{k_1^3 k_2^3 k_3^3}
\ee
is 99\% correlated to the exact bispectrum of the operator $\dot\pi^3$.

In this paper, we will concentrate on a ``physical'' approach to factorizability
which generalizes nicely to the trispectrum and also provides some physical interpretation.
The idea is that the Feynman diagram which one evaluates to compute a given bispectrum
automatically supplies a factorizable representation.
To illustrate this idea by example, consider the $\dot\pi^3$ bispectrum:
\be
\langle \zeta_{\k_1} \zeta_{\k_2} \zeta_{\k_3} \rangle' \propto \frac{1}{k_1k_2k_3(k_1+k_2+k_3)^3}
\ee
This bispectrum does not appear to be factorizable.
However, let us go back to the Feynman diagram which produced it:
\ba
\raisebox{-0.7cm}{
\begin{fmffile}{cubic_diagram}
\begin{fmfgraph*}(15,15)
\fmfleft{i1,i2}
\fmfright{o3}
\fmf{fermion}{v1,i1}
\fmf{fermion}{v1,i2}
\fmf{fermion}{v1,o3}
%\fmflabel{$k_1$}{i1}
%\fmflabel{$k_2$}{i2}
%\fmflabel{$k_3$}{o3}
\end{fmfgraph*}
\end{fmffile}
}
 & \propto & \int_{-\infty}^0 d\tau_E \, \tau_E^2 
       \left( \frac{e^{k_1\tau_E}}{k_1} \right)
       \left( \frac{e^{k_2\tau_E}}{k_2} \right)
       \left( \frac{e^{k_3\tau_E}}{k_3} \right) \label{eq:cubic_first_line} \\
    &=& \frac{2}{k_1k_2k_3(k_1+k_2+k_3)^3} \label{eq:cubic_second_line}
\ea
Here, $\tau_E$ is Wick-rotated conformal time, which we take throughout this paper to run from $\tau_E=-\infty$ to 0.
We see that the integrand in Eq.~(\ref{eq:cubic_first_line}) is factorizable in $k_1,k_2,k_3$.
This is not a coincidence: it arises from combinatorics of the Feynman diagram, since each factor 
corresponds to one external line.
If we do not evaluate the $\tau_E$ integral, but instead approximate it by a finite sum of
$\tau_E$ values, then we will obtain a factorizable $\zeta$-bispectrum.
As shown in~\cite{Smith:2006ud}, the number of terms in the sum can be kept manageable by 
sampling the integral with equal spacing in $\log|\tau_E|$.
This trick is general and shows that any CMB bispectrum which arises from a cubic diagram of the
combinatorial type shown in Eq.~(\ref{eq:cubic_first_line}) is factorizable, although the number
of terms in the CMB bispectrum may be large, since we get one term for every sampling
point needed to do the $(\tau_E, r)$ double integral.

\subsection{Factorizability for the trispectrum}

We would like to define a notion of factorizability for the
{\em trispectrum}, in order to make data analysis of primordial
trispectra computationally feasible.

We have just seen that in the bispectrum case, the notion of
factorizability derives from the combinatorics of
the Feynman diagram.  In the trispectrum case, the trispectrum
can come from either
a ``contact'' diagram with a quartic vertex,
or an ``exchange'' diagram with two cubic vertices:
\be
\raisebox{-0.7cm}{
\begin{fmffile}{quartic_diagram}
\begin{fmfgraph*}(20,15)
\fmfleft{i1,i2}
\fmfright{o3,o4}
\fmf{fermion}{v1,i1}
\fmf{fermion}{v1,i2}
\fmf{fermion}{v1,o3}
\fmf{fermion}{v1,o4}
\end{fmfgraph*}
\end{fmffile}
}
\hspace{1.5cm}
\raisebox{-0.7cm}{
\begin{fmffile}{exchange_diagram}
\begin{fmfgraph*}(25,15)
\fmfleft{i1,i2}
\fmfright{o3,o4}
\fmf{fermion}{v1,i1}
\fmf{fermion}{v1,i2}
\fmf{photon}{v1,v2}
\fmf{fermion}{v2,o3}
\fmf{fermion}{v2,o4}
\end{fmfgraph*}
\end{fmffile}
}  \label{eq:contact_and_exchange_diagrams}
\ee
Accordingly, we will define two different factorizability conditions
for the trispectrum, ``contact factorizability'' and ``exchange factorizability''.
We will give the precise definitions shortly, but there is one
feature which can be anticipated in advance.  
One might think (by analogy with the bispectrum case) that a
contact diagram always gives rise to a trispectrum of the form
\be
\langle \zeta_{\k_1} \zeta_{\k_2} \zeta_{\k_3} \zeta_{\k_4} \rangle'
   = \int_{-\infty}^0 d\tau_E \,\, \alpha(k_1) \beta(k_2) \gamma(k_3) \delta(k_4)
\ee
in which the integrand is a factorizable function of $k_i = |\k_i|$.
However, this is not fully general: if the quartic operator contains
spatial derivatives, then there will be additional factors $(\k_i\cdot\k_j)$,
as can be seen by inspection of Eqs.~(\ref{eq:gnlB}),~(\ref{eq:gnldpi4})
in the introduction.
In the bispectrum case, we would be able to get rid of a factor such as $(\k_1\cdot\k_2)$
by using the momentum-conserving delta function $\delta^3(\k_1+\k_2+\k_3)$
to write $(\k_1\cdot\k_2) = (k_3^2 - k_1^2 - k_2^2)/2$ and
reduce to the case in which the integrand is factorizable in $k_i = |\k_i|$.
In the trispectrum case, there is no analogous way to eliminate factors $(\k_i\cdot\k_j)$.
The consequence is that our definition of factorizability will contain non-scalar quantities,
i.e.~2D fields with spin $s > 0$.
This is a significant complication compared to the bispectrum case.
For reference, the mathematics of spin-$s$ fields is briefly reviewed in Appendix~\ref{app:spin_s}.

\subsection{Contact factorizability}
\label{ssec:contact_factorizability}

In this section we give the formal definition of contact factorizability.
We define an angular trispectrum $T^{\ell_1\ell_2\ell_3\ell_4}_{m_1m_2m_3m_4}$ to be ``contact factorizable'' if:
\ba
T^{\ell_1\ell_2\ell_3\ell_4}_{m_1m_2m_3m_4} &=& 
 \frac{1}{48} \sum_{I=1}^{\Nfact} A^I_{\ell_1} B^I_{\ell_2} C^I_{\ell_3} D^I_{\ell_4}
   \int d^2\n\, ({}_{\alpha_I}Y_{\ell_1m_1}(\n)) \, ({}_{\beta_I}Y_{\ell_2m_2}(\n)) \,
                  ({}_{\gamma_I}Y_{\ell_3m_3}(\n)) \, ({}_{\delta_I}Y_{\ell_4m_4}(\n)) \nn \\
 && + \mbox{ (23 perm.)} + \mbox{(c.c.)} \label{eq:contact_factorizable_def}
\ea
where $({}_sY_{\ell m})$ denotes a spin-$s$ spherical harmonic (Appendix~\ref{app:spin_s}),
$\alpha_I, \beta_I, \gamma_I, \delta_I$ are integer spins,
and $A^I_\ell, B^I_\ell, C^I_\ell, D^I_\ell$ are $\Nfact$-by-$\ellmax$ matrices.
We will assume that $\alpha_I + \beta_I + \gamma_I + \delta_I = 0$ for each $I$ so that the integrand
in Eq.~(\ref{eq:contact_factorizable_def}) has total spin zero.

Equivalently, we can define a contact factorizable trispectrum by the functional form of $Q_T[a]$.
\ba
Q_T[a] &=& \frac{1}{48} \sum_{I=1}^{\Nfact} \int d^2\n\,
  \Bigg( \sum_{\ell_1m_1} A^I_{\ell_1} a_{\ell_1m_1} ({}_{\alpha_I}Y_{\ell_1m_1}(\n)) \Bigg)
  \Bigg( \sum_{\ell_2m_2} B^I_{\ell_2} a_{\ell_2m_2} ({}_{\beta_I}Y_{\ell_2m_2}(\n)) \Bigg) \nn \\
&& \hspace{2.0cm} \times
  \Bigg( \sum_{\ell_3m_3} C^I_{\ell_3} a_{\ell_3m_3} ({}_{\gamma_I}Y_{\ell_3m_3}(\n)) \Bigg)
  \Bigg( \sum_{\ell_4m_4} D^I_{\ell_4} a_{\ell_4m_4} ({}_{\delta_I}Y_{\ell_4m_4}(\n)) \Bigg) + \mbox{c.c.}  \label{eq:cf_def_Q}
\ea
This definition of factorizability is useful because it is specific enough that
there is a fast algorithm for evaluating $Q[a]$ (by straightforward use of~Eq.(\ref{eq:cf_def_Q})
with fast spherical transforms), yet general enough that many
physically interesting trispectra are contact factorizable.

We illustrate this by example, by showing that the operator $\dot\sigma^2 (\partial\sigma)^2$ from 
the previous section is contact factorizable.  We start from the ``unintegrated'' form of the
trispectrum in the first line of Eq.~(\ref{eq:gnlB}):
\be
\langle \zeta_{\k_1} \zeta_{\k_2} \zeta_{\k_3} \zeta_{\k_4} \rangle_c =
  -\frac{13824}{325} \gnlB A_\zeta^3
  \int_{-\infty}^0 d\tau_E \, \tau_E^2 
     \frac{e^{\sum k_i\tau_E}}{k_1^3k_2^3k_3k_4}
    (1-k_1\tau_E) (1-k_2\tau_E) (\k_1\cdot\k_2) (2\pi)^3 \delta^3\Big(\sum\k_i\Big) + \mbox{5 perm.}
\ee
We plug this into the 3D$\rightarrow$2D projection formula~(\ref{eq:Q_projection})
to obtain an expression for $Q[a]$.  We replace the momentum-conserving delta function
$(2\pi)^3 \delta^3(\sum\k_i)$ by $\int d^3\r\, \exp(i\sum\k_i\cdot\r)$, and then replace
the dot product $(\k_1\cdot\k_2)$ by an appropriately placed pair of derivatives with
respect to $r$.  We obtain:
\ba
Q[a] &=& \frac{3456}{325} \gnlB A_\zeta^3 \int_{-\infty}^0 d\tau_E \, \tau_E^2 \, d^3\r \,
  \left( \frac{\partial}{\partial \r_i} \sum_{\ell_1m_1} \int \frac{d^3\k_1}{(2\pi)^3} 4\pi 
      (-i)^{\ell_1} \frac{(1-k_1\tau_E)e^{k_1\tau_E}}{k_1^3} \Delta_{\ell_1}(k_1) e^{i\k_1\cdot\r} a_{\ell_1m_1} Y_{\ell_1m_1}(\hk_1) \right) \nn \\
&& \hspace{4.65cm} \times \left( \frac{\partial}{\partial \r_i} \sum_{\ell_2m_2} \int \frac{d^3\k_2}{(2\pi)^3} 4\pi 
      (-i)^{\ell_2} \frac{(1-k_2\tau_E)e^{k_2\tau_E}}{k_2^3} \Delta_{\ell_2}(k_2) e^{i\k_2\cdot\r} a_{\ell_2m_2} Y_{\ell_2m_2}(\hk_2) \right) \nn \\
&& \hspace{4.65cm} \times \left( \sum_{\ell_3m_3} \int \frac{d^3\k_3}{(2\pi)^3} 4\pi (-i)^{\ell_3} \frac{e^{k_3\tau_E}}{k_3}
    \Delta_{\ell_3}(k_3) e^{i\k_3\cdot\r} a_{\ell_3m_3} Y_{\ell_3m_3}(\hk_3) \right) \nn \\
&& \hspace{4.65cm} \times \left( \sum_{\ell_4m_4} \int \frac{d^3\k_4}{(2\pi)^3} 4\pi (-i)^{\ell_4} \frac{e^{k_4\tau_E}}{k_4}
    \Delta_{\ell_4}(k_4) e^{i\k_4\cdot\r} a_{\ell_4m_4} Y_{\ell_4m_4}(\hk_4) \right)
\ea
Following a standard trick~\cite{Wang:1999vf},
the next step is to Rayleigh expand each exponential as $\exp(i\k_i\cdot\r) = 4\pi \sum_{\ell m} i^\ell j_\ell(k_ir) Y_{\ell m}(\hr) Y_{\ell m}^*(\hk)$
and do the angular parts of the $k$-integrals, obtaining:
\ba
Q[a] &=& \frac{3456}{325} \gnlB A_\zeta^3 \int_{-\infty}^0 d\tau_E \, \tau_E^2 \, d^3\r \,
   \left( \frac{\partial}{\partial\r_i} \sum_{\ell_1m_1} \int \frac{2 k_1^2 dk_1}{\pi} \frac{(1-k_1\tau_E)e^{k_1\tau_E}}{k_1^3}
           \Delta_{\ell_1}(k_1) j_{\ell_1}(k_1r) a_{\ell_1m_1} Y_{\ell_1m_1}(\hr) \right) \nn \\
&& \hspace{4.65cm} \times
   \left( \frac{\partial}{\partial\r_i} \sum_{\ell_2m_2} \int \frac{2 k_2^2 dk_2}{\pi} \frac{(1-k_2\tau_E)e^{k_2\tau_E}}{k_2^3}
           \Delta_{\ell_2}(k_2) j_{\ell_2}(k_2r) a_{\ell_2m_2} Y_{\ell_2m_2}(\hr) \right) \nn \\
&& \hspace{4.65cm} \times
   \left( \sum_{\ell_3m_3} \int \frac{2 k_3^2 dk_3}{\pi} \frac{e^{k_3\tau_E}}{k_3}
           \Delta_{\ell_3}(k_3) j_{\ell_3}(k_3r) a_{\ell_3m_3} Y_{\ell_3m_3}(\hr) \right) \nn \\
&& \hspace{4.65cm} \times
   \left( \sum_{\ell_4m_4} \int \frac{2 k_4^2 dk_4}{\pi} \frac{e^{k_4\tau_E}}{k_4}
           \Delta_{\ell_4}(k_4) j_{\ell_4}(k_4r) a_{\ell_4m_4} Y_{\ell_4m_4}(\hr) \right)
\ea
Next we split the dot product of gradients $\sum_i \frac{\partial f}{\partial r_i} \frac{\partial f}{\partial r_i}$,
as a sum of two terms: a term in which both derivatives act in the radial direction, and a term containing angular derivatives.
More formally, we can write:
\be
\sum_i \frac{\partial f(\r)}{\partial \r_i} \frac{\partial f(\r)}{\partial \r_i} = \left( \frac{\partial f(\r)}{\partial r} \right)^2
  + \frac{1}{r^2} (\sraise f)^* (\sraise f)
\ee
where $f(\r)$ is any scalar-valued function and $\sraise$ is the spin-raising operator (see Appendix~\ref{app:spin_s}).
Plugging this in we get:
\ba
Q[a] &=& \frac{384}{13} \gnlB \int_{-\infty}^0 d\tau_E  \int_0^\infty dr \int d^2\n \, \tau_E^2 r^2
   \left( \sum_{\ell m} \mu_\ell(\tau_E,r) a_{\ell m} Y_{\ell m}(\n) \right)^2 \nn \\
  && \hspace{2cm} \times \Bigg[ 
     \left( \sum_{\ell'm'} \nu_{\ell'}(\tau_E,r) a_{\ell'm'} Y_{\ell'm'}(\n) \right)^2 +
     \left| \sum_{\ell'm'} \omega_{\ell'}(\tau_E,r) a_{\ell'm'} ({}_1Y_{\ell'm'}(\n)) \right|^2
    \Bigg]  \label{eq:Q_dotpi2_dpi2_final}
\ea
where we have used the identity $\sraise Y_{\ell m} = \sqrt{\ell(\ell+1)} ({}_1Y_{\ell m})$ and defined
\ba
\mu_\ell(\tau_E,r) &=& \left( \frac{3}{5} \right)^{1/2} \int \frac{2k^2 dk}{\pi} e^{k\tau_E} k^{5/4} P_\zeta(k)^{3/4} \Delta_\ell(k) j_\ell(kr) \nn \\
\nu_\ell(\tau_E,r) &=& \left( \frac{3}{5} \right)^{1/2} \int \frac{2k^2 dk}{\pi} (1-k\tau_E) e^{k\tau_E} k^{1/4} P_\zeta(k)^{3/4} \Delta_\ell(k) j_\ell'(kr)  \nn \\
\omega_\ell(\tau_E,r) &=& 
  \frac{\sqrt{\ell(\ell+1)}}{r}
  \left( \frac{3}{5} \right)^{1/2} 
  \int \frac{2k^2 dk}{\pi} (1-k\tau_E) e^{k\tau_E} k^{-3/4} P_\zeta(k)^{3/4} \Delta_\ell(k) j_\ell(kr)  \label{eq:mnw_def}
\ea
As anticipated, $Q[a]$ is of contact factorizable form~(\ref{eq:contact_factorizable_def}) after replacing the
$(\tau_E,r)$ double integral by a finite sum.
This calculation generalizes to show that trispectra of the following
types are contact factorizable:
\begin{enumerate}
\item Any $\zeta$-trispectrum which is a product of functions $f(k_i)$
and any number of dot products of the form $(\k_i\cdot\k_j)$.
The local trispectrum in Eq.~(\ref{eq:gnlloc_def}) is an example.
In this case we obtain a factorizable representation by applying the projection
formula~(\ref{eq:Q_projection}) and approximating the $r$ integral by a finite sum.
\item Any trispectrum which arises from a local quartic operator in the
inflationary action.  We reduce to the previous case by writing the
$\zeta$-trispectrum as a time integral, and replacing the $(\tau_E, r)$
double integral by a finite sum.
\end{enumerate}
In Appendix~\ref{app:contact_factorizability}, we work this out explicitly for
the local, $(\dot\sigma^4)$, and $(\partial\sigma)^4$ trispectra.
The resulting formulas
(Eqs.~(\ref{eq:Q_gnlloc}),~(\ref{eq:Q_gnldotpi4}),~(\ref{eq:Q_gnldpi4}))
are useful for reference, and for numerical evaluation of trispectra
in our analysis pipelines.
In this appendix,
we also present our scheme for generalizing from the scale invariant case
(assumed above for simplicity) to the case of a power-law spectrum
$P_\zeta(k) = A_\zeta k^{n_s-4}$.

\subsection{Exchange factorizability}

As mentioned previously, there is another notion of factorizability 
for the trispectrum, ``exchange factorizability'', which arises from
exchange of a light particle during inflation, and also leads
to a computationally fast form for $Q[a]$.
It would be interesting to explore the phenomenology and data analysis
of exchange trispectra.
For example, quasi-single field inflation~\cite{Chen:2009zp,Bartolo:2010di,Baumann:2011nk,Green:2013rd}
should generate interesting continuous families of trispectra,
as the mass of the exchanged particle and the type of cubic operator are varied.
However in this paper, we will restrict attention to contact factorizable trispectra,
leaving this generalization for future work.
The main technical obstacle in generalizing the machinery of this paper to the
exchange factorizable case is developing an analogue of the optimization algorithm
in~\S\ref{sec:contact_optimization}.
In this section, we simply give the definition of exchange factorizability and
a few examples.

A CMB trispectrum is said to be exchange factorizable if the $Q$-symbol is given by:
\ba
Q_T[a] &=& \frac{1}{48} \sum_{I=1}^{N_1} \sum_{J=1}^{N_2} \sum_{\ell m}
             {\mathcal M}_\ell^{IJ}
\left( \int d^2\n\, ({}_{\alpha_I+\beta_I}Y_{\ell m}(\n))^*
  \left( \sum_{\ell_1m_1} A^I_{\ell_1} a_{\ell_1m_1} ({}_{\alpha_I}Y_{\ell_1m_1}(\n)) \right)
  \left( \sum_{\ell_2m_2} B^I_{\ell_2} a_{\ell_2m_2} ({}_{\beta_I}Y_{\ell_2m_2}(\n)) \right)
\right) \nn \\
&& \hspace{0.5cm} \times
\left( \int d^2\n'\, ({}_{\gamma_J+\delta_J}Y_{\ell m}(\n'))^*
  \left( \sum_{\ell_3m_3} C^J_{\ell_3} a_{\ell_3m_3} ({}_{\gamma_J}Y_{\ell_3m_3}(\n')) \right)
  \left( \sum_{\ell_4m_4} D^J_{\ell_4} a_{\ell_4m_4} ({}_{\delta_J}Y_{\ell_4m_4}(\n')) \right)
\right)^* + \mbox{c.c.}  \label{eq:exchange_factorizability}
\ea
An exchange trispectrum is parametrized by integer spins $\alpha_I$, $\beta_I$, $\gamma_J$, $\delta_J$
and coefficient arrays ${\mathcal M}_\ell^{IJ}$, $A^I_\ell$, $B^I_\ell$, $C^J_\ell$, $D^J_\ell$,
where $I=1,\cdots,N_1$ and $J=1,\cdots,N_2$.
Note that contact factorizability can be viewed as the special case of exchange factorizability
where $N_1=N_2$ and ${\mathcal M}_\ell^{IJ} = \delta_{IJ}$ (with no $\ell$ dependence),
so that the outer $\ell$-sum gives a delta function $\delta^2(\n-\n')$.
It is easy to see from the definition~(\ref{eq:exchange_factorizability})
that $Q_T[a]$ can be computed efficiently (more precisely, with cost 
$\bigoh(N_1\ellmax^3 + N_2\ellmax^3 + N_1N_2\ellmax^2)$)
by an appropriate sequence of fast spherical harmonic transforms.

The canonical example of an exchange trispectrum is the $\taunl$-trispectrum, defined previously in Eq.~(\ref{eq:taunl_def}).
To show that it is exchange factorizable, we first rewrite the $\zeta$-trispectrum as:
\be
\langle \zeta_{\k_1} \zeta_{\k_2} \zeta_{\k_3} \zeta_{\k_4} \rangle
  = \taunl \int \frac{d^3\q}{(2\pi)^3} P_\zeta(q) P_\zeta(k_2) P_\zeta(k_4) (2\pi)^6 \delta^3(\k_1+\k_2+\q) \delta^3(\k_3+\k_4-\q) + \mbox{(11 perm.)}
\ee
Following the calculation in the last section, 
we plug into the 3D$\rightarrow$2D projection formula~(\ref{eq:Q_projection}),
Rayleigh expand both delta functions,
and do the angular $q$ and $k$-integrals.
When the dust settles we get:
\ba
Q_T[a] &=& \frac{\taunl}{2} \int d^3\r\, d^3\r'
  \left( \int \frac{2q^2\,dq}{\pi} \sum_{\ell m} j_\ell(qr) j_\ell(qr') P_\zeta(q) Y_{\ell m}^*(\hr) Y_{\ell m}(\hr') \right) \nn \\
&& \hspace{2cm} \times
  \left( \int \frac{2k_1^2\,dk_1}{\pi} \sum_{\ell_1m_1} j_{\ell_1}(k_1r) \Delta_{\ell_1}(k_1) a_{\ell_1m_1} Y_{\ell_1m_1}(\hr) \right) \nn \\
&& \hspace{2cm} \times
  \left( \int \frac{2k_2^2\,dk_2}{\pi} \sum_{\ell_2m_2}  j_{\ell_2}(k_2r) \Delta_{\ell_2}(k_2) P_\zeta(k_2) a_{\ell_2m_2} Y_{\ell_2m_2}(\hr) \right) \nn \\
&& \hspace{2cm} \times
  \left( \int \frac{2k_3^2\,dk_3}{\pi} \sum_{\ell_3m_3} j_{\ell_3}(k_3r') \Delta_{\ell_3}(k_3) a_{\ell_3m_3} Y_{\ell_3m_3}(\hr') \right) \nn \\
&& \hspace{2cm} \times
  \left( \int \frac{2k_4^2\,dk_4}{\pi} \sum_{\ell_4m_4} j_{\ell_4}(k_4r') \Delta_{\ell_4}(k_4) P_\zeta(k_4) a_{\ell_4m_4} Y_{\ell_4m_4}(\hr') \right)
\ea
Using the standard notation~\cite{Komatsu:2003iq,Munshi:2009wy}:
\ba
\alpha_\ell(r) &=& \frac{5}{3} \int \frac{2 k^2 dk}{\pi} \Delta_\ell(k) j_\ell(kr) \hspace{1cm} \nn \\
\beta_\ell(r) &=& \frac{3}{5} \int \frac{2 k^2 dk}{\pi} \Delta_\ell(k) P_\zeta(k) j_\ell(kr)  \nn \\
F_\ell(r,r') &=& \int \frac{2 q^2 dq}{\pi} j_\ell(qr) j_\ell(qr') P_\zeta(q)  \label{eq:alpha_beta}
\ea
we can write this in the form:
\ba
Q_T[a] &=& \frac{\taunl}{2} \int r^2 dr \int r'^2 dr' \, \sum_{\ell m} F_\ell(r,r') \nn \\
&& \hspace{1cm} \times
     \left( \int d^2\n \,\, Y_{\ell m}^*(\n) 
        \left( \sum_{\ell_1m_1} \alpha_{\ell_1}(r) a_{\ell_1m_1} Y_{\ell_1m_1}(\hr) \right)
        \left( \sum_{\ell_2m_2} \beta_{\ell_2}(r) a_{\ell_2m_2} Y_{\ell_2m_2}(\hr) \right)
     \right) \nn \\
&& \hspace{1cm} \times
     \left( \int d^2\n' \,\, Y_{\ell m}^*(\n')
        \left( \sum_{\ell_3m_3} \alpha_{\ell_3}(r') a_{\ell_3m_3} Y_{\ell_3m_3}(\hr') \right)
        \left( \sum_{\ell_4m_4} \beta_{\ell_4}(r') a_{\ell_4m_4} Y_{\ell_4m_4}(\hr') \right)
     \right)^*
\ea
Comparing this expression with the definition~(\ref{eq:exchange_factorizability}), we see that
the $\taunl$ trispectrum is exchange factorizable.
This calculation generalizes to show that trispectra of the following
types are exchange factorizable:
\begin{enumerate}
\item Any $\zeta$-trispectrum which is a product of functions $f(k_i)$,
any number of dot products $(\k_i\cdot\k_j)$, and one factor of the form $f(|\k_1+\k_2|)$.
\item Any $\zeta$-trispectrum generated by an ``exchange'' diagram of the type
shown in the right side of Eq.~(\ref{eq:contact_and_exchange_diagrams}).
\end{enumerate}
Although our definition of exchange factorizability was constructed
with inflationary trispectra in mind, there are also interesting
non-primordial examples, for example gravitational lensing.
The lensing trispectrum $T^{\ell_1\ell_2\ell_3\ell_4}_{m_1m_2m_3m_4}$ can found in
Eq.~(76) of~\cite{Hu:2001fa}.
Plugging this into the definition~(\ref{eq:Q_def}) of $Q[a]$, we get the following
expression for $Q_T[a]$:
\ba
Q_T[a] &=& \frac{1}{8} \sum_{s=\pm 1} \sum_{s'=\pm 1} \sum_{\ell m} \ell(\ell+1) C_\ell^{\phi\phi} \nn \\
&& \hspace{1cm} \times
  \left(
     \int ({}_s Y_{\ell m})^* 
             \left( \sum_{\ell_1m_1} a_{\ell_1m_1} Y_{\ell_1m_1} \right)
             \left( \sum_{\ell_2m_2} \sqrt{\ell_2(\ell_2+1)} C_{\ell_2}^{TT} a_{\ell_2m_2} ({}_sY_{\ell_2m_2}) \right)     
  \right) \nn \\
&& \hspace{1cm} \times
  \left(
     \int ({}_{s'} Y_{\ell m})^* 
             \left( \sum_{\ell_3m_3} a_{\ell_3m_3} Y_{\ell_3m_3} \right)
             \left( \sum_{\ell_4m_4} \sqrt{\ell_4(\ell_4+1)} C_{\ell_4}^{TT} a_{\ell_4m_4} ({}_{s'}Y_{\ell_4m_4}) \right)     
  \right)^*  \label{eq:Q_lensing}
\ea
Comparing with the definition~(\ref{eq:exchange_factorizability}), we see that the
CMB lensing trispectrum is exchange factorizable.
In this case, there is no obstacle to applying the machinery in this paper
(since there are only a few terms in the trispectrum, we do not need an optimization
algorithm).
The pipelines we will develop in \S\ref{sec:pipelines} could be
used to give an optimal analysis of CMB lensing.
Such an analysis would be qualitatively similar to other lens reconstruction 
analyses (e.g.~\cite{Ade:2013tyw,Story:2014dwa,vanEngelen:2014zlh})
but different in its approach to minimizing bias due to errors in
modeling the two-point function.

\section{Fisher matrix algorithms}
\label{sec:fisher}

Consider an ideal CMB experiment with full sky coverage and
isotropic noise.  Such an experiment is completely specified
by its noise power spectrum $N_\ell$.
Given angular trispectra $T_1, \cdots, T_N$, the $N$-by-$N$ Fisher matrix is defined by
\be
F_{ij} = \frac{1}{4!} \sum_{\ell_im_i} \frac{(T_i)^{\ell_1\ell_2\ell_3\ell_4*}_{m_1m_2m_3m_4} (T_j)^{'\ell_1\ell_2\ell_3\ell_4}_{m_1m_2m_3m_4}}{(C_{\ell_1}+N_{\ell_1})
(C_{\ell_2}+N_{\ell_2}) (C_{\ell_3}+N_{\ell_3}) (C_{\ell_4}+N_{\ell_4})}  \label{eq:Fdef_isotropic}
\ee
and is interpreted as follows.
If the CMB trispectrum is assumed to be a linear combination $T = \sum_i g_{NL}^i T_i$ of
the trispectra $T_i$, and the coefficients $g_{NL}^i$ are jointly estimated using optimal estimators,
then the estimator covariance is the inverse Fisher matrix:
\be
\Cov(g_{NL}^i, g_{NL}^j) = (F^{-1})_{ij}
\ee
The Fisher matrix is a powerful tool for forecasting and analysis of parameter degeneracies.\footnote{Of 
course, the assumptions of full sky coverage and isotropic noise will not be satisfied for a real experiment,
but it is usually a good approximation to approximate the noise as isotropic, and account for sky
coverage by scaling $F \rightarrow \fsky F$.}
It will also play a central role in the trispectrum optimization algorithm which 
we will give in~\S\ref{sec:contact_optimization}.

Computing the Fisher matrix directly from the definition~(\ref{eq:Fdef_isotropic})
has computational cost $\bigoh(\ellmax^7)$  and is usually computationally prohibitive. 
In this section we will construct fast algorithms.

\subsection{Monte Carlo Fisher matrix algorithm}
\label{ssec:fisher_mc}

\par\noindent
A very simple fast algorithm for estimating the Fisher matrix is
to use the following Monte Carlo procedure:
\be
F(T,T') = \left\langle \sum_{\ell m} \frac{(\partial_{\ell m} Q_T[\tilde a])^* (\partial_{\ell m} Q_{T'}[\tilde a])}{C_\ell + N_\ell} \right\rangle_{\tilde a}  \label{eq:fisher_mc}
\ee
where $\langle \cdot \rangle_{\tilde a}$ denotes an expectation value
over Gaussian random fields $\tilde a_{\ell m}$ with power spectrum
$1/(C_\ell+N_\ell)$ (not power spectrum $C_\ell+N_\ell$).
The Monte Carlo error on the Fisher matrix is proportional to $1/\sqrt{\Nmc}$, 
where $\Nmc$ is the number of random realizations.
In practice we find that the proportionality coefficient is very favorable;
even one Monte Carlo realization is enough to approximate the Fisher matrix
to $\approx$10\% percent for the local trispectrum, or a few percent for the
trispectra generated by one of the quartic operators $\dot\sigma^4$, $\dot\sigma^2 (\partial_i\sigma)^2$,
or $(\partial_i \sigma)^2 (\partial_j \sigma)^2$.

The Monte Carlo algorithm in Eq.~(\ref{eq:fisher_mc}) may appear to be more
complicated than necessary, since one can give a simpler Monte Carlo algorithm
by simply estimating the variance of the all-sky optimal estimator for the
trispectrum $T$.
However, the fractional Monte Carlo error of this simpler algorithm would 
by the ``standard'' $\sqrt{2/\Nmc}$,
so it takes many random realizations to obtain a useful estimate of the Fisher
matrix.  For this reason we always use the algorithm~(\ref{eq:fisher_mc}) to estimate
the Fisher matrix by Monte Carlo.

This is our first example of a phenomenon which will recur throughout the paper:
there is an ``obvious'' Monte Carlo scheme which requires a large number of Monte Carlos,
and an alternate scheme which is significantly faster.
This phenomenon also occurs in the bispectrum context (e.g. Fig.~6 of~\cite{Smith:2006ud}),
where it was referred to as ``fast MC''.
We will see more examples shortly.

\subsection{Exact Fisher matrix algorithm for contact factorizable trispectra}
\label{ssec:fisher_exact}

In this section, we will present an exact (i.e.~non Monte Carlo based)
Fisher matrix algorithm, which assumes contact factorizable trispectra.
This is less generality than the Monte Carlo algorithm from the preceding section,
which only requires a fast algorithm for evaluating the $Q$-symbol.

Let $T$ be contact factorizable with spins $\alpha_I,\beta_I,\gamma_I,\delta_I$ 
and coefficients $A_{I\ell},B_{I\ell},C_{I\ell},D_{I\ell}$ with $I=1,2,\cdots,\Nfact$.
Likewise let $T'$ be contact factorizable with spins $\alpha'_J,\beta'_J,\gamma'_J,\delta'_J$ 
and coefficients $A'_{J\ell},B'_{J\ell},C'_{J\ell},D'_{J\ell}$ with $J=1,2,\cdots,\Nfact'$.

To obtain an exact expression for $F(T,T')$, we calculate as follows.  First write:
\be
F(T,T') = \Big\langle Q_T[\ta] Q_{T'}[\ta] \Big\rangle_{\fc}  \label{eq:f_exact_avg}
\ee
where $\ta$ is a Gaussian random field with power spectrum $1/(C_\ell+N_\ell)$
and $\langle \cdot \rangle_{\fc}$ denotes the fully connected part of the expectation value,
i.e.~the sum over Wick contractions in which all four contractions connect a factor of $Q_T[\ta]$
to a factor of $Q_{T'}[\ta]$.
We write the $Q$-symbols in the abbreviated form:
\ba
Q_T[\ta] &=& \frac{1}{48} \sum_{I=1}^{\Nfact} \int d^2\n \,
  M^{A_I}_{\alpha_I}(\n) M^{B_I}_{\beta_I}(\n) M^{C_I}_{\gamma_I}(\n) M^{D_I}_{\delta_I}(\n) + \mbox{c.c.} \nn \\
Q_{T'}[\ta] &=& \frac{1}{48} \sum_{J=1}^{\Nfact'} \int d^2\n' \,
  M^{A_J'}_{\alpha_J'}(\n') M^{B_J'}_{\beta_J'}(\n') M^{C_J'}_{\gamma_J'}(\n') M^{D_J'}_{\delta_J'}(\n')
  + \mbox{c.c.}  \label{eq:f_exact_qsym}
\ea
where we have introduced the following notation.
If $X = X_\ell$ is any $\ell$-dependent quantity and $s$ is an integer spin,
then $M^X_s$ is the spin-$s$ map defined by
\be
M^X_s\!(\n) = \sum_{\ell m} X_\ell \, \ta_{\ell m} \, ({}_sY_{\ell m}(\n))
\ee
We plug the $Q$-symbols in Eq.~(\ref{eq:f_exact_qsym}) into
the the expression~(\ref{eq:f_exact_avg}) for $F(T,T')$ and expand
the result as a sum of Wick contractions.
The contraction between two $M$ fields or their complex conjugates
is easy to calculate using the sum rule~(\ref{eq:sum_rule}) in Appendix~\ref{app:spin_s}:
\be
\wick{1}{<1M^X_s(\n) \,\, >1M^{X'}_{s'}(\n')} = (-1)^s \zeta^{XX'-}_{ss'}(\theta)
   \hspace{1cm}
\wick{1}{<1M^X_s(\n) \,\, >1M^{X'}_{s'}(\n')^*} = (-1)^s \zeta^{XX'+}_{ss'}(\theta)  \label{eq:MM_contraction}
\ee
where $\theta = \cos^{-1}(\n\cdot\n')$ is the angle 
between $\n,\n'$ and we have defined correlation functions
\be
\zeta^{XX'\pm}_{ss'}(\theta) = (\pm 1)^{s'} \sum_\ell \frac{2\ell+1}{4\pi} \frac{X_\ell X'_\ell}{C_\ell+N_\ell} d^\ell_{s,\mp s'}(\theta)
\ee
When the dust settles, we get the following explicit formula for $F(T,T')$:
\ba
F(T,T') &=& \frac{\pi^2}{144} \sum_{I=1}^{\Nfact} \sum_{J=1}^{\Nfact'} \sum_{\sigma = +,-}
   \int_{-1}^1 d(\cos\theta) \, \Big( 
         \zeta^{A_I,A'_J\sigma}_{\alpha_I\alpha_J}(\theta)
         \zeta^{B_I,B'_J\sigma}_{\beta_I\beta'_J}(\theta)
         \zeta^{C_I,C'_J\sigma}_{\gamma_I\gamma'_J}(\theta)
         \zeta^{D_I,D'_J\sigma}_{\delta_I\delta'_J}(\theta) + \mbox{23 perm.} \Big)
\ea
The integral can be evaluated exactly using Gauss-Legendre quadrature with $(2\ellmax+1)$ points,
since the integrand is a polynomial of degree $4\ellmax$.  For each quadrature point $\theta$,
one $\zeta$-function value $\zeta(\theta)$ can be computed with cost $\bigoh(\ellmax)$
using the recursion~(\ref{eq:wignerd_recursion}).  Thus the computational cost of computing
$F(T,T')$ is $\bigoh(\Nfact \Nfact' \ellmax^2)$.

Let us compare the computational cost of the exact algorithm in this section
with the Monte Carlo algorithm from the previous section (assuming contact
factorizable trispectra).
Suppose we have $\Ntr$ total trispectra (i.e.~the Fisher matrix being computed is $\Ntr$-by-$\Ntr$)
and each trispectrum has $\Nfact$ factorizable terms.
The exact algorithm has cost $\bigoh(\Ntr^2 \Nfact^2 \ellmax^2)$, and
the Monte Carlo algorithm has cost $\bigoh(\Nmc \Ntr \Nfact \ellmax^3 + \Nmc \Ntr^2 \ellmax^2)$.

Most interesting trispectra have factorizable representations with at most
a few hundred terms (see Table~\ref{tab:opt} below), and the exact algorithm is actually faster
due to the smaller power of $\ellmax$.
The Monte Carlo algorithm is useful in situations where the total number of
factorizable terms is very large.
For an example, see Appendix~\ref{app:integrals}, where we describe a Fisher matrix
based convergence test on numerical calculation of trispectra.
The Monte Carlo algorithm is also the only option for exchange factorizable trispectra.
For example, we will use the Monte Carlo Fisher matrix to compute the lensing bias to our
WMAP $g_{NL}$ estimates (see Eq.~(\ref{eq:lensing_bias}) below).

\section{Optimization algorithm for contact factorizable trispectra}
\label{sec:contact_optimization}

Consider the trispectrum generated by a quartic operator such as $\dot\sigma^4$.
So far, we have proposed a scheme for representing the tripsectrum
in contact factorizable form~(\ref{eq:contact_factorizable_def}), and
shown that this representation reduces the computational cost of data analysis from $\bigoh(\ellmax^7)$ to
$\bigoh(\Nfact \ellmax^3)$, by providing a fast algorithm for computing the $Q$-symbol $Q[a]$.
However, this is not quite enough to bring the computational cost fully under control,
since the number of terms $\Nfact$ in the factorizable representation can be very large.

For example, consider the trispectrum generated by the quartic operator $\dot\sigma^2 (\partial\sigma)^2$.
To represent it in factorizable form, we must approximate the double $(\tau_E, r)$ integral
in Eq.~(\ref{eq:Q_dotpi2_dpi2_final}) by a finite sum.
To accurately approximate the detailed $(\ell,m)$ dependence of the trispectrum, a huge
number of sampling points in the $(\tau_E,r)$ plane is required.
This issue is studied in detail in Appendix~\ref{app:integrals}.
As explained there, our sampling scheme has the property that the finite-sampled
trispectrum approximates the exact trispectrum in a controlled sense: there is an
end-to-end convergence test which shows that the two are nearly equal in the metric
defined by the Fisher matrix.
However, this requires many sampling points in the $(\tau_E,r)$ plane, e.g.~for the operator
$\dot\sigma^2 (\partial\sigma)^2$ and WMAP noise levels, we find that 31763 sampling
points are needed!

Fortunately, there is an optimization algorithm, first proposed for the bispectrum in~\cite{Smith:2006ud},
which can dramatically reduce the number of terms in the factorizable representation.
The input to the algorithm is a trispectrum which has been represented in contact factorizable
form with a large number $\Nin$ of terms.  We write:
\be
T_{\rm in} = \sum_{I=1}^{\Nin} T_I
\ee
where $T_I$ is the $I$-th term in the factorizable representation.
The output is an ``optimized'' representation obtained by linearly combining
a small subset of terms in the input representation.  Formally:
\be
T_{\rm out} = \sum_{J=1}^{\Nout} w_J T_{I_J}
\ee
with the subset $\{ T_{I_1}, T_{I_2}, \cdots \}$ of terms
and weights $w_J$ determined by the optimization algorithm.

The first step in the optimization algorithm is to compute the $\Nin$-by-$\Nin$
Fisher matrix $F_{IJ} = F(T_I,T_J)$ of individual terms in the input representation.
We use the exact Fisher matrix algorithm from~\S\ref{ssec:fisher_exact} to compute $F_{IJ}$.\footnote{In
this case, the exact Fisher matrix algorithm has computational cost $\bigoh(\Nin^2 \ellmax^2)$, whereas
the Monte Carlo Fisher matrix algorithm from~\S\ref{ssec:fisher_mc} has cost 
$\bigoh(\Nmc \Nin \ellmax^3 + \Nmc \Nin^2 \ellmax^2)$.
The exact algorithm turns out to be faster even for modest values of $\Nmc$.}
Once this matrix has been computed, the optimized representation $T_{\rm out}$ can then be computed  using a
purely formal linear algebra procedure described in~\S V.A of~\cite{Smith:2006ud}.
(This procedure was developed for purposes of optimizing the bispectrum, but
the Fisher matrix contains all the information needed for the optimization, and
once it has been computed it no longer matters whether the underlying objects are
bispectra or trispectra.)
The optimization algorithm guarantees that the input and output trispectra are
nearly equal, in the sense that 
\be 
F(T_{\rm in} - T_{\rm out}, T_{\rm in} - T_{\rm out}) < 10^{-5} F(T_{\rm in}, T_{\rm in})
\ee
where $F(T,T')$ denotes the Fisher matrix element.
This definition of ``nearly equal'' means that the two trispectra
cannot be distinguished observationally with statistical significance.
Because the Fisher matrix depends on the noise power spectrum, the
optimized trispectrum depends weakly on the noise properties of the experiment
being considered.

There is one more wrinkle: for the large input representations considered here
with $\Nin \gsim 10^4$, computing the Fisher matrix $F_{IJ}$ is a computational
bottleneck.
To get around this problem, we use a two-stage optimization algorithm as follows.
We divide the input representation into $M$ ``chunks'' of size $(\Nin/M)$, where
typically $M=16$ or 32, and optimize each chunk separately.
(Note that the total cost of optimizing all chunks is less than the cost of
optimizing their sum, since the exact Fisher matrix algorithm scales as $\Nfact^2$,
not $\Nfact$.)
We then combine the optimized chunks to obtain a semi-optimized representation of
the input bispectrum, and do a second pass of the optimization algorithm to obtain
the final optimized representation.

In Table~\ref{tab:opt}, we show results of applying the optimization
algorithm to the $\gnlloc$, $\dot\sigma^4$, $\dot\sigma^2 (\partial\sigma)^2$,
and $(\partial\sigma)^4$ trispectra.
It is seen that the optimization algorithm results in a dramatic
reduction in the size of the factorizable representation.
The optimized representations will be used throughout the rest of the paper.

\begin{table}[h]
\begin{center}
\begin{tabular}{|c|c|c|}  \hline
  Trispectrum   &  $\Nin$ & $\Nout$ \\  \hline\hline
 $\gnlloc$  &  960   &   16  \\
 $\dot\sigma^4$  &  31763  &  52  \\
 $\dot\sigma^2 (\partial\sigma)^2 $  &  63526  &  110  \\
 $(\partial\sigma)^4$  &  95289  &  141 \\ \hline
\end{tabular}
\end{center}
\caption{Number of factorizable terms $\Nin$ needed to represent each trispectrum
by ``brute-force'' replacement of integrals by finite sums (see Appendix~\ref{app:integrals} for details),
 and number of terms $\Nout$ obtained after running the optimization algorithm with WMAP noise levels.}
\label{tab:opt}
\end{table}

\section{Fisher matrix analysis of the trispectra $\dot\sigma^4$, $\dot\sigma^2 (\partial\sigma)^2$, and $(\partial\sigma)^4$}
\label{sec:shapes}

In this section, we will study correlations between the
trispectra $\{ \dot\sigma^4, \dot\sigma^2 (\partial\sigma)^2, (\partial\sigma)^4 \}$,
using the CMB Fisher matrix studied in \S\ref{sec:fisher}.

We note in passing that for primordial trispectra, there is an alternate,
simpler choice of Fisher matrix defined by:
\be
F(T_1,T_2) = \int \frac{d^3\k_1\,d^3\k_2\,d^3\k_3\,d^3\k_4}{(2\pi)^{12}}
   \frac{ \langle \zeta_{\k_1} \zeta_{\k_2} \zeta_{\k_3} \zeta_{\k_4} \rangle'_1
       \,\,\, \langle \zeta_{\k_1} \zeta_{\k_2} \zeta_{\k_3} \zeta_{\k_4} \rangle'_2}{P_\zeta(k_1) P_\zeta(k_2) P_\zeta(k_3) P_\zeta(k_4)}
   \,\, (2\pi)^3 \delta^3\Big(\sum\k_i\Big)  \label{eq:fisher3d}
\ee
This is the appropriate definition for an observer who sees all $\zeta$-modes
in a 3D volume (as opposed an observer who sees all CMB modes on a 2D sky).
In the bispectrum case, the 3D Fisher matrix and the 2D CMB Fisher matrix
tend to give nearly identical results in practice.
However, this need not be so for the trispectrum, since 3D$\rightarrow$2D projection
actually reduces the dimensionality of the parameter space.
The 3D $\zeta$-bispectrum and the 2D CMB bispectrum are both functions of three parameters
(assuming translation and rotation invariance in the 3D case, and rotation invariance in the 2D case).
In contrast, the 3D $\zeta$-trispectrum is a function of six parameters, but the
2D CMB trispectrum is a function of only five.
As a point of mathematical principle, this implies that there must exist examples of
$\zeta$-trispectra which are weakly correlated in 3D, but become highly correlated when
projected to the CMB.
For this reason, we have used the CMB Fisher matrix throughout this section rather
than the simpler 3D Fisher matrix~(\ref{eq:fisher3d}), but we actually find that the
two Fisher matrices agree well for the trispectra under consideration.

Let us recall the Fisher matrix analysis for the bispectrum which gives rise
to the parameters $\fnleq$ and $\fnlorth$~\cite{Senatore:2009gt}.
There are two cubic operators to consider, $\dot\pi^3$ and $\dot\pi(\partial\pi)^2$.
These generate bispectra which are nonidentical, but correlated at the $\approx$0.9 level.
This level of correlation is not so large that the two operators can be treated as
indistinguishable, but is large enough that orthogonalization is convenient~\cite{Senatore:2009gt}.
We therefore apply a linear transformation in the parameter space $(\dot\pi^3, \dot\pi(\partial\pi)^2)$
to define (approximately) decorrelated observables $\fnleq, \fnlorth$.

Analogously, for the trispectrum, the three quartic operators 
$\{ \dot\sigma^4, \dot\sigma^2 (\partial\sigma)^2, (\partial\sigma)^4 \}$ generate three distinct trispectra.
Using the exact Fisher matrix algorithm from~\S\ref{ssec:fisher_exact}, the correlation matrix
between these trispectra is found to be:
\be
\left( \begin{array}{ccc}
  1  &  0.9484  &  0.7558  \\
0.9484  &  1  &  0.9083  \\
0.7558  &  0.9083  &  1
\end{array} \right)  \label{eq:corr_matrix}
\ee
From this Fisher matrix, it can be shown that any of the three trispectra is highly correlated to
a linear combination of the other two.
For example, the trispectrum $\dot\sigma^2 (\partial\sigma)^2$ is 99.2\% correlated to a linear combination
of $\dot\sigma^4$ and $(\partial\sigma)^4$.
Therefore, we will not treat $\dot\sigma^2 (\partial\sigma)^2$ as a new trispectrum which is independent
of the other two.
More concretely,
we can convert $\gnlB$ into the following effective values of $\gnldotpi4$ and $\gnldpi4$:
\be
(\gnldotpi4)_{\rm eff} = 0.620 \gnlB \hspace{1.5cm} (\gnldpi4)_{\rm eff} = 0.0936 \gnlB  \label{eq:gnlB_eff}
\ee
Note that we have choose our trispectrum basis to simply be the coefficients of the operators
$\dot\sigma^4$ and $(\partial\sigma)^4$, rather than orthogonalizing as in the case of the bispectrum.
This somewhat simplifies the analysis and interpretation, but it should be kept in mind that
the two operators are $\approx$75\% correlated.

The local trispectra $\gnlloc$ is not particularly correlated to any of the quartic operator trispectra
$(\dot\sigma^4, \dot\sigma^2 (\partial\sigma)^2, (\partial\sigma)^4)$.
This can be understood by noting that the local trispectrum gets most of its signal-to-noise
from the squeezed limit $k_1 \ll \min(k_2,k_3,k_4)$, whereas the other
trispectra vanish in the squeezed limit.

% If we use Eq.~(\ref{eq:gnlB_eff}) to approximately convert $\gnlB$ to a linear combination of the 
% other two parameters, we get:
% \be
% \gnldotpi4 A_\zeta = \left( 0.00814 \frac{H^4}{\Lambda_1^4} - 0.00729 \frac{H^4}{\Lambda_2^4} \right)
%   \hspace{1cm}
% \gnldpi4 A_\zeta = \left( -0.00110 \frac{H^4}{\Lambda_2^4} + 0.0310 \frac{H^4}{\Lambda_3^4} \right)  \label{eq:gnl_mf_matrix}
% \ee

Throughout the preceding Fisher matrix analysis, we have used WMAP noise levels.
If we use Planck noise levels instead, the results are qualitatively unchanged but
the numerics are slightly different.
The correlation matrix between the $\dot\sigma^4$,
$\dot\sigma^2 (\partial_i\sigma)^2$, $(\partial_i\sigma)^2 (\partial_j\sigma)^2$,
and $(\partial_i\sigma)^2 (\partial_j\sigma)^2$ trispectra is:
\be
\left( \begin{array}{ccc}
   1      &   0.9113   &   0.6142  \\
  0.9113  &     1      &   0.8572  \\
  0.6142  &   0.8572   &      1
\end{array} \right)  \hspace{1.5cm} \mbox{(Planck noise)}
\ee
The $\dot\sigma^2 (\partial_i\sigma)^2$ shape is 98.6\% correlated to a linear combination
of the other two shapes.  The coefficients which convert $\gnlB$ to effective values of
$\gnldotpi4$ and $\gnldpi4$ are:
\be
(\gnldotpi4)_{\rm eff} = 0.597 \gnlB \hspace{1.5cm} (\gnldpi4)_{\rm eff} = 0.0914 \gnlB  \hspace{1cm} \mbox{(Planck noise)}
\ee

\section{Analysis pipelines}
\label{sec:pipelines}

In this section, we develop an analysis pipeline for estimating the
amplitude of a trispectrum $T$ for a realistic CMB experiment.
We will actually develop two analysis pipelines
which are appropriate for different sets of assumptions.

In some experiments, it is computationally feasible to multiply
a harmonic-space map by the operator $C^{-1}$ which appears in the
optimal trispectrum estimator (Eq.~(\ref{eq:optimal_estimator})).
For example, this is possible for WMAP, since the noise model is
simple: it is an excellent approximation to treat the noise covariance
as diagonal in the pixel domain.
In~\S\ref{ssec:optimal_pipeline} below, we develop an optimal pipeline
for such experiments.

In other experiments, it is infeasible to multiply a map by
$C^{-1}$, either because this is too computationally slow, or because
the noise model is too complicated.
The case we have in mind is Planck, although we will not attempt a
Planck trispectrum analysis in this paper.
The foreground-cleaned maps used for non-Gaussianity analysis by the Planck
collaboration~\cite{Ade:2013ydc} have a noise covariance which in principle
is determined precisely by the scan strategy, timestream noise properties, and
foreground cleaning method.
However, pixel-pixel correlations
are important, exact multiplication of a map by $N^{-1}$ is likely 
to be as expensive as full map-making, and multiplication by $C^{-1}$
(requiring iterated multiplication by $N^{-1}$) is likely prohibitive.
Fortunately, we can still proceed by implementing a filter which approximates
but is not precisely equal to $C^{-1}$.
Another feature of the Planck analysis is that making Monte Carlo simulations
of foreground cleaned maps is expensive.
A common set of Monte Carlo simulations is shared between the
Planck trispectrum analysis, bispectrum analysis, and other analyses, 
but it is impractical to make new simulations specifically for the trispectrum
pipeline.

With these considerations in mind, in~\S\ref{ssec:puremc_pipeline} below,
we propose a ``pure MC'' pipeline which compares the trispectrum of the
data to the trispectrum of an external set of Monte Carlo simulations,
using a filter which is not necessarily equal to $C^{-1}$.

An important property of the pure MC pipeline is that it does not assume
that the simulations are Gaussian.
For example, we might use lensed CMB simulations, which have a nonzero
trispectrum.
In this case, the pure MC pipeline is constructed so that it estimates
the trispectrum of the data in excess of the simulations, i.e.~lensing
bias will automatically be subtracted from the estimated trispectrum.

We have not worked out how to remove lensing bias in the optimal pipeline,
since our immediate goal is to use the optimal pipeline to analyze WMAP, 
where lensing is a small effect.
In cases where lensing bias is small, we can accurately approximate it
using a Fisher matrix based estimate; see discussion near Eq.~(\ref{eq:lensing_bias})
below.

In a case where $C^{-1}$ is affordable but lensing bias is large,
currently our only way of obtaining optimal error bars with reliable
lensing bias subtraction would
be to run the pure MC pipeline with $C^{-1}$ filtering
rather than running the optimal pipeline.
This has one disadvantage: the optimal pipeline is much faster to converge
than the pure MC pipeline, since we can use the assumption of Gaussian
simulations to give a ``fast MC'' algorithm.
The ultimate pipeline would combine the fast convergence of the optimal pipeline
and bias subtraction properties of the pure MC pipeline, but we defer construction
of such a pipeline to future work.

\subsection{Optimal pipeline}
\label{ssec:optimal_pipeline}

In this section we describe our first pipeline: an optimal pipeline
which can be applied to experiments where $C^{-1}$ filtering is practical.
Although more general, this pipeline was developed with WMAP in mind.
Let us state our assumptions explicitly:
\begin{enumerate}
\item The observed CMB $a_{\ell m} = s_{\ell m} + n_{\ell m}$ is the sum of
the true sky signal $s_{\ell m}$ and a Gaussian noise realization $n_{\ell m}$.
(Our convention here is that $a_{\ell m}$ denotes the beam-deconvolved map.)
\item Given a harmonic-space map $b_{\ell m}$, computing $(C^{-1}b)_{\ell m}$
is computationally feasible.
\item It is also computationally feasible to randomly generate a signal + noise
realization, i.e.~a Gaussian random map $b_{\ell m}$ with covariance matrix $C$.
\end{enumerate}
Throughout this section we will use the abbreviated notation
\be
\ta_{\ell m} = C^{-1}_{\ell m, \ell' m'} a_{\ell' m'}
\ee
As shown previously (Eq.~(\ref{eq:optimal_estimator_mc})), the optimal estimator is $\E[a] = (1/F) \E_0[a]$,
where $\E_0[a]$ can be computed as a Monte Carlo average over Gaussian signal+noise realizations $b$:
\be
\E_0[a] = \left\langle Q[\ta,\ta,\ta,\ta] - 6 Q[\ta,\ta,\tb,\tb] + Q[\tb,\tb,\tb,\tb] \right\rangle_b  \label{eq:opt_E0_mc}
\ee
The quantity $F$ was defined previously in Eq.~(\ref{eq:F_harmonic}).
It determines both the normalization of the estimator and its variance.
More precisely, $\Var(\E) = 1/F$ or equivalently $\Var(\E_0) = F$.

Since evaluating $\E_0$ by Monte Carlo is straightforward given our assumptions,
the only issue in the optimal pipeline is an algorithm for computing $F$.
This involves some nontrivial computational challenges, as we now explain.

Since $F = \Var(\E_0)$, one natural approach is to evaluate $\E_0$ on an ensemble of
Gaussian simulations and estimate the variance to get $F$.
Unfortunately, if implemented naively, the computational cost of this approach 
is $\bigoh(\Nmc^2)$, not $\bigoh(\Nmc)$!
This is due to a curious property of the estimator~(\ref{eq:opt_E0_mc}): if we want
to evaluate the estimator on a new realization $a$, we must recompute the Monte Carlo
average $\langle Q[\ta,\ta,\tb,\tb] \rangle$ ``from scratch'' by looping over
random realizations $b$ with computational cost $\bigoh(\Nmc)$.

One idea for reducing the computational cost from $\bigoh(\Nmc^2)$ to $\bigoh(\Nmc)$
is to group Monte Carlo simulations into pairs $(b_1,b_1'), (b_2,b_2'), \cdots$.
We then express $F$ as a Monte Carlo average involving only expressions which can
be computed from a single pair, for example $Q[b_i,b_i,b_i,b_i]$ or $Q[b_i,b_i,b'_i,b'_i]$,
but not $Q[b_i,b_i,b_j,b_j]$.

A second, more technical idea for reducing computational cost
is to use Monte Carlo averages involving $(\partial Q)$,
which converge much faster than averages involving $Q$.
This was noted previously in our discussion of the isotropic all-sky Fisher matrix (\S\ref{ssec:fisher_mc}).

Combining these ideas, we express $F$ as the following Monte Carlo average over pairs $(b,b')$:
\ba
F &=& \frac{1}{32} \left\langle 
                (\partial_{\ell m} Q[\tb,\tb,\tb]) C^{-1}_{\ell m,\ell'm'} (\partial_{\ell'm'} Q[\tb,\tb,\tb]) 
              + (\partial_{\ell m} Q[\tb',\tb',\tb']) C^{-1}_{\ell m,\ell'm'} (\partial_{\ell'm'} Q[\tb',\tb',\tb']) 
       \right\rangle \nn \\
 && \hspace{1cm}  + \frac{9}{32} \left\langle 
           (\partial_{\ell m} Q[\tb,\tb,\tb']) C^{-1}_{\ell m,\ell'm'} (\partial_{\ell'm'} Q[\tb,\tb,\tb'])
         + (\partial_{\ell m} Q[\tb,\tb',\tb']) C^{-1}_{\ell m,\ell'm'} (\partial_{\ell'm'} Q[\tb,\tb',\tb']) \right\rangle \nn \\
 && \hspace{1cm}  - \frac{3}{16} \left\langle
           (\partial_{\ell m} Q[\tb,\tb,\tb]) C^{-1}_{\ell m,\ell'm'} (\partial_{\ell'm'} Q[\tb,\tb',\tb'])
         + (\partial_{\ell m} Q[\tb',\tb',\tb']) C^{-1}_{\ell m,\ell'm'} (\partial_{\ell'm'} Q[\tb,\tb,\tb'])
     \right\rangle \,.  \label{eq:optimal_pipeline_F_final}
\ea
The specific choice of coefficients $(1/32,9/32,-3/16)$ is motivated in Appendix~\ref{app:magic_coefficients}.

It is sometimes useful to know the ``error on the error'', i.e.~the statistical error on our
estimate of $F$ due to the finite number of Monte Carlos.
We estimate this straightforwardly, since $F$ is an average over pairs $(b,b')$, so we can
estimate its uncertainty from the scatter between pairs.
The estimator for $F$ given in Eq.~(\ref{eq:optimal_pipeline_F_final}) has been designed to 
minimize this scatter, and in practice we do not need many Monte Carlos to get convergence.

As previously mentioned, we emphasize that the optimal pipeline assumes Gaussian
statistics and in particular does not subtract lensing bias.
Note that simply using lensed simulations in the optimal estimator~(\ref{eq:opt_E0_mc})
does not correctly remove lensing bias.
For WMAP the lensing bias is small, but in a case where it is large and must be
subtracted accurately, then the only option is the pure MC pipeline which we
present next.

\subsection{Pure MC pipeline}
\label{ssec:puremc_pipeline}

We now describe our second pipeline, a pipeline which operates on an external
ensemble of Monte Carlo simulations.
We start by choosing a filter which can be applied to the data to produce a
harmonic-space map $\ta_{\ell m}$.
To obtain near-optimal statistical errors, the filter should be chosen
to approximate $C^{-1}$ filtering as closely as possible.
For example, to analyze Planck data, we could use the same filtering used
for the bispectrum analysis~\cite{Ade:2013ydc}: we start with foreground-cleaned
maps in pixel space, inpaint the mask, transform to harmonic space, and multiply by
$1/(C_\ell + N_\ell)$, where $N_\ell$ is a sky-averaged noise power spectrum.
This filter is suboptimal in principle, since it is not precisely equal to
$C^{-1}$, but has been shown to be near-optimal for Planck, at least for
the bispectrum.

Let us state the assumptions of our ``pure MC'' pipeline:
\begin{enumerate}
\item The observed sky is specified as a filtered harmonic-space map $\ta_{\ell m}$,
and we want to compare its trispectrum to a set of external simulations, also specified
as filtered harmonic-space maps $\tb_{\ell m}^{(1)}, \tb_{\ell m}^{(2)}, \cdots$.
(In this section, we use tildes to denote any map which has been processed by the filter.)
\item The observed sky is a sum of signal and noise components $\ta_{\ell m} = \ts_{\ell m} + \tn_{\ell m}$
(and likewise for the simulations).  The filtered signal $\ts_{\ell m}$ is related to the true
CMB sky by a linear operator $T$, i.e.~$\ts_{\ell m} = (T s)_{\ell m}$, and the signal and noise
are statistically independent.
\item For each simulation, we know the underlying CMB realization $s_{\ell m}$ which was used.
An important feature of the pure MC pipeline is that we do not assume that either the CMB or noise realizations
used in the simulations are Gaussian.
If the simulations are non-Gaussian, then the trispectrum estimator will return an estimate of the trispectrum
amplitude in excess of any trispectrum which is in the simulations.
This is very convenient in practice.
For example, if the CMB realizations may be lensed, and the noise realizations include residual foregrounds,
then lensing and foreground contributions to the trispectrum will automatically be subtracted.
\item Given an arbitrary CMB map $b_{\ell m}$, there is a fast algorithm for computing $(Tb)_{\ell m}$.
Typically this will involve convolving with a beam or instrumental response, taking the spherical
transform to pixel space, then applying the same filter which was applied to the data.
\end{enumerate}
Our pipeline will use an estimator of the form $\E[\ta] = (1/F_N) \E_0[\ta]$, where the unnormalized estimator is defined by the Monte Carlo average:
\be
\E_0[a] = Q[\ta,\ta,\ta,\ta] - 6 \left\langle Q[\ta,\ta,\tb,\tb] \right\rangle_b 
   - \left\langle Q[\tb,\tb,\tb,\tb] \right\rangle_b + 6 \left\langle Q[\tb,\tb,\tb',\tb'] \right\rangle_{b,b'}  \label{eq:puremc_E0}
\ee
and the normalization $F_N$ will be specified shortly (Eq.~(\ref{eq:puremc_FN}) below).

It is easy to verify two key properties of this estimator.
First, its expectation value over the simulations vanishes: $\langle \E_0[b] \rangle = 0$.
This means that $\E_0[a]$ measures the trispectrum of the data relative to the
trispectrum of the simulations, as desired.
Second, if the two-point function of the simulations does not perfectly match the
two-point function of the data, due to slightly incorrect cosmological parameters
or noise model, then the bias on $\E_0$ will be second order.
As shown previously in~\S\ref{sec:estimator}, this is an important property of the
optimal estimator, and we would like to preserve it in our pure MC pipeline.

The last term in the estimator~(\ref{eq:puremc_E0}) is a double Monte Carlo average over 
pairs of simulations $(b,b')$.
This is necessary because we are not assuming Gaussian simulations.
If simulations were Gaussian, then we could use the relation
$\langle Q[\tb,\tb,\tb,\tb] \rangle_b = 3 \langle Q[\tb,\tb,\tb',\tb'] \rangle_{b,b'}$
to rewrite the last term as an average over single simulations $b$.
Note that if we assumed Gaussianity, and also assumed optimal filtering (i.e.~$\ta = C^{-1}a$)
then the estimator $\E_0$ would reduce to the optimal estimator studied previously.

In our pure MC pipeline, we would like to compute the unnormalized estimator $\E_0[a]$,
the estimator normalization $F_N$, and the variance $F_V = \Var(\E_0)$ by Monte Carlo.
We would also like to compute the ``error on the error'', i.e.~the uncertainty in $F_V$
when we estimate it by Monte Carlo.
Let us now discuss each of these in turn.

Considering first the estimator normalization $F_N$, one can show that it is given by the
following Monte Carlo average over pairs of simulations $(b_{\ell m} ,b'_{\ell m})$.
\ba
F_N &=& \Bigg\langle
   \frac{1}{32} (\partial_{\ell m} Q[\tb,\tb,\tb]) (T \partial Q[s,s,s])_{\ell m}
      + \frac{9}{32} (\partial_{\ell m} Q[\tb,\tb,\tb']) (T \partial Q[s,s,s'])_{\ell m} \nn \\
  && \hspace{1cm} 
  - \frac{3}{32} (\partial_{\ell m} Q[\tb,\tb,\tb]) (T \partial Q[s,s',s'])_{\ell m}
  - \frac{3}{32} (\partial_{\ell m} Q[\tb,\tb',\tb']) (T \partial Q[s,s,s])_{\ell m}
  + \left( \parbox{0.98cm}{$b \leftrightarrow b'$ \\ $s \leftrightarrow s'$} \right) \Bigg\rangle_{b,b'}  \label{eq:puremc_FN}
\ea
where $(s_{\ell m}, s'_{\ell m})$ denotes the underlying CMB realizations used
in simulations $(b,b')$.
The specific choice of coefficients here was motivated in the previous section
(see discussion near Eq.~(\ref{eq:optimal_pipeline_F_final})).
This expression is a ``fast MC'' scheme, in the sense that the fractional
error in $F_N$ is much better than $\sqrt{2/\Nmc}$.

Next we consider the estimator variance $F_V = \Var(\E_0)$.
Note that in the optimal pipeline from the previous section
we had $F_N = F_V$, but in the pure MC pipeline where the filter need not
be precisely equal to $C^{-1}$, one has $F_N \ne F_V$ in general.
A short calculation gives the following general expression for $F_V$:
\be
F_V
   = \left\langle \left( Q[\tb,\tb,\tb,\tb] - 6 Q[\tb,\tb,\tb',\tb'] \right) \left( Q[\tb,\tb,\tb,\tb] - 6 Q[\tb,\tb,\tb'',\tb''] \right) \right\rangle
         - \left\langle Q[\tb,\tb,\tb,\tb] - 6 Q[\tb,\tb,\tb',\tb'] \right\rangle^2
\ee
To write this in a slightly different way,
let us temporarily imagine that we have computed the quantity
\be
Q_{ij} = Q[\tb_i,\tb_i,\tb_j,\tb_j]
\ee
for every pair of simulations $(\tb_i, \tb_j)$.  Note that computing every $Q_{ij}$ has 
computational cost $\bigoh(\Nmc^2)$, which is something we are trying to avoid,
but we will address this shortly.
Then the following estimator has expectation value $\langle \hF_V \rangle = F_V$:
\ba
\hF_V &=& \frac{1}{\Nmc} \sum_i Q_{ii}^2 - \frac{12}{N_2} \sum_{\{ij\}} Q_{ii} Q_{ij} + \frac{36}{N_3} \sum_{\{ijk\}} Q_{ij} Q_{ik}  \\
  &&  \hspace{0.5cm}
      - \frac{1}{N_2} \sum_{\{ij\}} Q_{ii}^2 Q_{jj}^2 + \frac{12}{N_3} \sum_{\{ijk\}} Q_{ii} Q_{jk} 
       - \frac{36}{N_4} \sum_{\{ijkl\}} Q_{ij} Q_{kl} \nn
\ea
where we have defined $N_k = \Nmc(\Nmc-1)\cdots(\Nmc-k+1)$,
and $\sum_{\{ijk\cdots\}}$ denotes a sum over {\em distinct} indices $i,j,k,\cdots$
between 1 and $\Nmc$.
This can be simplified slightly by defining
\be
R_{ij} = \left\{ \begin{array}{cl}
   6 Q_{ij} - Q_{ii} - Q_{jj}  &  \mbox{if $i\ne j$}  \\
      0   & \mbox{if $i=j$}
\end{array} \right.
\ee
A short calculation then shows that $\hF_V$ simplifies to:
\be
\hF_V = \frac{1}{N_3} \sum_{\{ijk\}} R_{ij} R_{ik} - \frac{1}{N_4} \sum_{\{ijkl\}} R_{ij} R_{kl}  \label{eq:puremc_FV}
\ee
We would also like to estimate the ``error on the error'', i.e.~the variance of $\hF_V$.
It is easy to see that the following estimator has expectation value $\Var(\hF_V)$:
\be
\hSigma = (\hF_V)^2 - \frac{1}{N_6} \sum_{\{ijklmn\}} R_{ij} R_{ik} R_{lm} R_{ln}
       + \frac{2}{N_7} \sum_{\{ijklmno\}} R_{ij} R_{ik} R_{lm} R_{no}
       - \frac{1}{N_8} \sum_{\{ijklmnop\}} R_{ij} R_{kl} R_{mn} R_{op}  \label{eq:puremc_sigma}
\ee
Evaluating $\hSigma$ in this form has $\bigoh(\Nmc^8)$ computational cost, which may be a problem in practice.
In Appendix~\ref{app:fv} we give an equivalent expression with lower cost.

Summarizing results in this section so far, we have expressed the unnormalized estimator~$\E_0$,
the normalization $F_N$, the variance $F_V$, and the ``error on the error'' $\Sigma$ as Monte Carlo
averages (Eqs.~(\ref{eq:puremc_E0}),~(\ref{eq:puremc_FN}),~(\ref{eq:puremc_FV}),~(\ref{eq:puremc_sigma})).
These expressions contain double Monte Carlo averages over pairs of simulations $(b,b')$,
naively leading to $\bigoh(\Nmc^2)$ computational cost.
We avoid this as follows.
We divide the ensemble of $\Nmc$ simulations into $(\Nmc/M)$ subsets containing $M$ simulations each.
For each such subset, we evaluate 
Eqs.~(\ref{eq:puremc_E0}),~(\ref{eq:puremc_FN}),~(\ref{eq:puremc_FV}),~(\ref{eq:puremc_sigma})
using only the $M$ simulations in the subset (replacing $\Nmc$ when it appears by $M$ of course).
We then average over all subsets to obtain our final estimates for $\E_0, F_N, F_V$, and $\Sigma$.
This reduces computational cost from $\bigoh(\Nmc^2)$ to $\bigoh(\Nmc M)$.

For a fixed total number of Monte Carlo simulations $\Nmc$, the larger we choose $M$, the more accurate our estimate
for the variance $F_V$ will be, but the computational cost of the trispectrum pipeline will also increase.
Note that we must choose $M \ge 8$ in order for the expression~(\ref{eq:puremc_sigma}) for $\hSigma$ to make sense.
We have found that $M=16$ or $M=32$ is usually a good compromise.

\section{WMAP results and interpretation}
\label{sec:wmap}

We conclude this paper by constraining the parameters $\gnlloc$, $\gnldotpi4$, 
and $\gnldpi4$ from WMAP data.

In WMAP, the $C^{-1}$-filtering operation
is computationally feasible, so we can use the optimal pipeline 
from~\S\ref{ssec:optimal_pipeline}.
We implement $C^{-1}$-filtering using the multigrid conjugate
gradient algorithm from Appendix A of~\cite{Smith:2007rg}.
The $C^{-1}$ filter optimally combines data from the six V-band
and W-band WMAP channels (V1,V2,W1,W2,W3,W4), and incorporates the
kq75 sky mask by assigning infinite variance to masked pixels.
The filter marginalizes foreground templates, sky monopoles,
and dipoles in an analogous way, by assigning infinite variance
to the appropriate pixel-space modes, independently in each of 
the six channels.
These details of the filtering are the same as the optimal bispectrum
analysis in the WMAP nine-year results paper~\cite{Bennett:2012zja}
(see also~\cite{Smith:2009jr}).

We ran the optimal pipeline with $\Nmc=2048$ Monte Carlo simulations, which turned
out to be overkill: the ``error on the error'' due to the finite number of simulations
was 0.4\% for the local $g_{NL}$, or 0.05\% for the other two trispectra.
The constraints from our pipeline are:
\ba
\gnlloc &=& (-3.71 \pm 2.19) \times 10^5  \nn \\
\gnldotpi4 &=& (-2.32 \pm 3.09) \times 10^6  \label{eq:wmap_gnl1} \\
\gnldpi4 &=& (-9.07 \pm 6.33) \times 10^5 \nn
\ea
No statistically significant deviation from Gaussian statistics is seen.
This is the first constraint on the $(\partial\sigma)^4$ trispectrum.
Constraints on the other two trispectra have been previously reported as follows.

The optimal estimator for $\gnlloc$ has also been implemented in~\cite{Sekiguchi:2013hza},
where the constraint $\gnlloc = (-3.3 \pm 2.2) \times 10^5$ was obtained.
This agrees nearly perfectly with our result in Eq.~(\ref{eq:wmap_gnl1}):
the error bars are identical and the central
values differ by 0.2$\sigma$.  This is expected since the optimal estimator contains no free
parameters.
The $\gnldotpi4$ trispectrum was studied in~\cite{Fergusson:2010gn},
where the constraint $\gnldotpi4 = (-2.88 \pm 6.94)$ was obtained.\footnote{The
parameter $t_{NL}^{\rm equil}$ defined in~\cite{Fergusson:2010gn} is related
to our $\gnldotpi4$ by $t_{NL}^{\rm equil} = (27/25) \gnldotpi4$.}
The smaller error bar in Eq.~(\ref{eq:wmap_gnl1}) is partly due to our use
of WMAP9 data rather than WMAP5, and partly due to our use of the optimal
estimator.

The $g_{NL}$ central values in Eq.~(\ref{eq:wmap_gnl1}) are slightly biased by
gravitational lensing.  We can study the bias semianalytically using the Fisher
matrix.  Since the lensing trispectrum $T_{\rm lens}$ is not contact factorizable, 
we cannot use the exact Fisher matrix algorithm (\S\ref{ssec:fisher_exact}), but since 
$T_{\rm lens}$ is exchange factorizable, we can still use the Monte Carlo Fisher matrix 
algorithm (\S\ref{ssec:fisher_mc}).

Using a Fisher matrix forecast with WMAP9 noise levels, we find that the
correlation coefficients of the lensing trispectrum with the local, $\dot\sigma^4$, 
and $(\partial\sigma)^4$ trispectra are 0.02, 0.15, and 0.14 respectively,
and the total signal-to-noise of the CMB lensing trispectrum is only 2.1.
This makes it intuitively clear that the lensing bias is small.

To quantify this better, we can estimate the lensing bias to each of the $g_{NL}$
parameters semianalytically as follows.  For any primordial trispectrum $T$, we
approximate the bias as $\Delta g_{NL} = F(T,T_{\rm lens}) / F(T,T)$.
This gives the following estimates for lensing bias:
\be
\Delta\gnlloc = 9.24 \times 10^3
   \hspace{1cm}
\Delta\gnldotpi4 = 8.82 \times 10^5
   \hspace{1cm}
\Delta\gnldpi4 = 1.71 \times 10^5  \label{eq:lensing_bias}
\ee
which are 0.04$\sigma$, 0.3$\sigma$, and 0.3$\sigma$ shifts respectively.
Subtracting lensing bias, our final ``bottom line'' trispectrum constraints are:
\ba
\gnlloc &=& (-3.80 \pm 2.19) \times 10^5  \nn \\
\gnldotpi4 &=& (-3.20 \pm 3.09) \times 10^6  \label{eq:wmap_gnl2} \\
\gnldpi4 &=& (-10.8 \pm 6.33) \times 10^5 \nn
\ea
The above semianalytic prescription for lensing bias is approximate, since it
is only valid to lowest order in $C_\ell^{\phi\phi}$, and also approximates
the noise covariance of the survey as all-sky isotropic.
Note that since the CMB lensing trispectrum is not scale-invariant, the preceding
Fisher matrix based results are specific to WMAP noise levels, and lensing
will be a larger effect in Planck.
For WMAP, where lensing is small, the semianalytic bias correction is adequate,
but for Planck a more accurate treatment will be necessary.

Each ``bottom line'' trispectrum constraint in Eq.~(\ref{eq:wmap_gnl2})
is a constraint on a single $g_{NL}$ parameter assuming that the other
$g_{NL}$-parameters are zero.
We also consider the case of a joint constraint on the parameters 
$(\gnldotpi4, \gnldpi4)$, which are 75\% correlated.
It is convenient to introduce the vector notation $g_i = (\gnldotpi4, \gnldpi4)$.
Let $F_{ij}$ be the two-by-two Fisher matrix, which can be constructed as follows.
The diagonal is given by $F_{ii} = 1/\sigma_i^2$, where $\sigma_i$ is the $g_{NL}$ statistical error in Eq.~(\ref{eq:wmap_gnl2}).
The off-diagonal is then given by $F_{12} = r_{12} F_{11}^{1/2} F_{22}^{1/2}$, where $r_{12} = 0.7558$ is the correlation
coefficient from Eq.~(\ref{eq:corr_matrix}).
From this procedure we obtain the Fisher matrix:
\be
F_{ij} = \left( \begin{array}{cc}
  1.05  &  3.86  \\
  3.86  &  25.0
\end{array} \right) \times 10^{-13}
\ee
Let $\hat g_i = (-3.20 \times 10^6, -1.08 \times 10^6)$ be the vector of $g_{NL}$ estimates
appearing in Eq.~(\ref{eq:wmap_gnl2}).
Now for a given parameter vector $g_i$, we define a trispectrum $\chi^2$-value by:\footnote{To
derive this $\chi^2$, we note that the unnormalized trispectrum estimator $F_{ii} \hat g_i$ has
expectation value $F_{ij} g_j$ and covariance matrix 
$\mbox{Cov}(F_{ii}\hat g_i, F_{jj} \hat g_j) = F_{ij}$.}
\be
\chi^2(g) = ((Fg)_i - F_{ii} \hat g_i) F_{ij}^{-1} ((Fg)_j - F_{jj} \hat g_j)  \label{eq:chi2_def}
\ee
This $\chi^2$ can be thresholded to obtain constraints in various parameter spaces of interest.
For example, we can plot error ellipses in the $(\gnldotpi4, \gnldpi4)$-plane,
showing the off-diagonal correlation (Fig.~2).  The 68\% and 95\% regions
are obtained by thresholding at $\chi^2 = 2.279$ and $\chi^2=5.991$ respectively,
as appropriate for a $\chi^2$ random variable with two degrees of freedom.

\begin{figure}
\begin{center}
\includegraphics[width=9cm]{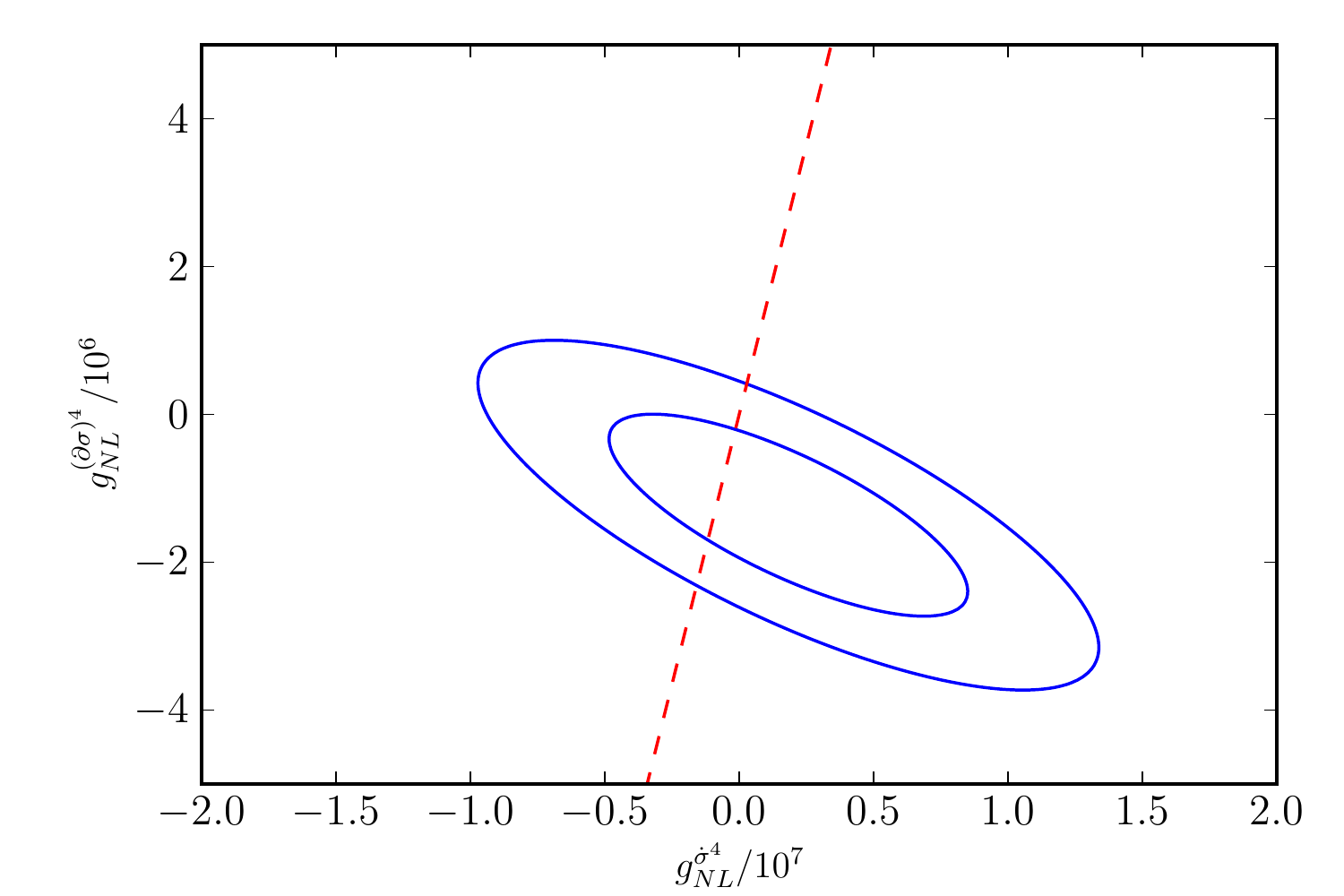}
\caption{68\% and 95\% confidence regions in the $(\gnldotpi4, \gnldpi4)$ plane,
with the Lorentz invariant model in Eq.~(\ref{eq:S_lorentz}) shown as the dashed line.}
\end{center}
\label{fig:gnl_plane}
\end{figure}

We conclude with some brief physical interpretation.
In single-field inflation, only the quartic operator $\dot\pi^4$ is 
allowed by the symmetries to induce a large trispectrum without generating an even larger bispectrum.
Its coefficient $M_4^4$ in the action~(\ref{eq:Spi}) is related to $\gnldotpi4$ by Eq.~(\ref{eq:gnldotpi4_sf}).
Using our WMAP constraint from Eq.~(\ref{eq:wmap_gnl2}), we get the following constraint on $M_4$:
\be
(-2.47 \times 10^{15}) < \frac{M_4^4 c_s^3}{H^4} < (7.86 \times 10^{14}) \hspace{1cm} \mbox{(95\% CL)}  \label{eq:limit-single}
\ee
We can develop a more intuitive understanding of this limit by noticing the following two facts~\cite{Senatore:2010jy}. 
First, in the case where in single field inflation this operator leads to observable non-Gaussianities, the speed of sound is not expected to be parametrically smaller than one: $c_s\lesssim 1$. Second, the unitarity bound of the theory, $\Lambda_U$, scales as 
$\Lambda_U^4 \sim 16\pi^2 (\dot H M_{\rm Pl}^2)^2 M_4^{-4}$. 
Therefore the limit in Eq.~(\ref{eq:limit-single}) translates to
$g_{NL}^{\dot\pi^4} A_\zeta\sim (H^4/\Lambda_U^4) \lesssim 10^{-3}$, 
which is of order the inverse square root of the number of signal-dominated modes in WMAP, 
as expected from a Fisher matrix analysis.

In multifield inflation,
the limits above can be translated into limits on the ratios $H/(2\pi\Lambda_i)$
between the de Sitter temperature $H/(2\pi)$ during inflation and the scale suppressing the higher 
dimension operators in the action~(\ref{eq:multi}),
in this way effectively mapping cosmological information into constraints of 
parameters of a fundamental Lagrangian.~\footnote{Some readers might wonder 
why we call the effective field theory of inflation as a fundamental theory. 
Even though it is just an effective theory valid up to a scale of order $\Lambda$, 
this energy scale is still extremely large, and indeed quite fundamental. 
Furthermore, since we are not probing directly the energy scale active during 
inflation, but we do this only indirectly though the CMB and the LSS, the 
effective field theory of inflation is the only theory we are testing through 
observations, unless we make further assumptions.}
Generally, our results show that $(H/(2\pi\Lambda))^4$ must be smaller than $10^{-2}$ or $10^{-3}$.
This tells us that the quartic interactions are still largely unconstrained, 
and new cosmological probes, such as Large Scale Structure surveys, 
will be required to significantly improve the limits.
For example, if $\Lambda_2$ and $\Lambda_3$ are assumed to be zero, then
our WMAP constraint on $\gnldotpi4$ in Eq.~(\ref{eq:wmap_gnl2})
gives the following constraint on $\Lambda_1$: 
\be
-8.09 \times 10^{-3} < \frac{1}{(2\pi)^4}\frac{H^4}{\Lambda_1^4} < 2.57 \times 10^{-3} \hspace{1cm} \mbox{(95\% CL)}
\ee

As another example, consider a one-parameter space consisting of the Lorentz-invariant quartic interaction:
\be
S = \int d^4x\, \sqrt{-g}
  \left( \frac{1}{2} (\partial_\mu \sigma)^2 + \frac{1}{\Lambda^4} (\partial_\mu \sigma)^2 (\partial_\nu \sigma)^2 \right)  \label{eq:S_lorentz}
\ee
For a given value of $\Lambda$,
we use Eq.~(\ref{eq:gnl_multifield}) to compute $g_{NL}$ coefficients,
use Eq.~(\ref{eq:gnlB_eff}) to absorb $\gnlB$ into the values of $\gnldotpi4$ and $\gnldpi4$,
then compute $\chi^2$ using Eq.~(\ref{eq:chi2_def}).
Thresholding this $\chi^2$ in Eq.~(\ref{eq:chi2_def}) at $\Delta\chi^2=4$,
we obtain the 95\% confidence limits:
\be
-4.42 \times 10^{-4} < \frac{1}{(2\pi)^4}\frac{H^4}{\Lambda^4} < 4.00 \times 10^{-5}
  \hspace{1cm} \mbox{(95\% CL)}
\ee

\section{Discussion}

\par\noindent
The main conclusions of this paper are as follows:
\begin{itemize}
\item To lowest order in the derivative expansion, the quartic operators
  allowed by the symmetries of inflation are $\dot\sigma^4$, $\dot\sigma^2 (\partial\sigma)^2$,
  and $(\partial\sigma)^4$.  In single-field inflation, only the $\dot\sigma^4$ operator
  is allowed by the symmetries to induce a large trispectrum without generating an even larger bispectrum,
  but multifield inflation allows an arbitrary linear combination
  of the three operators.  A Fisher matrix analysis shows that there is one near-degeneracy
  between these three operators, which we can use to approximate the $\dot\sigma^2 (\partial\sigma)^2$
  trispectrum as a linear combination of $\dot\sigma^4$ and $(\partial\sigma)^4$.
\item Based on this analysis, we propose the parameter space 
  ($\gnlloc$, $\gnldotpi4$, $\gnldpi4$)
  as a starting point for analyzing inflationary 4-point signals.  This is roughly
  analogous to the parameter space ($\fnlloc$, $\fnleq$, $\fnlorth$) for the 3-point
  function.  It will be interesting to explore 4-point signals beyond these leading 
  ones.  In particular, ``exchange'' trispectra arising from cubic operators and
  exchange of a light field during inflation are not included in this parameter
  space, and would be interesting to study in future work.
\item We propose two factorizability conditions for the trispectrum, contact
  factorizability and exchange factorizability, and study the contact factorizable
  case in detail.  We argue that, in order to apply to operators with spatial derivatives
  such as $(\partial\sigma)^4$, the definition of factorizability for the trispectrum must
  include higher-spin fields.
\item For each of our trispectra, we write the CMB trispectrum as either a single
  integral over a radial coordinate $r$ (in the case of the local trispectrum) or a double
  integral over $(\tau_E,r)$ using a Feynman diagram (in the case of the quartic
  operator trispectra).  By approximating the integral by a finite sum, we represent
  the trispectrum as a sum of a large number of factorizable terms, then apply
  an optimization algorithm to obtain a compact factorizable representation.
\item We emphasize that the final compact representation obtained in this way
  approximates the exact trispectrum in a controlled sense: the two are nearly
  equal in the metric defined by the Fisher matrix.  This is because our integration
  scheme includes an end-to-end Fisher matrix based convergence test 
  (Appendix~\ref{app:integrals}), and the optimization algorithm is also
  guaranteed to converge in the Fisher matrix metric.
\item We develop a toolkit of algorithms which can be applied to factorizable
  trispectra, including estimator evaluation, non-Gaussian simulation 
  (Appendix~\ref{app:ngsim}), and Fisher matrix calculation.
\item We develop an optimal trispectrum pipeline and apply it to WMAP, finding
  consistency with Gaussian statistics.  We also develop a ``pure MC'' pipeline
  which scales to Planck.  The optimal pipeline can be used if $C^{-1}$ is computationally
  affordable and lensing bias is small enough to be estimated semianalytically.  The
  pure MC pipeline relaxes both of these assumptions but is slower to converge.
\item The tools we have developed in this paper are sufficient to analyze the local,
 $\dot\sigma^4$, $\dot\sigma^2 (\partial\sigma)^2$, and $(\partial\sigma)^4$ trispectra in WMAP
 and Planck.  However there are a few directions in which our machinery might be
 improved.  In order to analyze exchange trispectra, one would need to generalize
 the optimization algorithm from \S\ref{sec:contact_optimization}.  It would be
 interesting to improve the sensitivity of our pipeline to noise modeling, as
 suggested in \S\ref{ssec:alternate_approach}.  Finally, in cases where $C^{-1}$ 
 is affordable, we currently have to choose between the fast convergence of
 the optimal pipeline (\S\ref{ssec:optimal_pipeline}) and precise calculation
 of lensing bias in the pure MC pipeline (\S\ref{ssec:puremc_pipeline}).
\end{itemize}

\section*{Acknowledgements}

Research at Perimeter Institute is supported by the Government of Canada
through Industry Canada and by the Province of Ontario through the Ministry of Research \& Innovation.
Some computations were performed on the GPC cluster at the Scinet HPC Consortium.
SciNet is funded by the Canada Foundation for Innovation under the auspices of Compute Canada,
the Government of Ontario, and the University of Toronto.
KMS was supported by an NSERC Discovery Grant.
LS is supported by by DOE Early Career Award DE-FG02-12ER41854 and the National Science Foundation under
PHY-1068380.
MZ is supported in part by the NSF grants AST-0907969, PHY-1213563 and AST-1409709.

\bibliographystyle{h-physrev}
\bibliography{optimal_trispectrum}

\begin{thebibliography}{10}

\bibitem{Lyth:2002my}
D.~H. Lyth, C.~Ungarelli, and D.~Wands,
\newblock Phys.Rev. {\bf D67}, 023503 (2003), astro-ph/0208055.

\bibitem{Zaldarriaga:2003my}
M.~Zaldarriaga,
\newblock Phys.Rev. {\bf D69}, 043508 (2004), astro-ph/0306006.

\bibitem{Komatsu:2003iq}
E.~Komatsu, D.~N. Spergel, and B.~D. Wandelt,
\newblock Astrophys.J. {\bf 634}, 14 (2005), astro-ph/0305189.

\bibitem{Senatore:2010wk}
L.~Senatore and M.~Zaldarriaga,
\newblock JHEP {\bf 1204}, 024 (2012), 1009.2093.

\bibitem{Creminelli:2005hu}
P.~Creminelli, A.~Nicolis, L.~Senatore, M.~Tegmark, and M.~Zaldarriaga,
\newblock JCAP {\bf 0605}, 004 (2006), astro-ph/0509029.

\bibitem{Senatore:2009gt}
L.~Senatore, K.~M. Smith, and M.~Zaldarriaga,
\newblock JCAP {\bf 1001}, 028 (2010), 0905.3746.

\bibitem{Behbahani:2014upa}
S.~R. Behbahani, M.~Mirbabayi, L.~Senatore, and K.~M. Smith,
\newblock (2014), 1407.7042.

\bibitem{Fergusson:2009nv}
J.~Fergusson, M.~Liguori, and E.~Shellard,
\newblock Phys.Rev. {\bf D82}, 023502 (2010), 0912.5516.

\bibitem{Ade:2013ydc}
Planck Collaboration, P.~Ade {\em et~al.},
\newblock (2013), 1303.5084.

\bibitem{Carrasco:2012cv}
J.~J.~M. Carrasco, M.~P. Hertzberg, and L.~Senatore,
\newblock JHEP {\bf 1209}, 082 (2012), 1206.2926.

\bibitem{Babich:2004gb}
D.~Babich, P.~Creminelli, and M.~Zaldarriaga,
\newblock JCAP {\bf 0408}, 009 (2004), astro-ph/0405356.

\bibitem{Maldacena:2002vr}
J.~M. Maldacena,
\newblock JHEP {\bf 0305}, 013 (2003), astro-ph/0210603.

\bibitem{Senatore:2012wy}
L.~Senatore and M.~Zaldarriaga,
\newblock JCAP {\bf 1208}, 001 (2012), 1203.6884.

\bibitem{Baldauf:2011bh}
T.~Baldauf, U.~Seljak, L.~Senatore, and M.~Zaldarriaga,
\newblock JCAP {\bf 1110}, 031 (2011), 1106.5507.

\bibitem{Vielva:2009jz}
P.~Vielva and J.~Sanz,
\newblock Mon.Not.Roy.Astron.Soc. {\bf 404}, 895 (2010), 0910.3196.

\bibitem{Smidt:2010ra}
J.~Smidt {\em et~al.},
\newblock Phys.Rev. {\bf D81}, 123007 (2010), 1004.1409.

\bibitem{Fergusson:2010gn}
J.~Fergusson, D.~Regan, and E.~Shellard,
\newblock (2010), 1012.6039.

\bibitem{Hikage:2012bs}
C.~Hikage and T.~Matsubara,
\newblock Mon.Not.Roy.Astron.Soc. {\bf 425}, 2187 (2012), 1207.1183.

\bibitem{Sekiguchi:2013hza}
T.~Sekiguchi and N.~Sugiyama,
\newblock JCAP {\bf 1309}, 002 (2013), 1303.4626.

\bibitem{Wang:1999vf}
L.-M. Wang and M.~Kamionkowski,
\newblock Phys.Rev. {\bf D61}, 063504 (2000), astro-ph/9907431.

\bibitem{Smith:2006ud}
K.~M. Smith and M.~Zaldarriaga,
\newblock Mon.Not.Roy.Astron.Soc. {\bf 417}, 2 (2011), astro-ph/0612571.

\bibitem{Bucher:2009nm}
M.~Bucher, B.~Van~Tent, and C.~S. Carvalho,
\newblock Mon.Not.Roy.Astron.Soc. {\bf 407}, 2193 (2010), 0911.1642.

\bibitem{Donzelli:2012ts}
S.~Donzelli, F.~K. Hansen, M.~Liguori, D.~Marinucci, and S.~Matarrese,
\newblock Astrophys.J. {\bf 755}, 19 (2012), 1202.1478.

\bibitem{Byun:2013jba}
J.~Byun and R.~Bean,
\newblock JCAP {\bf 1309}, 026 (2013), 1303.3050.

\bibitem{Das:2011ak}
S.~Das {\em et~al.},
\newblock Phys.Rev.Lett. {\bf 107}, 021301 (2011), 1103.2124.

\bibitem{Das:2013zf}
S.~Das {\em et~al.},
\newblock JCAP {\bf 1404}, 014 (2014), 1301.1037.

\bibitem{vanEngelen:2012va}
A.~van Engelen {\em et~al.},
\newblock Astrophys.J. {\bf 756}, 142 (2012), 1202.0546.

\bibitem{Story:2014dwa}
K.~Story {\em et~al.},
\newblock (2014), 1412.4760.

\bibitem{Ade:2013tyw}
Planck Collaboration, P.~Ade {\em et~al.},
\newblock Astron.Astrophys. {\bf 571}, A17 (2014), 1303.5077.

\bibitem{Cheung:2007st}
C.~Cheung, P.~Creminelli, A.~L. Fitzpatrick, J.~Kaplan, and L.~Senatore,
\newblock JHEP {\bf 0803}, 014 (2008), 0709.0293.

\bibitem{Senatore:2010jy}
L.~Senatore and M.~Zaldarriaga,
\newblock JCAP {\bf 1101}, 003 (2011), 1004.1201.

\bibitem{Chen:2009bc}
X.~Chen, B.~Hu, M.-x. Huang, G.~Shiu, and Y.~Wang,
\newblock JCAP {\bf 0908}, 008 (2009), 0905.3494.

\bibitem{Arroja:2009pd}
F.~Arroja, S.~Mizuno, K.~Koyama, and T.~Tanaka,
\newblock Phys.Rev. {\bf D80}, 043527 (2009), 0905.3641.

\bibitem{Bennett:2012zja}
WMAP, C.~Bennett {\em et~al.},
\newblock Astrophys.J.Suppl. {\bf 208}, 20 (2013), 1212.5225.

\bibitem{Bartolo:2010di}
N.~Bartolo, M.~Fasiello, S.~Matarrese, and A.~Riotto,
\newblock JCAP {\bf 1009}, 035 (2010), 1006.5411.

\bibitem{Creminelli:2010qf}
P.~Creminelli, G.~D'Amico, M.~Musso, J.~Norena, and E.~Trincherini,
\newblock JCAP {\bf 1102}, 006 (2011), 1011.3004.

\bibitem{Bartolo:2013eka}
N.~Bartolo, E.~Dimastrogiovanni, and M.~Fasiello,
\newblock JCAP {\bf 1309}, 037 (2013), 1305.0812.

\bibitem{Arroja:2013dya}
F.~Arroja, N.~Bartolo, E.~Dimastrogiovanni, and M.~Fasiello,
\newblock JCAP {\bf 1311}, 005 (2013), 1307.5371.

\bibitem{Sasaki:1995aw}
M.~Sasaki and E.~D. Stewart,
\newblock Prog.Theor.Phys. {\bf 95}, 71 (1996), astro-ph/9507001.

\bibitem{Starobinsky:1986fxa}
A.~A. Starobinsky,
\newblock JETP Lett. {\bf 42}, 152 (1985).

\bibitem{Sasaki:1998ug}
M.~Sasaki and T.~Tanaka,
\newblock Prog.Theor.Phys. {\bf 99}, 763 (1998), gr-qc/9801017.

\bibitem{Lee:2005bb}
H.-C. Lee, M.~Sasaki, E.~D. Stewart, T.~Tanaka, and S.~Yokoyama,
\newblock JCAP {\bf 0510}, 004 (2005), astro-ph/0506262.

\bibitem{Chen:2009we}
X.~Chen and Y.~Wang,
\newblock Phys.Rev. {\bf D81}, 063511 (2010), 0909.0496.

\bibitem{Chen:2009zp}
X.~Chen and Y.~Wang,
\newblock JCAP {\bf 1004}, 027 (2010), 0911.3380.

\bibitem{Creminelli:2011mw}
P.~Creminelli,
\newblock Phys.Rev. {\bf D85}, 041302 (2012), 1108.0874.

\bibitem{Maldacena:2011nz}
J.~M. Maldacena and G.~L. Pimentel,
\newblock JHEP {\bf 1109}, 045 (2011), 1104.2846.

\bibitem{Bernardeau:2002jy}
F.~Bernardeau and J.-P. Uzan,
\newblock Phys.Rev. {\bf D66}, 103506 (2002), hep-ph/0207295.

\bibitem{Bernardeau:2002jf}
F.~Bernardeau and J.-P. Uzan,
\newblock Phys.Rev. {\bf D67}, 121301 (2003), astro-ph/0209330.

\bibitem{Seery:2006vu}
D.~Seery, J.~E. Lidsey, and M.~S. Sloth,
\newblock JCAP {\bf 0701}, 027 (2007), astro-ph/0610210.

\bibitem{Huang:2006eha}
X.~Chen, M.-x. Huang, and G.~Shiu,
\newblock Phys.Rev. {\bf D74}, 121301 (2006), hep-th/0610235.

\bibitem{Seery:2006js}
D.~Seery and J.~E. Lidsey,
\newblock JCAP {\bf 0701}, 008 (2007), astro-ph/0611034.

\bibitem{Byrnes:2006vq}
C.~T. Byrnes, M.~Sasaki, and D.~Wands,
\newblock Phys.Rev. {\bf D74}, 123519 (2006), astro-ph/0611075.

\bibitem{Bernardeau:2007xi}
F.~Bernardeau and T.~Brunier,
\newblock Phys.Rev. {\bf D76}, 043526 (2007), 0705.2501.

\bibitem{Arroja:2008ga}
F.~Arroja and K.~Koyama,
\newblock Phys.Rev. {\bf D77}, 083517 (2008), 0802.1167.

\bibitem{Seery:2008ax}
D.~Seery, M.~S. Sloth, and F.~Vernizzi,
\newblock JCAP {\bf 0903}, 018 (2009), 0811.3934.

\bibitem{Engel:2008fu}
K.~T. Engel, K.~S. Lee, and M.~B. Wise,
\newblock Phys.Rev. {\bf D79}, 103530 (2009), 0811.3964.

\bibitem{Huang:2009xa}
Q.-G. Huang,
\newblock JCAP {\bf 0905}, 005 (2009), 0903.1542.

\bibitem{Gao:2009gd}
X.~Gao and B.~Hu,
\newblock JCAP {\bf 0908}, 012 (2009), 0903.1920.

\bibitem{Kawakami:2009iu}
E.~Kawakami, M.~Kawasaki, K.~Nakayama, and F.~Takahashi,
\newblock JCAP {\bf 0909}, 002 (2009), 0905.1552.

\bibitem{Mizuno:2009mv}
S.~Mizuno, F.~Arroja, and K.~Koyama,
\newblock Phys.Rev. {\bf D80}, 083517 (2009), 0907.2439.

\bibitem{Bartolo:2009kg}
N.~Bartolo, E.~Dimastrogiovanni, S.~Matarrese, and A.~Riotto,
\newblock JCAP {\bf 0911}, 028 (2009), 0909.5621.

\bibitem{ValenzuelaToledo:2009nq}
C.~A. Valenzuela-Toledo and Y.~Rodriguez,
\newblock Phys.Lett. {\bf B685}, 120 (2010), 0910.4208.

\bibitem{Huang:2010ab}
Q.-G. Huang,
\newblock JCAP {\bf 1007}, 025 (2010), 1004.0808.

\bibitem{Izumi:2010wm}
K.~Izumi and S.~Mukohyama,
\newblock JCAP {\bf 1006}, 016 (2010), 1004.1776.

\bibitem{Gao:2010xk}
X.~Gao and C.~Lin,
\newblock JCAP {\bf 1011}, 035 (2010), 1009.1311.

\bibitem{Leblond:2010yq}
L.~Leblond and E.~Pajer,
\newblock JCAP {\bf 1101}, 035 (2011), 1010.4565.

\bibitem{Langlois:2010fe}
D.~Langlois and T.~Takahashi,
\newblock JCAP {\bf 1102}, 020 (2011), 1012.4885.

\bibitem{Meyers:2011mm}
J.~Meyers and N.~Sivanandam,
\newblock Phys.Rev. {\bf D84}, 063522 (2011), 1104.5238.

\bibitem{Agullo:2011aa}
I.~Agullo, J.~Navarro-Salas, and L.~Parker,
\newblock JCAP {\bf 1205}, 019 (2012), 1112.1581.

\bibitem{Elliston:2012wm}
J.~Elliston, L.~Alabidi, I.~Huston, D.~Mulryne, and R.~Tavakol,
\newblock JCAP {\bf 1209}, 001 (2012), 1203.6844.

\bibitem{Anderson:2012em}
G.~J. Anderson, D.~J. Mulryne, and D.~Seery,
\newblock JCAP {\bf 1210}, 019 (2012), 1205.0024.

\bibitem{Renaux-Petel:2013wya}
S.~Renaux-Petel,
\newblock JCAP {\bf 1307}, 005 (2013), 1302.6978.

\bibitem{Abolhasani:2013zya}
A.~A. Abolhasani, R.~Emami, J.~T. Firouzjaee, and H.~Firouzjahi,
\newblock JCAP {\bf 1308}, 016 (2013), 1302.6986.

\bibitem{Renaux-Petel:2013ppa}
S.~Renaux-Petel,
\newblock JCAP {\bf 1308}, 017 (2013), 1303.2618.

\bibitem{Leung:2013rza}
G.~Leung, E.~R. Tarrant, C.~T. Byrnes, and E.~J. Copeland,
\newblock JCAP {\bf 1308}, 006 (2013), 1303.4678.

\bibitem{Fasiello:2013dla}
M.~Fasiello,
\newblock JCAP {\bf 1312}, 033 (2013), 1303.5015.

\bibitem{Byrnes:2013qjy}
C.~T. Byrnes, S.~Nurmi, G.~Tasinato, and D.~Wands,
\newblock Europhys.Lett. {\bf 103}, 19001 (2013), 1306.2370.

\bibitem{Hu:2001fa}
W.~Hu,
\newblock Phys. Rev. {\bf D64}, 083005 (2001), astro-ph/0105117.

\bibitem{Regan:2010cn}
D.~M. Regan and E.~P.~S. Shellard,
\newblock Phys. Rev. {\bf D82}, 023520 (2010), 1004.2915.

\bibitem{Sherwin:2010ge}
B.~D. Sherwin and S.~Das,
\newblock (2010), 1011.4510.

\bibitem{Plaszczynski:2012ej}
S.~Plaszczynski, A.~Lavabre, L.~Perotto, and J.-L. Starck,
\newblock Astron.Astrophys. {\bf 544}, A27 (2012), 1201.5779.

\bibitem{Namikawa:2012pe}
T.~Namikawa, D.~Hanson, and R.~Takahashi,
\newblock Mon.Not.Roy.Astron.Soc. {\bf 431}, 609 (2013), 1209.0091.

\bibitem{Lewis:1999bs}
A.~Lewis, A.~Challinor, and A.~Lasenby,
\newblock Astrophys.J. {\bf 538}, 473 (2000), astro-ph/9911177.

\bibitem{Baumann:2011nk}
D.~Baumann and D.~Green,
\newblock Phys.Rev. {\bf D85}, 103520 (2012), 1109.0292.

\bibitem{Green:2013rd}
D.~Green, M.~Lewandowski, L.~Senatore, E.~Silverstein, and M.~Zaldarriaga,
\newblock JHEP {\bf 1310}, 171 (2013), 1301.2630.

\bibitem{Munshi:2009wy}
D.~Munshi {\em et~al.},
\newblock Mon.Not.Roy.Astron.Soc. {\bf 412}, 1993 (2011), 0910.3693.

\bibitem{vanEngelen:2014zlh}
ACT Collaboration, A.~van Engelen {\em et~al.},
\newblock (2014), 1412.0626.

\bibitem{Smith:2007rg}
K.~M. Smith, O.~Zahn, and O.~Dor\'e,
\newblock Phys. Rev. {\bf D76}, 043510 (2007), 0705.3980.

\bibitem{Smith:2009jr}
K.~M. Smith, L.~Senatore, and M.~Zaldarriaga,
\newblock JCAP {\bf 0909}, 006 (2009), 0901.2572.

\bibitem{Smith:2004up}
K.~M. Smith, W.~Hu, and M.~Kaplinghat,
\newblock Phys.Rev. {\bf D70}, 043002 (2004), astro-ph/0402442.

\bibitem{Li:2006pu}
C.~Li, T.~L. Smith, and A.~Cooray,
\newblock Phys.Rev. {\bf D75}, 083501 (2007), astro-ph/0607494.

\bibitem{Liguori:2003mb}
M.~Liguori, S.~Matarrese, and L.~Moscardini,
\newblock Astrophys. J. {\bf 597}, 57 (2003), astro-ph/0306248.

\bibitem{Hanson:2009kg}
D.~Hanson, K.~M. Smith, A.~Challinor, and M.~Liguori,
\newblock Phys.Rev. {\bf D80}, 083004 (2009), 0905.4732.

\bibitem{Lewis:2005tp}
A.~Lewis,
\newblock Phys. Rev. {\bf D71}, 083008 (2005), astro-ph/0502469.

\bibitem{Hinshaw:2012aka}
WMAP, G.~Hinshaw {\em et~al.},
\newblock Astrophys.J.Suppl. {\bf 208}, 19 (2013), 1212.5226.

\bibitem{Creminelli:2006rz}
P.~Creminelli, L.~Senatore, M.~Zaldarriaga, and M.~Tegmark,
\newblock JCAP {\bf 0703}, 005 (2007), astro-ph/0610600.

\bibitem{Seljak:1996is}
U.~Seljak and M.~Zaldarriaga,
\newblock Astrophys.J. {\bf 469}, 437 (1996), astro-ph/9603033.

\bibitem{steed}
A.~Barnett, D.~Feng, J.~Steed, and L.~Goldfarb,
\newblock Computer Physics Communications {\bf 8}, 377 (1974).

\bibitem{Hanson:2010rp}
D.~Hanson, A.~Challinor, G.~Efstathiou, and P.~Bielewicz,
\newblock Phys.Rev. {\bf D83}, 043005 (2011), 1008.4403.

\end{thebibliography}

\appendix

\section{Wigner $d$-functions and spin-$s$ spherical harmonics}
\label{app:spin_s}

In this appendix, we briefly review
properties of the Wigner $d$-function $d^\ell_{ss'}(\theta)$
and spin-$s$ spherical harmonics $({}_sY_{\ell m})$
which will be used in the text.

For integers $s,s'$,
the Wigner $d$-function $d^\ell_{ss'}(\theta)$ 
is defined for $\ell \ge \max(|s|,|s'|)$
and satisfies the orthogonality condition:
\be
\int_{-1}^1 d(\cos\theta) \, d^{\ell_1}_{ss'}(\theta) \, d^{\ell_2}_{ss'}(\theta) = \frac{2}{2\ell+1} \delta_{\ell_1\ell_2}
\ee
as well as the identity:
\be
d^\ell_{-s,-s'}(\theta) = d^\ell_{s's}(\theta) = (-1)^{s+s'} d^\ell_{ss'}(\theta)  \label{eq:wignerd_flip}
\ee
The Wigner $d$-functions can be computed using the recursion relation
\be
\alpha^{\ell+1}_{ss'}(\theta) d^{\ell+1}_{ss'}(\theta)
   - (2\ell+1) \left(\cos\theta - \frac{ss'}{\ell(\ell+1)} \right) d^\ell_{ss'}(\theta)
   + \alpha^\ell_{ss'}(\theta) d^{\ell-1}_{ss'}(\theta) = 0 \hspace{1cm} (\ell \ge \max(|s|,|s'|))  \label{eq:wignerd_recursion}
\ee
where $\alpha^\ell_{ss'} = \sqrt{(\ell^2-s^2)(\ell^2-(s')^2)} / \ell$.
If $s' \ge |s|$, then initial conditions for the recursion are given by:
\be
d^{s'}_{ss'}(\theta) = \sqrt{\frac{(2s')!}{(s'-|s|)! (s'+|s|)!}}
   \left( \frac{1+\cos\theta}{2} \right)^{(s'+s)/2}
   \left( \frac{1-\cos\theta}{2} \right)^{(s'-s)/2}  \hspace{1cm} (s' \ge |s|)
\ee
Initial conditions for arbitrary $(s,s')$ can be obtained through use of Eq.~(\ref{eq:wignerd_flip}).

A spin-$s$ field $({}_sf)$ is a field whose value at a point $\n$ depends on a choice $\{ \e_1, \e_2 \}$ of local orthonormal
frame at $\n$, such that under a change of frame $(\e_1 \pm i\e_2) \rightarrow e^{\pm i\alpha} (\e_1 \pm i\e_2)$, the field value transforms
as $({}_sf) \rightarrow e^{-is\alpha} ({}_sf)$.

The spin-raising operator $\sraise$ and spin-lowering operator $\slower$ are defined by:
\be
\sraise = m_a \nabla^a  \hspace{1cm}  \slower = m_a^* \nabla^a
\ee
where we have defined the spin-1 vector field $m_a = (\e_1 - i\e_2)$.
The spin-raising and spin-lowering operators
transform a spin-$s$ field to fields of spin $(s+1)$ and $(s-1)$ respectively.

The spin-$s$ spherical harmonics $({}_sY_{\ell m})$ are an orthonormal basis for spin-$s$ fields on the full sky,
defined for $\ell \ge |s|$ and $-\ell\le m\le \ell$.
The spin-raising and spin-lowering operators act on the $({}_sY_{\ell m})$ by:
\be
\sraise({}_sY_{\ell m}) = \sqrt{(\ell-s)(\ell+s+1)} ({}_{s+1}Y_{\ell m})
  \hspace{1cm}
\slower({}_sY_{\ell m}) = -\sqrt{(\ell+s)(\ell-s+1)} ({}_{s-1}Y_{\ell m})
\ee
The spin-$s$ harmonics satisfy the identity:
\be
({}_sY_{\ell m})^* = (-1)^{s+m} ({}_{-s}Y_{\ell,-m})
\ee
and are related to Wigner $d$-functions by the following sum rule:
\be
\sum_{m=-\ell}^\ell ({}_sY_{\ell m}(\n)) \, ({}_{s'}Y_{\ell m}(\n'))^*
  = (-1)^s \, \frac{2\ell+1}{4\pi} \, d^{\ell}_{ss'}(\theta)  \label{eq:sum_rule}
\ee
where $\theta = \cos^{-1}(\n\cdot\n')$ is the angle between unit vectors $\n,\n'$.
The sum rule applies in the ``two-point'' frame where the local frame vectors $\e_1,\e'_1$ 
at points $\n,\n'$ have been chosen to point along the geodesic connecting the two points.
(Note that the LHS of the sum rule depends on the choice of frames, but the RHS does not, 
so it must be understood that the sum rule applies only in the two-point frame.)

\section{Non-Gaussian simulations}
\label{app:ngsim}

In \cite{Smith:2006ud}, an algorithm was proposed for simulating a random, weakly non-Gaussian CMB realization with prescribed power spectrum
and bispectrum.  It is straightforward to generalize this algorithm to simulate a CMB realization with prescribed power spectrum $C_\ell$
and trispectrum $T^{\ell_1\ell_2\ell_3\ell_4}_{m_1m_2m_3m_4}$.  

First, a definition.  For any angular trispectrum $T$, define the symmetric matrix $\tT_{\ell\ell'}$ by:
\be
\tT_{\ell\ell'} = \sum_{mm'} (-1)^{m+m'} T^{\ell\ell\ell'\ell'}_{m(-m)m'(-m')} \,. \label{eq:Tll_def}
\ee
This matrix arises in several contexts.  First, if $a_{\ell m}, b_{\ell m}$ are all-sky Gaussian random fields,
then $\tT_{\ell\ell'}$ appears in the following expectation values:
\ba
\langle Q_T[a] \rangle 
  &=& \frac{1}{8} \sum_{\ell\ell'} \tT_{\ell\ell'} C_\ell^{aa} C_{\ell'}^{aa}  \label{eq:tT_ex1} \\
\Big\langle b_{\ell'm'}^* \, \partial_{\ell m} Q_T[a] \Big\rangle 
  &=& \left( \frac{C_\ell^{ab}}{2(2\ell+1)} \sum_{\ell''} \tT_{\ell\ell''} C_{\ell''}^{aa} \right) \delta_{\ell\ell'} \delta_{mm'}  \label{eq:tT_ex2}
\ea
Second, if the CMB is non-Gaussian, then the estimated CMB power spectrum $\hC_\ell = (2\ell+1)^{-1} \sum_m a_{\ell m}^* a_{\ell m}$
contains a term proportional to $\tT_{\ell\ell'}$:
\be
\Cov(\hC_\ell, \hC_{\ell'}) = \frac{2 C_\ell^2}{2\ell+1} \delta_{\ell\ell'} + \frac{\tT_{\ell\ell'}}{(2\ell+1)(2\ell'+1)}  \label{eq:ngcov}
\ee
Note that non-Gaussian power spectrum covariance due to the gravitational
lensing trispectrum has been studied extensively (e.g.~\cite{Smith:2004up,Li:2006pu}); Eq.~(\ref{eq:ngcov})
generalizes to an arbitrary trispectrum.

% The following identity will be useful:
% \be
% \sum_m (-1)^m T^{\ell\ell\ell_3\ell_4}_{m(-m)m_3m_4} = \frac{(-1)^{m_3}}{2\ell_3+1} \tT^{\ell\ell_3} \delta_{\ell_3\ell_4} \delta_{m_3(-m_4)}  \label{eq:magic_identity}
% \ee
% This follows from rotation invariance of the trispectrum and we sketch a proof as follows.  The left-hand side is invariant under combined rotations of the $(\ell_3,m_3)$
% and $(\ell_4,m_4)$ indices, since it is obtained by contracting the rotation-invariant trispectrum $T^{\ell_1\ell_2\ell_3\ell_4}_{m_1m_2m_3m_4}$ with the
% invariant tensor $(-1)^{m_1} \delta_{\ell\ell_1} \delta_{\ell_1\ell_2} \delta_{m_1,-m_2}$.
% There is a theorem that a rotation invariant tensor with two pairs of $(\ell,m)$ indices must be diagonal, i.e.~the left-hand side of~(\ref{eq:magic_identity})
% is of the form $(-1)^{m_3} A^{\ell\ell_3} \delta_{\ell_3\ell_4} \delta_{m_3,-m_4}$.
% To determine $A$, we multiply both sides by $(-1)^{m_3} \delta_{\ell_3\ell_4} \delta_{m_3,-m_4}$ and sum over $\ell_4,m_3,m_4$, obtaining
% $\tT^{\ell\ell_3} = (2\ell_3+1) A^{\ell\ell_3}$.  This completes the proof.

Our non-Gaussian simulation algorithm is as follows.
We first simulate a {\em Gaussian} field $a^G_{\ell m}$ with power
spectrum $C_\ell$, and then define the non-Gaussian field $a_{\ell m}^{NG}$ by:
\be
a^{NG}_{\ell m} = a^G_{\ell m} + \frac{1}{4} \partial_{\ell m} Q[a_{\ell m}^G/C_\ell] 
   - \frac{1}{8} \sum_{\ell'} \frac{\tT_{\ell\ell'}}{(2\ell+1)C_\ell C_{\ell'}} a^G_{\ell m}  \label{eq:simulation_algorithm}
\ee
A short calculation shows that the power spectrum and four-point function of the simulated field are given by:
\be
\langle a_{\ell m}^{NG*} a_{\ell m}^{NG} \rangle = C_\ell + \bigoh(T^2) \hspace{1.5cm}
\langle a_{\ell_1 m_1}^{NG} a_{\ell_2 m_2}^{NG} a_{\ell_3 m_3}^{NG} a_{\ell_4m_4}^{NG} \rangle_c = T^{\ell_1\ell_2\ell_3\ell_4}_{m_1m_2m_3m_4} + \bigoh(T^2)
\ee
where $\bigoh(T^2)$ denotes contributions which are second-order in the trispectrum $T^{\ell_1\ell_2\ell_3\ell_4}_{m_1m_2m_3m_4}$.
The last term in Eq.~(\ref{eq:simulation_algorithm}) has been included in order to avoid an order-$\bigoh(T)$ correction to the power spectrum.
(We note that odd $(2N+1)$-point correlation functions of $a_{\ell m}^{NG}$ are zero, and even $(2N)$-point connected 
correlation functions are of order $\bigoh(T^{N-1})$.)

To apply the simulation algorithm, we need to compute $\tT_{\ell\ell'}$.
In the case where $T$ is contact factorizable we can do this using a method
similar to the exact Fisher matrix algorithm from~\S\ref{ssec:fisher_exact}.
We write $Q_T[a]$ in the abbreviated form:
\be
Q_T[a] = \frac{1}{48} \sum_{I=1}^{\Nfact} \int d^2\n \,
  M^{A_I}_{\alpha_I}(\n) M^{B_I}_{\beta_I}(\n) M^{C_I}_{\gamma_I}(\n) M^{D_I}_{\delta_I}(\n) + \mbox{c.c.}
\ee
where we have defined $M^X_s(\n) = \sum_{\ell m} X_\ell a_{\ell m} ({}_sY_{\ell m}(\n))$.
We can compute the expectation value $\langle Q_T[a] \rangle$ using Wick's theorem
and the contraction
\be
\wick{1}{<1M^X_s(\n) \,\, >1M^{X'}_{s'}(\n)} = (-1)^s \sum_\ell \frac{2\ell+1}{4\pi} X_\ell X'_\ell C_\ell^{aa} \delta_{s,-s'}
\ee
obtaining:
\ba
\langle Q_T[a] \rangle
  &=& \frac{1}{48} \sum_{I=1}^{\Nfact} \sum_{\ell\ell'} \frac{(2\ell+1)(2\ell'+1)}{4\pi} C_\ell^{aa} C_{\ell'}^{aa}
     \Big[
         (-1)^{\alpha_I+\gamma_I} A^I_\ell B^I_\ell C^I_{\ell'} D^I_{\ell'} \delta_{\alpha_I,-\beta_I} \delta_{\gamma_I,-\delta_I} \nn \\
&& \hspace{0.5cm}
       + (-1)^{\alpha_I+\beta_I} A^I_\ell B^I_{\ell'} C^I_\ell D^I_{\ell'} \delta_{\alpha_I,-\gamma_I} \delta_{\beta_I,-\delta_I}
       + (-1)^{\alpha_I+\beta_I} A^I_\ell B^I_{\ell'} C^I_{\ell'} D^I_{\ell} \delta_{\alpha_I,-\delta_I} \delta_{\beta_I,-\gamma_I}
       \Big] + \mbox{c.c.}
\ea
Comparing with Eq.~(\ref{eq:tT_ex1}) we can read off an expression for $\tT_{\ell\ell'}$
(note that we symmetrize in $\ell,\ell'$):
\ba
\tT_{\ell\ell'} &=& \frac{(2\ell+1)(2\ell'+1)}{48\pi} \sum_{I=1}^{\Nfact}
     \Big[
         (-1)^{\alpha_I+\gamma_I} A^I_\ell B^I_\ell C^I_{\ell'} D^I_{\ell'} \delta_{\alpha_I,-\beta_I} \delta_{\gamma_I,-\delta_I}
       + (-1)^{\alpha_I+\beta_I} A^I_\ell B^I_{\ell'} C^I_\ell D^I_{\ell'} \delta_{\alpha_I,-\gamma_I} \delta_{\beta_I,-\delta_I} \nn \\
&& \hspace{3.5cm}
       + (-1)^{\alpha_I+\beta_I} A^I_\ell B^I_{\ell'} C^I_{\ell'} D^I_{\ell} \delta_{\alpha_I,-\delta_I} \delta_{\beta_I,-\gamma_I}
       \Big] + \Big(\ell \leftrightarrow \ell'\Big) + \mbox{c.c.}
\ea
This algorithm for computing $\tT_{\ell\ell'}$ can be generalized to the exchange
factorizable case, using the same strategy of computing $\langle Q_T[a] \rangle$
with Wick's theorem, but we omit the details.
Note that the matrix $\tT_{\ell\ell'}$ may be useful outside the context of non-Gaussian
simulations, since it appears in the non-Gaussian power spectrum covariance~(\ref{eq:ngcov}).

This generic simulation algorithm formally generates a non-Gaussian field whose power spectrum $C_\ell$ 
and trispectrum $T$ are prescribed, up to contributions of order $\bigoh(T^2)$.
A significant caveat is that for some trispectrum shapes, these $\bigoh(T^2)$ contributions 
can be large even for modest levels of non-Gaussianity.
We have not experimented much with the simulation algorithm, but
based on experience with the analogous bispectrum algorithm,
we expect it it will work well for shapes which do not have large squeezed
limits, for example the $\dot\sigma^4$, $\dot\sigma^2 (\partial\sigma)^2$, and $(\partial\sigma)^4$
shapes.

The local trispectra $\gnlloc$ and $\taunl$ have large squeezed limits, so
we do not expect our generic simulation algorithm to work well in these cases.
One alternate approach is to simulate 3D fields at the end of inflation, 
apply the relevant local operation (either $\zeta = \zeta_G + (9/25) \gnlloc \zeta_G^3$
or $\zeta = \zeta_G + \taunl^{1/2} \zeta_G \sigma$), and then apply the
CMB transfer function to generate $a_{\ell m}$'s~\cite{Liguori:2003mb}.
Another approach is to reweight terms in the generic algorithm to avoid
infrared divergences in specific cases (see discussion in the appendix of~\cite{Hanson:2009kg}).

The gravitational lensing trispectrum~(\ref{eq:Q_lensing}) is another
example of a shape with a large squeezed limit, where we do not expect
our generic simulation algorithm to work well.  In this case the best
approach is to simply simulate the lensing deflection $T(\n) \rightarrow T(\n + \nabla\phi(\n))$
directly~\cite{Lewis:2005tp}.

\section{Factorizable representations for $\gnlloc$, $\gnldotpi4$, $\gnldpi4$}
\label{app:contact_factorizability}

\par\noindent
In \S\ref{ssec:contact_factorizability}, we calculated the factorizable representation
explicitly for the trispectrum generated by the quartic operator $\dot\sigma^2 (\partial\sigma)^2$.
In this appendix, we do the same for the $\gnlloc$ trispectrum, and the trispectra
generated by the operators $\dot\sigma^4$ and $(\partial\sigma)^4$.

\subsection{$\gnlloc$ shape}

\par\noindent
First we consider the $\gnlloc$ shape.
The $\zeta$-trispectrum was given previously in Eq.~(\ref{eq:gnlloc_zeta_trispectrum}):
\be
\langle \zeta_{\k_1} \zeta_{\k_2} \zeta_{\k_3} \zeta_{\k_4} \rangle_c
= \left( \frac{54}{25} \gnlloc P_\zeta(k_2) P_\zeta(k_3) P_\zeta(k_4) + \mbox{3 perm.} \right) (2\pi)^3 \delta^3\Big(\sum\k_i\Big)
\ee
Following the previous calculation in~\S\ref{ssec:contact_factorizability},
we plug into the projection formula~(\ref{eq:Q_projection}),
replace the delta function $\delta^3(\sum\k_i)$ by $\int d^3\r \exp(i\sum\k_i\cdot\r)$,
and do the angular parts of the $\k$ integrals, obtaining:
\ba
Q_T[a] &=& \frac{9}{25} \gnlloc \int_0^\infty r^2 \, dr \int d^2\n\,
           \left( \prod_{i=1}^4 \int \frac{2k_i^2\,dk_i}{\pi} \sum_{\ell_im_i} j_{\ell_i}(k_ir) 
                 \Delta_{\ell_i}(k_i) a_{\ell_im_i} Y_{\ell_im_i}(\n) \right)
           P_\zeta(k_2) P_\zeta(k_3) P_\zeta(k_4) \nn \\
 &=& \gnlloc \int r^2 dr \int d^2\n\, 
  \Bigg( \sum_{\ell m} \alpha_\ell(r) a_{\ell m} Y_{\ell m}(\n) \Bigg)
  \Bigg( \sum_{\ell'm'} \beta_{\ell'}(r) a_{\ell'm'} Y_{\ell'm'}(\n) \Bigg)^3  \label{eq:Q_gnlloc}
\ea
where the functions $\alpha_\ell(r), \beta_\ell(r)$ were defined previously in Eq.~(\ref{eq:alpha_beta}).
Comparing with the definition~(\ref{eq:cf_def_Q}), we see that after replacing the $r$
integral by a finite sum, the trispectrum is contact factorizable with all spins
equal to zero.

\subsection{$\gnldotpi4$ shape}

\par\noindent
Next we consider the quartic operator $\dot\sigma^4$.
The $\zeta$-trispectrum was given previously in Eq.~(\ref{eq:gnldotpi4}):
\be
\langle \zeta_{\k_1} \zeta_{\k_2} \zeta_{\k_3} \zeta_{\k_4} \rangle_c
   = \frac{9216}{25} A_\zeta^3 \gnldotpi4 \int_{-\infty}^0
        d\tau_E \, \tau_E^4
          \left( \prod_{i=1}^4 \frac{e^{k_i\tau_E}}{k_i} \right) 
          (2\pi)^3 \delta^3\Big(\sum\k_i\Big)
\ee
Using the same method of calculation as in~\S\ref{ssec:contact_factorizability},,
we find the following expression for $Q_T[a]$:
\be
Q_T[a] = \frac{384}{25} A_\zeta^3 \gnldotpi4 \int_{-\infty}^0 d\tau_E \int_0^\infty dr \, \tau_E^4 r^2 \int d^2\n
  \left( \prod_{i=1}^4 \sum_{\ell_im_i} \int \frac{2k_i^2\,dk_i}{\pi} j_{\ell_i}(k_ir) \Delta_{\ell_i}(k_i)
      \frac{e^{k_i\tau_E}}{k_i} a_{\ell_im_i} Y_{\ell_im_i}(\n) \right)
\ee
Using the notation $\mu_\ell(\tau_E,r)$ defined previously in Eq.~(\ref{eq:mnw_def}),
we rewrite this as:
\be
Q_T[a] = \frac{128}{3} \gnldotpi4 \int_{-\infty}^0 d\tau_E \int_0^\infty dr \, \tau_E^4 r^2 \int d^2\n
  \left( \sum_{\ell m} \mu_\ell(\tau_E,r) a_{\ell m} Y_{\ell m}(\n) \right)^4   \label{eq:Q_gnldotpi4}
\ee
After replacing the $(\tau_E,r)$ double integral by a finite sum, $Q_T[a]$
is of contact factorizable form~(\ref{eq:contact_factorizable_def}).

\subsection{$\gnldpi4$ shape}

\par\noindent
Finally we consider the case of a quartic operator $(\partial_i \sigma)^2 (\partial_j \sigma)^2$.
The $\zeta$-trispectrum was given previously in Eq.~(\ref{eq:gnldpi4}):
\be
\langle \zeta_{\k_1} \zeta_{\k_2} \zeta_{\k_3} \zeta_{\k_4} \rangle_c
  = \frac{82944}{2575} \gnldpi4 A_\zeta^3 \int_{-\infty}^0 d\tau_E
              \left( \prod_{i=1}^4 \frac{(1-k_i\tau_E)e^{k_i\tau_E}}{k_i^3} \right) 
              \bigg( (\k_1\cdot\k_2)(\k_3\cdot\k_4) + \mbox{2 perm.} \bigg)
              (2\pi)^3 \delta^3\Big(\sum\k_i\Big)
\ee
In this case we find the following expression for $Q_T[a]$:
\be
Q_T[a] = \frac{1152}{103} \gnldpi4 \int_{-\infty}^0 d\tau_E \int d^3\r  \,
     \frac{\partial F(\tau_E,\r)}{\partial\r_i}
     \frac{\partial F(\tau_E,\r)}{\partial\r_i}
     \frac{\partial F(\tau_E,\r)}{\partial\r_j}
     \frac{\partial F(\tau_E,\r)}{\partial\r_j}
\ee
where we have defined
\be
F(\tau_E,\r) = \left( \frac{3}{5} \right)^{1/2} A_\zeta^{3/4}
   \int \frac{2k^2\,dk}{\pi} \sum_{\ell m} \frac{(1-k\tau_E) e^{k\tau_E}}{k^3} j_\ell(kr) \Delta_\ell(k) a_{\ell m} Y_{\ell m}(\r)
\ee
As in~\S\ref{ssec:contact_factorizability} we can split the gradients into radial and tangential terms using the identity:
\be
\sum_i \frac{\partial F(\r)}{\partial \r_i} \frac{\partial F(\r)}{\partial \r_i} = \left( \frac{\partial F(\r)}{\partial r} \right)^2
  + \frac{1}{r^2} (\sraise F)^* (\sraise F)
\ee
where the real and tangential derivatives can be written in terms of the 
functions $\nu,\omega$ introduced previously in Eq.~(\ref{eq:mnw_def}):
\be
\frac{\partial F(\tau_E,\r)}{\partial r} = \sum_{\ell m} \nu_\ell(\tau_E,r) a_{\ell m} Y_{\ell m}(\hr)
  \hspace{1cm}
\sraise F(\tau_E,\r) = r \sum_{\ell m} \omega_\ell(\tau_E,r) a_{\ell m} ({}_1Y_{\ell m}(\hr))
\ee
Plugging this in we get the representation of $Q_T[a]$ in contact factorizable form:
\be
Q_T[a] = \frac{1152}{103} \gnldpi4 \int_{-\infty}^0 d\tau_E \int_0^\infty dr \, r^2 \int d^2\n \,
       \left[
           \left( \sum_{\ell m} \nu_{\ell}(\tau_E,r) a_{\ell m} Y_{\ell m}(\n) \right)^2
         + \left| \sum_{\ell m} \omega_{\ell}(\tau_E,r) a_{\ell m} ({}_1Y_{\ell m}(\n)) \right|^2 \,\,
       \right]^2  \label{eq:Q_gnldpi4}
\ee

\subsection{Deviation from scale invariance}
\label{app:ns}

We have now obtained explicit factorizable representations for the shapes
$\{ \gnlloc, \gnldotpi4, \gnldpi4 \}$, but have assumed a scale invariant power
spectrum $P_\zeta(k) = A_\zeta k^{-3}$ throughout.  To compute these shapes in the
WMAP or Planck cosmologies, we need to generalize slightly to the case of a power-law 
spectrum $P_\zeta(k) = A_\zeta k^{n_s-4}$.

In fact, our factorizable representations have been written in such a way that
they generalize to an arbitrary power spectrum, by simply plugging it in whenever
$P_\zeta(k)$ appears in the definitions 
(Eqs.~(\ref{eq:mnw_def}),~(\ref{eq:alpha_beta}))
of the functions
$\alpha_\ell(k)$, $\beta_\ell(k)$, $\mu_\ell(k)$, $\nu_\ell(\tau_E,k)$,
and $\omega_\ell(\tau_E,k)$.

This prescription has several nice properties.
First, it gives the correct trispectrum in the local case, 
i.e.~when $\zeta = \zeta_G + \gnlloc \zeta_G^3$ with arbitrary $P_\zeta(k)$.
Second, it is the analogue
of the prescription which is commonly used for the bispectrum (e.g.~Eqs.~(51)--(53)
of~\cite{Hinshaw:2012aka}).
Finally, for a power-law spectrum $P_\zeta(k) = A_\zeta k^{n_s-4}$, the
$\zeta$-trispectrum scales under dilations as:
\be
\langle \zeta_{\lambda\k_1} \zeta_{\lambda\k_2} \zeta_{\lambda\k_3} \zeta_{\lambda\k_4} \rangle' = 
  \lambda^{3(n_s-4)} 
\langle \zeta_{\k_1} \zeta_{\k_2} \zeta_{\k_3} \zeta_{\k_4} \rangle'
\ee
In the case of trispectra other than the local one, the deviation from scale invariance 
of the bispectrum and trispectrum cannot be reconstructed from the tilt of the power 
spectrum~\cite{Creminelli:2006rz}.
The above parametrization is however the closer guess to the actual dependence we can expect,
and it is the correct one for the local case.

\section{Numerical calculation of trispectra}
\label{app:integrals}

\par\noindent
We have now written down factorizable representations for 
the local, $\dot\sigma^4$, and $(\partial\sigma)^4$ trispectra.
In this appendix we discuss computational issues in
calculating these trispectra numerically.
The chain of steps is:
\begin{enumerate}
\item We precompute the CMB transfer function $\Delta_\ell(k)$
  on a grid of $k$-values.
\item Each trispectrum shape is represented either as a
  single integral over $r$ (in the case of the local shape, Eq.~(\ref{eq:Q_gnlloc}))
  or double integral over $(\tau_E,r)$ (in the case of the $\dot\sigma^4$ and 
  $(\partial\sigma)^4$ shapes, Eqs.~(\ref{eq:Q_gnldotpi4}) and~(\ref{eq:Q_gnldpi4})).
  We choose a finite sampling for this integral, in order to obtain a factorizable
  representation with a finite number of terms.
\item For each point in the $(\tau_E,r)$ plane, we compute the functions 
  $\alpha_\ell(r)$, $\beta_\ell(r)$, $\mu_\ell(\tau_E,r)$, $\nu_\ell(\tau_E,r)$, 
  and $\omega_\ell(\tau_E,r)$ appearing in the factorizable representation
  by evaluating the appropriate $k$-integral (Eqs.~(\ref{eq:mnw_def}),~(\ref{eq:alpha_beta})).
\end{enumerate}
Let us consider each of these steps in detail, starting with 
the CMB transfer function $\Delta_\ell(k)$.
The transfer function can be written as a line-of-sight integral~\cite{Seljak:1996is}:
\be
\Delta_\ell(k) = \int d\chi\, S(\chi,k) j_\ell(k\chi)
\ee
We obtain the source function $S(\chi,k)$ from CAMB~\cite{Lewis:1999bs}
and evaluate the above integral using equal spacing $\Delta\chi$.
We compute the transfer function up to maximum wavenumber $k_{\rm max}$
on a grid of $k$-values defined using the $k$-dependent
step size $\Delta k = \min( \epsilon k, \kappa_0 )$.
This sampling scheme switches from equal spacing in $\log(k)$ at low-$k$
to equal spacing in $k$ at high-$k$.
We choose the following default values for the parameters just defined:
\be
(\Delta\chi, k_{\rm max}, \epsilon, \kappa_0) = (0.5, 5000 r_{horiz}^{-1}, 2 \times 10^{-3}, 3 \times 10^{-5})
\ee
Next consider discretization of the $(\tau_E,r)$ integral (step 2 above).
We discretize the outer $\tau_E$ integral using equally spaced points in $\log|\tau_E|$
from initial time $\tau_{Ei}$ to final time $\tau_{Ef}$.  Our default parameter values are:
\be
(\tau_{Ei}, \tau_{Ef}, \Delta\log|\tau_E|) = \left(-10^6 \mbox{ Mpc}, -\frac{50 \mbox{ Mpc}}{\ellmax}, \frac{\log(10)}{3} \right)
\ee
For each $\tau_E$, we discretize the inner $r$ integral as follows.
Let $r_{\rm rec}$ and $r_{\rm horiz}$ be the comoving distance to recombination
and the causal horizon respectively.
The integral formally goes to $r=\infty$, but the integrand decays beyond $r_{\rm horiz}$,
with characteristic decay scale given by $|\tau_E|$ plus the sound horizon.
Therefore, we integrate from $r=0$ to
$r_{\rm max} = r_{\rm horiz} + \rho_0 + \alpha |\tau_E|$, with default parameter values
\be
(\rho_0, \alpha) = (2000 \mbox{ Mpc}, 10)
\ee
We sample the $r$ integral with spacing $(\Delta r)_1$ from $r=0$ to $r_{\rm rec}$,
spacing $(\Delta r)_2$ from $r_{\rm rec}$ to $r_{\rm horiz}$, and spacing
$(\Delta r)_1$ from $r_{\rm horiz}$ to $r_{\rm max}$, where
\be
(\Delta r)_1 = \max\big(\rho_1, \beta|\tau_E| \big)
  \hspace{1cm}
(\Delta r)_2 = \max\big(\rho_2, \beta|\tau_E| \big)
\ee
with default parameter values:
\be
(\rho_1, \rho_2, \beta) = (50 \mbox{ Mpc}, 5 \mbox{ Mpc}, 0.1)
\ee
In the case where the trispectrum is represented as a single integral
over $r$, rather than a double integral over $(\tau_E,r)$, we use
the $r$-sampling for $\tau_E=0$.

Finally, consider evaluation of $k$-integrals (step 3 above).
We sample the integrals at the same values of $k$ where the transfer function is
computed as described previously.
The $k$-integrals  include factors 
of either the spherical Bessel function $j_\ell(x)$
or its derivative $j'_\ell(x)$.
We precompute $j_\ell(x)$ using Steed's algorithm~\cite{steed}
on a regularly spaced grid with $\Delta x = 0.2$
and interpolate to arbitrary $x$.
To evaluate $j_\ell'(x)$ we use the identity:
\be
j_\ell'(x) = \frac{\ell}{2\ell+1} j_{\ell-1}(x) - \frac{\ell+1}{2\ell+1} j_{\ell+1}(x) \hspace{1cm} \mbox{(for $\ell \ge 1$).}
\ee
This concludes our description of the numerics.
To verify that numerical errors are fully controlled, we use the following end-to-end convergence test.
In the above discussion we defined tolerance parameters controlling the accuracy of the integration.
We compute an ``improved'' trispectrum using more conservative values of tolerance parameters as follows:
\ba
&& (\tau_{Ei}, \tau_{Ef}, \Delta\log|\tau_E|, \rho_0, \rho_1, \rho_2, \beta, k_{\rm max}, \epsilon, \kappa_0, \Delta\chi, \Delta x) \nn \\
  && \hspace{1cm} \rightarrow 
      \left( -10\tau_{Ei}, -\frac{\tau_{Ef}}{10}, \frac{2}{3} \log|\tau_E|, 2\rho_0, \frac{\rho_1}{2}, 
       \frac{\rho_2}{2}, \frac{\beta}{2}, 2 k_{\rm max}, \frac{\epsilon}{2}, \frac{\kappa_0}{2}, \frac{\Delta\chi}{2}, \frac{\Delta x}{2} \right)
\ea
and also adjusting several parameters in CAMB.
We then verify that the original and improved trispectra are nearly equal,
in the metric defined by the Fisher matrix.
For this comparison, we do not optimize the trispectra, since the number of
terms $N_{\rm fact}$ in the improved trispectrum will be very large, and the
optimization algorithm will be too slow, but computing the Fisher matrix is
still affordable using the Monte Carlo algorithm from~\S\ref{ssec:fisher_mc}.
This end-to-end test is a complete check that any numerical errors in our
calculation of the trispectrum are not observationally important.

\section{Constructing the estimator $\hF$}
\label{app:magic_coefficients}

\par\noindent
In the optimal pipeline (\S\ref{ssec:optimal_pipeline})
we introduced the following estimator:
\ba
\hF &=& \frac{\alpha}{2} \left\langle 
                (\partial_{\ell m} Q[\tb,\tb,\tb]) C^{-1}_{\ell m,\ell'm'} (\partial_{\ell'm'} Q[\tb,\tb,\tb]) 
              + (\partial_{\ell m} Q[\tb',\tb',\tb']) C^{-1}_{\ell m,\ell'm'} (\partial_{\ell'm'} Q[\tb',\tb',\tb']) 
       \right\rangle \nn \\
 && \hspace{1cm}  + \frac{\beta}{2} \left\langle 
           (\partial_{\ell m} Q[\tb,\tb,\tb']) C^{-1}_{\ell m,\ell'm'} (\partial_{\ell'm'} Q[\tb,\tb,\tb'])
         + (\partial_{\ell m} Q[\tb,\tb',\tb']) C^{-1}_{\ell m,\ell'm'} (\partial_{\ell'm'} Q[\tb,\tb',\tb']) \right\rangle \nn \\
 && \hspace{1cm}  + \frac{\gamma}{2} \left\langle
           (\partial_{\ell m} Q[\tb,\tb,\tb]) C^{-1}_{\ell m,\ell'm'} (\partial_{\ell'm'} Q[\tb,\tb',\tb'])
         + (\partial_{\ell m} Q[\tb',\tb',\tb']) C^{-1}_{\ell m,\ell'm'} (\partial_{\ell'm'} Q[\tb,\tb,\tb'])
     \right\rangle \,  \label{eq:abc_def}
\ea
with coefficients $(\alpha, \beta, \gamma) = (1/16, 9/16, -3/8)$.
In this appendix we explain how these coefficients were chosen.

We use a ``contraction'' notation in which each line
between factors of $T$ denotes one factor of $C^{-1}_{\ell m,\ell'm'}$
contracted with the appropriate indices.
For example, the quantity $F$ defined in Eq.~(\ref{eq:F_harmonic}) could be denoted:
\be
\contraction[4ex]{\hspace{36pt}}{\pT}{\hspace{36pt}}{\pT}
\contraction[3ex]{\hspace{40pt}}{\pT}{\hspace{28pt}}{\pT}
\contraction[2ex]{\hspace{44pt}}{\pT}{\hspace{20pt}}{\pT}
\contraction{\hspace{48pt}}{\pT}{\hspace{12pt}}{\pT}
F = \frac{1}{4!} \Big( \hspace{5pt} T \hspace{25pt} T \hspace{5pt} \Big)
= \frac{1}{4!} \sum_{\ell_im_i\ell'_im'_i} T^{\ell_1\ell_2\ell_3\ell_4}_{m_1m_2m_3m_4}
  C^{-1}_{\ell_1m_1,\ell'_1m'_1}
  C^{-1}_{\ell_2m_2,\ell'_2m'_2}
  C^{-1}_{\ell_3m_3,\ell'_3m'_3}
  C^{-1}_{\ell_4m_4,\ell'_4m'_4}
T^{\ell'_1\ell'_2\ell'_3\ell'_4}_{m'_1m'_2m'_3m'_4}
\ee
and, as another example:
\be
\contraction[2ex]{\hspace{14pt}}{\pT}{\hspace{20pt}}{\pT}
\contraction{\hspace{18pt}}{\pT}{\hspace{12pt}}{\pT}
\contraction{\hspace{6pt}}{\pT}{\hspace{-4pt}}{\pT}
\contraction{\hspace{46pt}}{\pT}{\hspace{-4pt}}{\pT}
\Big( \hspace{5pt} T \hspace{25pt} T \hspace{5pt} \Big) \hspace{2pt}
  = \sum_{\ell_i m_i\ell_i'm_i'} 
 T^{\ell_1\ell_1'\ell_3\ell_4}_{m_1m_1'm_3m_4}
  C^{-1}_{\ell_1m_1,\ell'_1m'_1}
  C^{-1}_{\ell_2m_2,\ell'_2m'_2}
  C^{-1}_{\ell_3m_3,\ell'_3m'_3}
  C^{-1}_{\ell_4m_4,\ell'_4m'_4}
T^{\ell_2\ell'_2\ell'_3\ell'_4}_{m'_1m'_2m'_3m'_4}
\ee
Our estimator $\hF$ should have the property that $\langle \hF \rangle = F$.
It is easy to calculate the expectation value of Eq.~(\ref{eq:abc_def}), obtaining:
\be
\contraction[4ex]{\hspace{83pt}}{\pT}{\hspace{36pt}}{\pT}
\contraction[3ex]{\hspace{87pt}}{\pT}{\hspace{28pt}}{\pT}
\contraction[2ex]{\hspace{91pt}}{\pT}{\hspace{20pt}}{\pT}
\contraction{\hspace{95pt}}{\pT}{\hspace{12pt}}{\pT}
\contraction[2ex]{\hspace{240pt}}{\pT}{\hspace{20pt}}{\pT}
\contraction{\hspace{244pt}}{\pT}{\hspace{12pt}}{\pT}
\contraction{\hspace{232pt}}{\pT}{\hspace{-4pt}}{\pT}
\contraction{\hspace{272pt}}{\pT}{\hspace{-4pt}}{\pT}
\langle \hF \rangle 
= \left( \frac{\alpha}{6} + \frac{\beta}{18} \right) 
     \Big( \hspace{5pt} T \hspace{25pt} T \hspace{5pt} \Big)
+ \left( \frac{\alpha}{4} + \frac{\beta}{36} + \frac{\gamma}{12} \right)
     \Big( \hspace{5pt} T \hspace{25pt} T \hspace{5pt} \Big)
\ee
We see that $\langle \hF \rangle = F$ if the coefficients $(\alpha,\beta,\gamma)$ satisfy the constraints:
\be
\frac{\alpha}{6} + \frac{\beta}{18} = \frac{1}{24} \hspace{1.5cm} \frac{\alpha}{4} + \frac{\beta}{36} + \frac{\gamma}{12} = 0  \label{eq:hF_constraints}
\ee
These constraints do not fully determine $(\alpha,\beta,\gamma)$;
there is a 1-parameter family of solutions.
We noticed empirically that the choice $(\alpha,\beta,\gamma) = (1/16, 9/16, -3/8)$
nearly minimizes the variance $\Var(\hF)$, and even a small change in these coefficients
results in a dramatically larger value of $\Var(\hF)$.
We subsequently found a semianalytic explanation for this phenomenon as follows.
The variance $\Var(\hF)$ is a sixteen-point function which can be expanded using Wick's
theorem as a sum of many terms.
One of these terms is:
\be
\contraction[2ex]{\hspace{144pt}}{\pT}{\hspace{20pt}}{\pT}
\contraction{\hspace{148pt}}{\pT}{\hspace{12pt}}{\pT}
\contraction{\hspace{136pt}}{\pT}{\hspace{-4pt}}{\pT}
\contraction[2ex]{\hspace{176pt}}{\pT}{\hspace{17pt}}{\pT}
\contraction{\hspace{180pt}}{\pT}{\hspace{9pt}}{\pT}
\contraction[2ex]{\hspace{205pt}}{\pT}{\hspace{20pt}}{\pT}
\contraction{\hspace{209pt}}{\pT}{\hspace{12pt}}{\pT}
\contraction{\hspace{237pt}}{\pT}{\hspace{-4pt}}{\pT}
\Var(\hF) \supset \left( 12\alpha + \frac{4}{3} \beta + 4\gamma \right)^2
     \Big( \hspace{5pt} T \hspace{25pt} T \hspace{5pt} \Big)
     \Big( \hspace{5pt} T \hspace{25pt} T \hspace{5pt} \Big)
\ee
We can speculate that this term will dominate $\Var(\hF)$, since
it ``maximally factors'' in the sense defined in~\cite{Hanson:2010rp}.
If we set this term to zero by imposing the constraint
$12\alpha + (4/3)\beta + 4\gamma = 0$ in addition to the constraints~(\ref{eq:hF_constraints}),
then we obtain the coefficients $(\alpha,\beta,\gamma) = (1/16, 9/16, -3/8)$.

\section{Efficient evaluation of $\hF_V$ and $\Var(\hF_V)$}
\label{app:fv}

In the pure MC pipeline (\S\ref{ssec:puremc_pipeline}), we gave expressions for estimators $\hF_V$ and $\hSigma$,
used to estimate the statistical error on $g_{NL}$, and the ``error on the error''.
As given (in Eqs.~(\ref{eq:puremc_FV}) and~(\ref{eq:puremc_sigma})), these expressions have computational 
cost $\bigoh(\Nmc^4)$ and $\bigoh(\Nmc^8)$ respectively.
In this appendix we give mathematically equivalent expressions with cost $\bigoh(\Nmc^2)$ and $\bigoh(\Nmc^3)$.

We decompose $R_{ij} = S_{ij} + T_i + T_j + U$, where we have defined
\be
U = \frac{\sum_{ij} R_{ij}}{N(N-1)}\,  \hspace{1cm}
T_i = \frac{\sum_j R_{ij}}{N-2} - \frac{\sum_{jk} R_{jk}}{N(N-2)} \, \hspace{1cm}
S_{ij} = \left\{ \begin{array}{cl} R_{ij} - T_i - T_j - U & \mbox{if $i\ne j$} \\
         0 & \mbox{if $i=j$} \end{array} \right.
\ee
This is the unique decomposition $R_{ij} = S_{ij} + T_i + T_j + U$ (for $i\ne j$)
satisfying $\sum_i T_i = \sum_i S_{ij} = S_{ii} = 0$.
In terms of the new variables $S,T,U$, a long computer algebra assisted calculation
gives the following alternate forms for $\hF_V$ and $\hSigma$:
\ba
\hF_V &=& \frac{\sum_i T_i^2}{N-1} - \frac{\sum_{ij} S_{ij}^2}{N(N-2)(N-3)} \nn \\
\hSigma &=& 
 \bigg( \frac{(N-1)^2}{N_4^2} - \frac{N^2-9N+26}{N_8} \bigg) \bigg( \sum_{ij} S_{ij}^2 \bigg)^2
    + 6 \frac{N-3}{N_6} \bigg( \sum_{ij} S_{ij}^2 \bigg) \bigg( \sum_i T_i^2 \bigg)
    - \frac{N^2-3}{(N-1)N_4} \bigg( \sum_i T_i^2 \bigg)^2 \nn \\
  && 
          + \frac{13N^2-73N+162}{N_8} \bigg( \sum_{ijk} S_{ij}^2 S_{jk}^2 \bigg)
          + 4\frac{3N^2-7N-2}{N_8} \bigg( \sum_{ijk} S_{ij}^2 S_{ik} S_{jk} \bigg)
          - 2\frac{5N^2-17N+18}{N_8} \bigg( \sum_{ij} S_{ij}^4 \bigg) \nn \\
  &&
          - 2\frac{N^2-5N+10}{N_8} \bigg( \sum_{ijkl} S_{ij} S_{jk} S_{kl} S_{il} \bigg)
          + 16\frac{N(N-2)}{N_7} \bigg( \sum_{ij} S_{ij}^3 T_j \bigg)
          - 12\frac{N^2-5N+10}{N_7} \bigg( \sum_{ijk} S_{ij}^2 S_{jk} T_k \bigg) \nn \\
  &&
          - 8\frac{N(N-2)}{N_7} \bigg( \sum_{ijk} S_{ij} S_{jk} S_{ik} T_k \bigg)
          + 4\frac{N^2-4N+7}{N_6} \bigg( \sum_{ijk} S_{ij} S_{jk} T_i T_k \bigg)
          - 2\frac{5N^2-13N+12}{N_6} \bigg( \sum_{ij} S_{ij}^2 T_j^2 \bigg) \nn \\
  &&
          - 4\frac{(N+3)(N-2)}{N_6} \bigg( \sum_{ij} S_{ij}^2 T_i T_j \bigg)
	  + 4\frac{N(N-1)}{N_5} \bigg( \sum_{ij} S_{ij} T_i T_j^2 \bigg)
          + \frac{N(N-1)}{N_4} \bigg( \sum_i T_i^4 \bigg)  \label{eq:fast_FV}
\ea

\end{document}